\documentclass[12pt]{article}
\usepackage{mathtools}
\usepackage{natbib}
\usepackage{graphics}
\usepackage{graphicx}
\usepackage{setspace} 
\usepackage[portrait, margin=1.25in]{geometry}
\usepackage{amsmath}  
\usepackage{amsfonts}
\usepackage{indentfirst}
\usepackage{array}
\usepackage{booktabs}
\usepackage{multirow}
\usepackage{booktabs}
\usepackage{tabularx}
\usepackage{tabulary}
\usepackage{dcolumn}
\usepackage{tabularht}
\usepackage{verbatim}
\usepackage{graphicx}
\usepackage{natbib}
\usepackage{booktabs,caption}
\usepackage[flushleft]{threeparttable}
\usepackage{enumitem}   
\usepackage{placeins}
\usepackage{authblk}
\usepackage{subcaption}
\usepackage{relsize}

\usepackage{xcolor}

\definecolor{dark-maroon}{HTML}{5D0F0D}
\definecolor{navyblue}{HTML}{0A2044}

\definecolor{purple}{HTML}{5601A3}
\definecolor{navy}{HTML}{0D3D56}
\definecolor{ruby}{HTML}{9a2515}
\definecolor{alice}{HTML}{107895}
\definecolor{daisy}{HTML}{EBC944}
\definecolor{coral}{HTML}{F26D21}
\definecolor{kelly}{HTML}{829356}
\definecolor{cranberry}{HTML}{E64173}
\definecolor{jet}{HTML}{131516}
\definecolor{asher}{HTML}{555F61}
\definecolor{slate}{HTML}{314F4F}

\usepackage{hyperref}
\hypersetup{
    colorlinks= true,
    citecolor= dark-maroon,
    linkcolor= dark-maroon,
    filecolor= dark-maroon,      
    urlcolor= dark-maroon,
}
\makeatletter
\DeclareRobustCommand{\pdot}{\mathbin{\mathpalette\pdot@\relax}}
\newcommand{\pdot@}[2]{%
  \ooalign{%
    $\m@th#1\circ$\cr
    \hidewidth$\m@th#1\cdot$\hidewidth\cr
  }%
}
\makeatother

\DeclareMathOperator*{\argmin}{arg\,min}

\begin{document}

\onehalfspacing
\title{Orthogonalized Synthetic Controls}
\author[$\dagger$]{\footnotesize Joseph Fry\footnote{Rutgers University, 75 Hamilton St, 08901, New Brunswick, New Jersey, United States. \par\nopagebreak   \textit{E-mail address:} \texttt{jf1372@economics.rutgers.edu}. I would like to thank Adam McCloskey, Carlos Martins-Filho, and Xiaodong Liu for helpful comments and feedback.}}
  
\date{\today}
\maketitle

\begin{small}
    
\begin{abstract}\footnotesize
    When conducting inference for the average treatment effect on the treated with a Synthetic Control Estimator, the vector of control weights is a nuisance parameter that is often constrained, high-dimensional, and may be only partially identified even when the average treatment effect on the treated is point-identified. All three of these features of a nuisance parameter can lead to failure of asymptotic normality for the estimate of the parameter of interest when using standard methods. I provide a new method that yields asymptotic normality for an estimate of average treatment effects, even when all three complications are present. This is accomplished by first estimating the control weights and any other nuisance parameters using a regularization penalty to achieve identification, and then estimating average treatment effects using moment conditions that are orthogonalized with respect to the nuisance parameters. Additionally, I extend results from the fixed-smoothing literature to provide tests that control size without requiring consistent standard errors. I present high-level sufficient conditions applicable to the traditional Synthetic Control Estimator as well as other weighting-based panel data methods, and verify them in an example involving instrumental variables. 
\end{abstract}
\end{small}

\section{Introduction}\label{s: Intro}

\cite{AandG} and \cite{Abadie2010} introduced the Synthetic Control Estimator (SCE) for estimating treatment effects when there is a single unit that becomes and stays treated and a large pool of never-treated units that can be used as control. Since then, it has been expanded to other settings and a number of modified versions of the method have been proposed, although they generally involve a first step where weights, and potentially other nuisance parameters, are estimated and then a second step where these are used to estimate averages of treatment effects, such as the average treatment effect on the treated (ATT). In the original setting with a single treated unit, the first step is to construct a weighted average of the control units, or a Synthetic Control (SC) unit, and the second step involves taking the difference between the SC and the treated unit's outcomes in the time periods after treatment. However, an important challenge to conducting inference for the ATT is that the vector of control weights is a nuisance parameter, which is constrained, high-dimensional, and may be partially identified even when the ATT is point-identified. In the context of method of moments estimation, if the moment conditions jointly identify a subvector of the parameters, then a standard estimation method such as General Method of Moments (GMM) is generally a consistent estimator for that subvector. However, if the remaining subvector is not point-identified, then this remaining parameter cannot be consistently estimated. As a result, the estimates for the identified subvector (e.g., the ATT) are not asymptotically normal (see, for example, \cite{AndrewsCheng2012}), which significantly complicates inference. Additionally, if a nuisance parameter is at the boundary of the parameter space, then this can also result in our estimates for the parameter of interest not being asymptotically normal (see, for example, \cite{Andrews1999} and \cite{ConstrainedMEstimation}). Lastly, if the full vector is high-dimensional, this can also complicate standard asymptotic normality results even when the subvector we would like to perform inference on is low-dimensional. Given a set of moment conditions that identify a vector of averages treatment effects such as the ATT, I propose what I refer to as the Orthogonalized Synthetic Control Estimator, that aims to overcome these complications simultaneously to obtain an estimate for an identified parameter of interest that is asymptotically normal even when a nuisance parameter is partially identified, on the boundary of the parameter space, and high-dimensional.
\par 
Several papers in the literature have provided moment conditions that can point-identify the ATT. I use a version of the method that uses linear instrumental variables moment conditions, as in  \cite{fry2024} and \cite{SyntheticIV}, as my working example. However, the formal results are general enough to cover cases where we would like to conduct inference for a different point-identified low-dimensional vector of average treatment effects and where the nuisance parameter vector includes multiple sets of weights to cover other cases, such as the staggered treatment adoption settings and cases where time periods are also weighted as discussed in section \ref{s: extensions}. In the first step of the procedure, regularized estimates of parameters are found by minimizing a penalty function subject to the constraint that the sample moment conditions are close to zero. The primary purpose of this penalty function is to make the estimated nuisance parameter converge to a unique element of the identified set, although there are other considerations in choosing the penalty which are discussed in section \ref{s: Regularized Estimator}. The second step is to use Neyman orthogonalization to construct a set of moment conditions that are orthogonal with respect to the nuisance parameter. This involves introducing a new parameter to take linear combinations of the moment conditions, which will be approximately orthogonal to the original nuisance parameter. After the nuisance parameters are plugged into the orthogonal moment conditions, a method of moments estimator yields asymptotically normal estimates of the ATT or other parameter of interest. 
\par
 Several other inference methods have been proposed for SCEs including the placebo method of \cite{Abadie2010}, the subsampling method of \cite{Li2020}, the cross-fitting t-test method of \cite{t-test2022}, the conformal inference method of \cite{ConformalInference}, and the End-of-Sample Instability Test, originally introduced by \cite{Andrews2003} and applied to SCE by \cite{CaoDowd2019}. Additionally, \cite{CARVALHO2018} and \cite{SyntheticIV} both provide asymptotic normality for estimates of the ATT using their versions of the SCE. \cite{Arkhangelskyetal2021} introduce the Synthetic Difference-in-Differences (SDID) estimator, which is a Difference-in-Differences style method that involves weighting across both control units and pre-treatment time periods. They establish consistency and asymptotic normality of the estimated ATT. However, the results of \cite{Arkhangelskyetal2021} and \cite{SyntheticIV} rely on the Euclidean norm of the control weights converging to zero at a sufficiently fast rate, and \cite{CARVALHO2018} and \cite{t-test2022} impose conditions for their estimators to be asymptotically unbiased that I do not. In sections \ref{s: EmpiricalApplication} and \ref{s: extensions}, I discuss these alternatives further, as well as compare my method with many of these approaches in simulations and show that it usually controls size and has high power relative to the other methods that control size.
\par
This work is also connected to the broader literature on method of moments estimation with panel data, including seminal work by \cite{Chamberlain1992} and \cite{Holtz-EakinNewey1988}, as well as advancements by \cite{AHN2001}, \cite{ChamberlainMorerira2009}, and others. Work on the fixed effect estimation by \cite{Freyberger2017}, \cite{Lancaster2002}, \cite{Hahn2011}, and \cite{Bester2009} is conceptually related, particularly when estimated fixed effects are inconsistent. Some of these papers employ related solutions, such as penalizing nuisance parameters and orthogonalizing estimates with respect to nuisance parameters, to address the inconsistency of these estimates. However, here the potential inconsistency in nuisance parameter estimates arises from them being partially identified when the outcomes have a low-rank factor structure, rather than the incidental parameters problem. A large body of work considers inference in cases where a set of moment equations only partially identifies a vector of parameters (e.g., \cite{ChernozhukovHongTamer2007} and \cite{RomanoShaikh2010}). \cite{Hansen1996} offers a method for inference when a nuisance parameter is unidentified under the null hypothesis, but it requires simulating the sampling distribution of the estimated nuisance parameter. \cite{CHAUDHURI2011} provide an inference method for GMM estimators when a nuisance parameter may be weakly identified, and \cite{AndrewsCheng2012} propose an inference method for extremum estimators when a subvector may be weakly identified, which is generalized by \cite{HanMcCloskey2019}. However, these methods are generally not suited to SC applications, as they do not allow the parameter vector to be high-dimensional. Also, when only the nuisance parameters are partially identified, we can focus on conducting the estimation in a way that allows for inference to be conducted simply such as using t-test and F-tests. On the other hand, methods that conduct inference on the whole identified set have the advantage that they can be used when both the parameter of interest and the nuisance parameter are partially identified.
\par 
This work also extends the literature on the Neyman Orthogonalized Score. The technique has a long history dating back to \cite{Neyman1959}, and several recent papers have used it as a way to achieve asymptotic normality after obtaining an estimate of a high-dimensional nuisance parameter using a regularization penalty or machine learning technique (e.g., \cite{belloni2018highdimensional}, \cite{NingLiu2017}, \cite{ChernozhukovHansenSpindler2015}, \cite{BelloniChernozhukovHansen2014}, and \cite{ChernozhukovDoubleDebiased}). Many of these methods estimate the nuisance parameter with LASSO. While this can be a powerful technique when the nuisance parameter is high-dimensional but has a sparse, point-identified value, if the nuisance parameter is partially identified, the LASSO penalty may often be insufficient for the estimates to converge to a specific vector. I show in section \ref{s: Regularized Estimator} that combining linear constraints with a different penalty can be a method to make use of sparsity without relying on point-identification. This helps extend the literature by showing how the Neyman Orthogonalized Score can be applied in cases where the nuisance parameter is partially identified, and I also give results that combine this Neyman Orthogonality with the fixed-smoothing asymptotics literature in section \ref{s: VarianceEstimation}. \cite{ChernozhukovDoubleDebiased}'s Neyman Near-Orthogonal Score is most closely related to my approach, as they use a regularized estimate of the parameter that is introduced to orthogonalize the moment conditions and choose its value so that the moment conditions are nearly orthogonal. 
\par 
In section \ref{s: Orthogonalization}, I discuss the construction and use of the orthogonalized moment conditions. I discuss this first to show what properties we want the regularized estimates of the nuisance parameters to satisfy. I then show when the regularized estimates of parameters satisfy these conditions in section \ref{s: Regularized Estimator}, and I show in section \ref{s: VarianceEstimation} how to conduct valid inference even when the variance cannot be consistently estimated. In section \ref{s: EmpiricalApplication}, I apply my method by extending the work of \cite{Andersson2019} who estimates the effect of Sweden's carbon pricing policies on CO2 emissions. My method finds their results statistically significant at commonly used levels, whereas several other methods do not. I also conduct simulations to compare the performance of my inference method with existing approaches for SCEs. Lastly, in section \ref{s: extensions} I discuss other implementations of the method, including a version for a staggered adoption setting, and its connection to related estimators.

\section{Identification and Orthogonalization}\label{s: Orthogonalization}

\subsection{Neyman Orthogonalization}

I assume that the researcher has a set of moment conditions they wish to use that are a function of a vector of parameters $\theta = (\beta, \delta)$ where the subvector $\beta \in B \subseteq \mathbb{R}^p$ is the parameter of interest and $\delta \in D_n \subseteq \mathbb{R}^J$ is a nuisance parameter. Researchers are most commonly interested in conducting inference on the ATT and the weights on the control units are a nuisance parameter, in which case $\beta$ will be the ATT and $\delta$ will include all the control weights and any other nuisance parameters that appear in the moment conditions. In formal results for SCE, the number of time periods is usually allowed to grow to infinity. However, there are relevant distinctions in the sizes of different groups of time periods (e.g., both the number of time periods before and after treatment for a given unit may be relevant). To be flexible on this issue for my general asymptotic results, I assume there is a sequence of panel data sets with the number of time periods diverging to infinity, where the time periods in each panel are partitioned into a fixed number of blocks indexed starting at zero $b = 0,1,...$ with indices for time periods in the $b$th block in $\mathcal{T}_b$. Each moment condition involves taking a sample average over one of these blocks, and for the $q$th moment condition, I use $b_q$ to denote the block of time periods it averages over. I let  $n$ denote the minimum number of time periods across these blocks and focus on an asymptotic regime where $n \rightarrow \infty$, meaning that the number of time periods in each block is becoming large. For the example discussed in subsection \ref{s: SC}, a single unit is untreated for $|\mathcal{T}_0| = T_0$ time periods and then treated for $|\mathcal{T}_1| = T_1$ time periods, and the method will use moment conditions that take averages of the sets of pre-treatment and post-treatment time periods. Therefore, $q_b = 0$ for the moment conditions using pretreatment time periods and $q_b =1$ for moment conditions using post-treatment and 
$n$ will be equal to $\min \{T_0, T_1\}$ and both $T_0$ and $T_1$ will need to be growing for the asymptotic results to hold. Breaking the data into two blocks of time periods is a natural choice in the traditional SC setting, but allowing for more blocks lets the formal results cover other methods, like the cross-fitting method of \cite{t-test2022}, and other treatment regimes such as cases with staggered adoption as discussed in section \ref{s: extensions}.
\par 
I let $\Theta_n = B \times D_n$ equal the parameter space for the entire vector. I use the subscript $n$ to indicate that the parameter space for $\delta$ may change as the sample size grows. This is particularly relevant when $\delta$ is high-dimensional, so $J$ can grow with $n$. Then I let $g(\theta)$ and $\hat{g}(\theta)$ denote $Q$-dimensional vectors for the population and sample moment conditions,\footnote{Note that since I wish to allow for cases where observations are not identically distributed, I allow $g$ to change with the sample size as well but omit the subscript $n$ to avoid notational confusion.} meaning 

$$\hat{g}(\theta) 
 = 
 \begin{pmatrix}
    \sum_{t \in \mathcal{T}_{b_1}} g_{1,t}(\theta)/T_{b_1} \\
    ... \\
    \sum_{t \in \mathcal{T}_{b_{Q-1}}} g_{Q-1,t}(\theta)/T_{b_{Q-1}} \\
     \sum_{t \in \mathcal{T}_{b_{Q}}} g_{Q,t}(\theta)/T_{b_{Q}}
 \end{pmatrix} \text{ and } g(\theta) 
 = 
 \begin{pmatrix}
    E[\sum_{t \in \mathcal{T}_{b_1}} g_{1,t}(\theta)/T_{b_1}] \\
    ... \\
   E[ \sum_{t \in \mathcal{T}_{b_{Q-1}}} g_{Q-1,t}(\theta)/T_{b_{Q-1}}] \\
    E[ \sum_{t \in \mathcal{T}_{b_{Q}}} g_{Q,t}(\theta)/T_{b_{Q}}]
 \end{pmatrix}.$$
\par 
I am interested in the case where these moment conditions jointly identify the parameter of interest $\beta$, but may only partially identify $\delta$. The true value of the parameter of interest and the identified set of the nuisance parameter may also change with the sample size, so I denote the identified set as
\begin{equation}\label{Identification of beta}
       \Theta_{0,n} \coloneqq \{\theta \in \Theta_n: g(\theta) = 0\} = \{\beta_{0,n} \} \times D_{0,n},
\end{equation}
where $\beta_{0,n}$ is the true value of $\beta$ and $D_{0,n}$ is the identified set for $\delta$. Since $\beta_{0,n}$ generally corresponds to the average of a sequence of dynamic treatment effects, we want to allow that average to change as our sample size changes. Using this set of the moment conditions, the orthogonalized moment conditions are given by
\begin{equation} \phantomsection\label{eq: OrthogonalizedScore}
    M(\theta, \eta) = \eta g(\theta),
\end{equation}
where $\eta \in H \subseteq \mathbb{R}^{m \times Q}$ is an additional nuisance parameter and $m$ is the number of orthogonalized moment conditions.\footnote{I focus on cases where $Q$ is fixed. When the number of moment conditions is also large so that $vec(\eta)$ is high-dimensional, there may be a benefit to constraining $\eta$, similarly to the benefits of constraining $\delta$ when it is high-dimensional as discussed in Remark \hyperref[R3.1]{3.1} below.} Because the orthogonalized moments are linear combinations of the original moment conditions, there is no reason to choose $m > Q$. Furthermore, if the $q$-th element of the original moment conditions $g_q(\theta)$ does not vary with $\beta$ at all, then $g_q(\beta,\delta_{0,n}) = 0$ for any $\delta_{0,n} \in D_{0,n}$.
Therefore, if $m$ is greater than the number of elements of $g$ which are nontrivial functions of $\beta$, the elements of $M(\beta,\delta_{0,n},\eta) = \eta g(\beta,\delta_{0,n})$ are linearly dependent functions of the elements of $g(\beta,\delta_{0,n})$, for any fixed values of $\eta$ and $\delta_{0,n} \in D_{0,n}$. This is relevant because the orthogonalized moment conditions are used to estimate $\beta$ after plugging in values of the nuisance parameters.  
\par 
The general procedure for the Orthogonalized SCE can be viewed as three steps. The first step in the estimation procedure is to obtain initial parameter estimates by picking values that minimize a penalty function among all values that make the sample moment conditions close to zero, as I discuss in section \ref{s: Regularized Estimator}:
$$(\hat{\beta},\hat{\delta},\hat{\eta}) = \argmin_{\beta, \delta \in D_n, \eta \in H} \hat{f}(\beta,\delta,\eta),$$
$$|| \hat{g}(\beta,\delta)||_\infty \leq \lambda_\delta, \quad ||\partial_\delta \eta \hat{g}(\beta,\delta)||_\infty \leq \lambda_\eta,$$
where $\hat{f}$ is some (possibly stochastic) penalty function, $\hat{g}$ are the sample versions of some moment conditions which identify $\beta$, and $\lambda_\delta$ and $\lambda_\eta$ are tuning parameters. After obtaining estimates of the nuisance parameters $\hat{\delta}$ and $\hat{\eta}$, they are plugged into the sample orthogonalized moments where
\begin{equation}\phantomsection\label{eq: SampleOrthogonalizedScore}
    \hat{M}(\beta,\hat{\delta},\hat{\eta}) =  \hat{\eta}\hat{g}(\beta,\hat{\delta}) .
\end{equation}
Lastly, the estimate of $\beta_{0,n}$ is obtained setting $\hat{M}(\beta,\hat{\delta},\hat{\eta})$ equal to zero when $m = p$, or using a GMM estimator:
$$\Tilde{\beta} = \argmin_{\beta} \hat{M}(\beta,\hat{\delta},\hat{\eta})'W_n \hat{M}(\beta,\hat{\delta},\hat{\eta}),$$
for some moment weighting matrix $W_n$. 
\par
One property we want $\eta_{0,n}$ to satisfy is for $M(\beta,\delta_{0,n},\eta_{0,n})$ to identify $\beta$, meaning that we will want $ \partial_\beta M(\theta_{0,n},\eta_{0,n}) = \eta_{0,n} \partial_\beta g(\theta_{0,n})$ to have rank $p$. One approach to guaranteeing this is to have the penalty function be an upper bound on the asymptotic variance of $\Tilde{\beta}$, since this expression diverges when the parameters approach values where $\beta$ is not identified by the orthogonal moment conditions. However, in all the applications discussed in subsection \ref{s: SC} and section \ref{s: extensions}, $\partial_\beta \hat{g}(\theta) = \partial_\beta g(\theta)$ and it does not vary with $\theta$. In these cases, it will simplify the method computationally to let $m = p$ and constrain $\eta$ so that $\eta_{0,n} \partial_\beta g(\theta_{0,n})$ is proportional to the identity matrix, such as letting
$$H = \{ \eta \in \mathbb{R}^{p \times Q}: \partial_\beta \eta  \hat{g}(\theta_{0,n})  - I_p = 0 \}.$$
As I illustrate in subsection \ref{sub: IV-SCE continued}, this directly imposes that identification of $\beta$ is preserved and simplifies the upper bound on the asymptotic variance of $\Tilde{\beta}$ being minimized.
\par
Let $\hat{\theta} = (\hat{\beta},\hat{\delta})$ and $\hat{\eta}$ be estimates from the first step, and suppose that the distance from our estimates to the values $\theta_{0,n} = (\beta_{0,n},\delta_{0,n})$ and $\eta_{0,n}$ is converging to zero as $n \rightarrow \infty$.\footnote{Since the dimension of $\delta$ may be growing, which metrics this holds under is key for the results below. Assumption \hyperref[Ass2.1]{2.1} contains the details on exactly which metrics this convergence must hold.} We want to choose $\eta_{0,n}$ so that $M(\theta_{0,n},\eta_{0,n})$ is insensitive to $\delta$. More precisely, we want the sequence of matrices $\eta_{0,n}$ to satisfy  
\begin{equation}\phantomsection\label{eq: Orthogonality Condition}
    \partial_\delta M(\theta_{0,n}, \eta_{0,n}) = 0. 
\end{equation}
Generally, there may exist many choices of $\eta$ that satisfy this condition, particularly if $rank(\partial_\delta g(\theta_{0,n})) < Q$, which may happen when $\delta$ is partially identified. Therefore, $\eta$ can also be a partially identified nuisance parameter. By construction, $\partial_{vec(\eta)} M(\theta_{0,n},\eta) = 0$ for any $\theta_{0,n} \in \Theta_{0,n}$ because $g(\theta_{0,n}) = 0$. Hence, $M(\theta_{0,n},\eta)$ is also orthogonal with respect to $\eta$. This allows $\eta$ to be estimated in the same manner as $\delta$, where the regularization penalty $\hat{f}(\theta,\eta)$ is used to make $(\hat{\theta},\hat{\eta})$ converge to a specific element of the set 
$$(\theta_{0,n},\eta_{0,n}) \in S_{0,n} := \{ (\theta,\eta) \in \Theta_{0,n} \times H : \partial_{\delta} M(\theta,\eta) = 0 \}.$$
\par 
Because of equation \eqref{eq: Orthogonality Condition}, under suitable conditions, the sample moment conditions are not sensitive to the values of the nuisance parameters when $\beta = \beta_{0,n}$ and $(\hat{\delta},\hat{\eta})$ is close to $(\delta_{0,n},\eta_{0,n})$. The assumption below gives high-level conditions under which $\hat{M}(\beta_{0,n}, \hat{\delta},\hat{\eta})$ is asymptotically equivalent to $\hat{M}(\beta_{0,n},\delta_{0,n},\eta_{0,n})$.

\textbf{Assumption 2.1}\phantomsection\label{Ass2.1} 
Let $(||\cdot||_{\mathcal{E}}, ||\cdot||_{\mathcal{D}})$ be a pair of norms such that $|x^T y| \leq ||x||_{\mathcal{E}} ||y||_{\mathcal{D}}$, specifically either $(||\cdot||_2, ||\cdot||_2)$ or $(||\cdot||_1, ||\cdot||_\infty)$. Let $r_J$ be a sequence representing the corresponding dimensional growth rate, where $r_J = \sqrt{J}$ for the $(L_2, L_2)$ norm pair and $r_J = \log(J)$ for the $(L_1, L_\infty)$ norm pair. As $n \rightarrow \infty$ while $p$ and $Q$ are fixed and either $J$ is fixed or $J \rightarrow \infty$, we have that:
\begin{enumerate}
    \item $||\hat{\delta} - \delta_{0,n}||_{\mathcal{E}} + ||\hat{\eta} - \eta_{0,n}||_{\mathcal{E}} = o_p\left(\frac{1}{r_J\log(n)}\right)$.\footnote{I use $||\cdot||_1$ to denote the element-wise $L_1$ norm and $||\cdot||_\infty$ to denote the element-wise $L^\infty$ norm. I use $||\cdot||_2$ to denote both the $L_2$ norm for vectors and the Frobenius norm for matrices.}
    \item $g(\theta)$ is twice continuously differentiable on $\Theta_n$ and for each $q \in \{1,...,Q\}$, 
    $||\partial_\delta g_q (\theta_{0,n})||_{\mathcal{D}} = O(r_J)$ and 
    $||\partial_\delta \hat{g}_q(\theta_{0,n}) - \partial_\delta g_q(\theta_{0,n})||_{\mathcal{D}} = O_p(r_J/\sqrt{n})$ where $\hat{g}_q$ is the $q$-th element of $\hat{g}$.
    \item There exists $\epsilon > 0$ such that $\sup_{\delta: ||\delta - \delta_{0,n}||_{\mathcal{E}} \leq \epsilon} \max eig(\partial_\delta^2 \hat{g}_{q}(\beta_{0,n},\delta)) = O_p(r_J) $ for each $q \in \{1,...,Q\}$, where $\hat{g}_q$ is the $q$-th element of $\hat{g}$ and $\max eig$ denotes the maximum eigenvalue of a matrix.
    \item Either $||\hat{\delta} - \delta_{0,n}||_{\mathcal{E}} = o_p\left(\frac{n^{-1/4}}{\sqrt{r_J}}\right)$ and $||\hat{\eta} - \eta_{0,n}||_{\mathcal{E}} = o_p\left(\frac{n^{-1/4}}{\sqrt{r_J}}\right)$ or $\hat{g}$ is linear in $\theta$ and $|| \hat{\eta}\partial_\delta \hat{g}(\theta)||_{\mathcal{D}} = O_p\left(\frac{r_J\log(n)}{\sqrt{n}}\right)$.
\end{enumerate}

The assumption is phrased to cover two separate cases: one using the $L_2$ norm where the control weights are allowed to be dense but $J$ is smaller than $n$, and one using the $L_1$ and $L_\infty$ norms where sparsity of the control weights will be needed to achieve the $L_1$ rates of convergence but $J$ may be growing much faster than $n$.\footnote{I show in Appendix B how Assumption \hyperref[Ass2.1]{2.1.2} holds with $r_J = \log (J)$ and $\mathcal{D} = \infty$ if the observations of the moment conditions are a triangular array that is $\alpha$-mixing with exponential speed, is mean-invariant, has uniformly bounded fourth moments, and has an exponential-type bound on the tails of their distributions, and if $J/n^\gamma \rightarrow 0$ for some $\gamma > 0$. This allows for cases where there is a significant degree of dependence across the observations and $J$ grows faster than $n$.}  Assumption \hyperref[Ass2.1]{2.1.3} imposes that there is a bound on how locally convex or concave $g$ is at $\theta_{0,n}$, which holds trivially when $g$ is linear in $\theta$. Assumption \hyperref[Ass2.1]{2.1.1} imposes that distances between $\hat{\delta}$ and $\delta_{0,n}$ as well as between $\hat{\eta}$ and $\eta_{0,n}$ are converging to zero at the given rates. For $\hat{\delta}$, the faster its dimension is growing, the faster its rate of convergence must be. Furthermore, an additional condition on its rate of convergence must be imposed when the moment conditions are a non-linear function of $\delta$, as shown in Assumption \hyperref[Ass2.1]{2.1.4}.\footnote{The reason why weaker conditions are needed when $\hat{g}$ is linear in $\delta$ is because in this case, making $\partial_\delta \hat{M}(\beta_{0,n},\delta_{0,n},\eta_{0,n})$ close to zero makes $\partial_\delta \hat{M}(\beta_{0,n}, \delta,\eta_{0,n})$ close to zero for any $\delta$. For nonlinear settings where it is hard to achieve a rate of convergence for the nuisance parameters that is faster than $n^{-1/4}$, \cite{mackey18} shows that making the moment conditions $h$-th order orthogonal can allow this condition to be weakened to $o_p(n^{-1/(2h+2)})$.}  I show how to obtain $\hat{\delta}$ and $\hat{\eta}$ so that Assumptions \hyperref[Ass2.1]{2.1.1} and \hyperref[Ass2.1]{2.1.4} are satisfied under plausible conditions in section \ref{s: Regularized Estimator}. Together with the orthogonality condition, this gives an adaptivity condition.

\textbf{Lemma 2.1 (Adaptivity Condition)}\phantomsection\label{L2.1} Suppose $(\beta_{0,n},\delta_{0,n}) \in  \Theta_{0,n}$, $\eta_{0,n}$ satisfies the orthogonality condition, and Assumption \hyperref[Ass2.1]{2.1} holds. Then as $n \rightarrow \infty$ with $p$ and $Q$ fixed with either $J$ fixed or $J \rightarrow \infty$,
\begin{equation}\label{eq: Adaptivity Condition}
    \sqrt{n}(\hat{M}(\beta_{0,n}, \hat{\delta},\hat{\eta}) - \hat{M}(\beta_{0,n},\delta_{0,n},\eta_{0,n})) = o_p(1).
\end{equation}

This adaptivity condition is useful for showing that an estimator of $\beta$ using $\hat{M}(\beta, \hat{\delta},\hat{\eta})$ is asymptotically equivalent to an estimator using $\hat{M}(\beta,\delta_{0,n},\eta_{0,n})$.

\subsection{IV Synthetic Control Example}\label{s: SC}
 
I now discuss how to implement the SCE using a set of instrumental variable moment conditions that identify the ATT when the data follow a linear factor model. Linear factor models, also known as interactive fixed-effect models, have been a common setting for exploring the properties of SCEs, beginning with \cite{Abadie2010}. I first focus on analyzing my method in this original context, where there is a single unit that becomes and stays treated. I index the units $\{0, 1,...,J\}$ where $i = 0$ is the treated unit and $\mathcal{J} = \{1,...,J\}$ denotes the set of control units. The control units never receive treatment, whereas the treated unit becomes and stays treated after a known point in time. Following the notation setup above, I denote the sets of indices for time periods prior to its treatment and after its treatment as $\mathcal{T}_0$ with $T_0 = |\mathcal{T}_0|$ and $\mathcal{T}_1$ with $T_1 = |\mathcal{T}_1|$ respectively.

\textbf{Condition 1 (Linear Factor Model)}\phantomsection\label{Ass1}  For all units $i \in \{0,...,J\}$ and time periods $t \in \mathcal{T}_0\bigcup \mathcal{T}_1$, outcomes follow a linear factor model with $R$ factors so that

\begin{equation*}
    Y_{it} = \beta_t d_{it} + f_t \mu_i  + \epsilon_{it}, 
\end{equation*}

where $d_{it}$ is an indicator function equal to 1 if and only if $i = 0$ and $t \in \mathcal{T}_1$ and equal to $0$ otherwise. The factor loadings $\mu_i$, dynamic treatment effects $\beta_t$, and treatment assignment $d_{it}$ are fixed, but the latent factors $f_t$ and idiosyncratic shocks $\epsilon_{it}$ are stochastic.
\par 
I let $\mu$ denote the $R \times (J+1)$ matrix of factor loadings with $\mu_i$ being its $i$-th column, $f$ denotes the $(T_0+T_1) \times R$ matrix of realizations of the factors with $f_t$ being is $t$-th row, and $\epsilon$ denote the $(J+1) \times (T_0 + T_1)$ matrix of idiosyncratic shocks. Additionally, I use the subscript $\mathcal{J}$ to denote the sub-matrix for only the units $j \in \mathcal{J}$ and the superscripts $pre$ and $post$ to denote the sub-matrices for only pre-treatment and post-treatment values respectively. I define the average treatment effect on the treated to be $\beta_{0,n} = \sum_{t \in \mathcal{T}_1} \beta_t /T_1$. If the idiosyncratic shocks are mean-zero and the SC has the same factor loadings as the treated unit (i.e., $\mu_0 = \mu_{\mathcal{J}}\delta$), then we can identify $\beta_{0,n}$ using the moment condition $\sum_{t \in \mathcal{T}_1} E[Y_{0t} - \beta - Y_{\mathcal{J},t}'\delta]/T_1 = 0$. Therefore, we want to use this moment condition plus a set of moment conditions that identify the set $D_{0,n} = \{\delta \in \Delta^J: \mu_0 = \mu_{\mathcal{J}}\delta \}$, where $\Delta^J \coloneqq \{\delta \geq 0 : ||\delta||_1 = 1 \}$ is the $J-1$-dimensional unit simplex.
\par 
Not only may there be many $\delta \in D_{0,n}$ so $\delta$ is partially identified, but analysis of the asymptotic distribution of the control weights is also complicated by the fact that it is often high-dimensional (relative to $T_0$ and $T_1$). Lastly, even when $J$ is fixed and $\delta$ is point-identified, $\sqrt{T_0}(\hat{\delta} - \delta_{0,n})$ generally has a non-standard asymptotic distribution due to the constraints $\hat{\delta} \in \Delta^J$.\footnote{This is illustrated by \cite{Li2020} who use the method of \cite{Andrews1999} to characterize the asymptotic distribution of the estimated average treatment effect.} Other inference methods, like \cite{CaoDowd2019} and \cite{Li2020}, also have the limitations of assuming that the control weights are point-identified and treating the number of units as fixed in their asymptotic results. One exception is \cite{Zhang_2023} who obtain a $\sqrt{n}$-consistent estimator while allowing $\delta$ to be partially identified, and another is \cite{t-test2022} whose method is discussed further below in subsection \ref{sub: Simulations}.
\par 

Several recent papers have discussed how to estimate the SC using a set of moment conditions, including \cite{fry2024}, \cite{Powell2021}, and \cite{shi2023}. A linear instrumental variables approach along the lines of \cite{fry2024}, \cite{shi2023}, or \cite{SyntheticIV} is the most straightforward to verify the conditions for. In this case, the moment conditions are
\begin{equation*}
     \hat{g}(\beta,\delta) = \begin{pmatrix}
    \sum_{t \in \mathcal{T}_0}Z_t(Y_{0t} - Y_{\mathcal{J},t}\delta)/T_0 \\
    \sum_{t \in \mathcal{T}_1} (Y_{0t} - \beta - Y_{\mathcal{J},t}\delta)/T_1
\end{pmatrix},
\end{equation*}
where $Z_t$ contains the values of $Q - 1$ instruments in pre-treatment time periods. In \cite{fry2024}, the vector of instruments is a constant and outcomes of other units that are not included in the set of controls but are also untreated in pre-treatment time periods. The exclusion restriction $E[Z_{qt}(Y_{0t} -  Y_{\mathcal{J},t}\delta)] = 0 $ holds for these other units in pre-treatment time periods when the idiosyncratic shocks are uncorrelated across units and the factors are uncorrelated with the idiosyncratic shocks. Intuitively, when the factors are responsible for the covariance across units, we can guarantee that the SC has the same exposure to latent factors as the treated unit by estimating the SC to have the same covariance with other units. In the empirical application of section \ref{s: EmpiricalApplication}, I provide a practical example for how the set of instruments can be chosen.\footnote{\cite{fry2024} also provides several model selection methods for splitting untreated units into a set of controls and set of instruments. However, because of the potential problems for inference that using a data-driven model selection procedure may introduce, I focus on cases where the units used as instruments are only the units that are known not to be valid controls but plausibly valid instruments. I discuss how this can be done in the empirical application below.} Other potentially valid choices of instruments exist, such as using lagged values of the outcome variable or using shift-share instruments like \cite{SyntheticIV}. Additionally, it may be possible to reframe other estimators, such as the Debiased OLS estimator of \cite{t-test2022} discussed below, as method of moments estimators, in which case the same orthogonalization technique could be employed.
\par 
Following my recommendation above, I set $m = 1$ so $\eta$ is $1 \times Q$, so there is only a single orthogonal moment condition, and I constrain the $Q$-th element of $\eta$ to equal one so that $\partial_\beta M(\theta,\eta) = -\eta_Q = -1$. Since there is only a single orthogonalized moment condition that is linear in $\beta$, we can obtain an estimate $\Tilde{\beta}$ simply by setting $\hat{M}(\beta,\hat{\delta},\hat{\eta})$ equal to zero. Therefore:
\begin{equation*}
    \Tilde{\beta} = \sum_{t \in \mathcal{T}_1}(Y_{0t} - Y_{\mathcal{J},t}\hat{\delta})/T_1 + \sum_{q=1}^{Q-1} \hat{\eta}_{q} \sum_{t \in \mathcal{T}_0} Z_{qt}(Y_{0t} - Y_{\mathcal{J},t}\hat{\delta})/T_0.
\end{equation*}
Note that for $\delta_{0,n} \in D_{0,n}$,
$$\hat{g}(\beta_{0,n},\delta_{0,n}) = \begin{pmatrix}
   \sum_{t \in \mathcal{T}_0} Z_t(\epsilon_{0t} - \epsilon_{\mathcal{J},t}\delta_{0,n})/T_0 \\
    \sum_{t \in \mathcal{T}_1}(\epsilon_{0t} - \epsilon_{\mathcal{J},t}\delta_{0,n})/T_1
\end{pmatrix}.$$
By verifying that the adaptivity condition holds in this case, we can show that
\begin{equation}\label{eq: IV-SCE Adaptivity}
\begin{split}
\sqrt{ \min \{T_0,T_1\}}(\Tilde{\beta} - \beta_{0,n}) = \sqrt{ \min \{T_0,T_1\}}\sum_{t \in \mathcal{T}_1}(\epsilon_{0t} - \epsilon_{\mathcal{J},t}\delta_{0,n,j})/T_1 \\\\
     + \sqrt{ \min \{T_0,T_1\}}\sum_{q=1}^{Q-1} \eta_{0,n,q} \sum_{t \in \mathcal{T}_0} Z_{qt}(\epsilon_{0t} - \epsilon_{\mathcal{J},t} \delta_{0,n,j})/T_0 + o_p(1).
\end{split}
\end{equation}
This means that the asymptotic variance of the ATT should only come from post-treatment averages of the idiosyncratic shocks and pre-treatment averages of the products of the instruments and the idiosyncratic shocks, with the relative contribution of these depending on the relative size of $T_0$ and $T_1$.

\section{Penalization Estimation of Weights}\label{s: Regularized Estimator}

I now discuss the estimation of the nuisance parameters. The adaptivity condition may hold for many possible values of $(\delta_{0,n},\eta_{0,n})$ where 
$$(\beta_{0,n},\delta_{0,n},\eta_{0,n}) \in S_{0,n} := \{ (\theta,\eta) \in \Theta_{0,n} \times H : \partial_{\delta} M(\theta,\eta) = 0 \}.$$
For some penalty function $f(\theta,\eta)$, I define the target nuisance parameters as
\begin{equation}\phantomsection\label{eq: OptimalNuisanceParameters}
    (\delta_{0,n},\eta_{0,n}) = \argmin_{(\beta_{0,n},\delta,\eta) \in S_{0,n}} f(\beta_{0,n},\delta,\eta). 
\end{equation}
The primary purpose of the penalty function is to select a unique pair of elements from the identified sets. Therefore, we want $f(\theta,\eta)$ to have a unique minimum on $S_{0,n}$. However, the penalty not only influences whether the estimated nuisance parameters converge to a particular element of the identified set, but also which element they converge to. One might hope to achieve a form of relative asymptotic efficiency by making the penalty function depend directly on the asymptotic variance of $\Tilde{\beta}$. However, given that it is challenging to consistently estimate the variance of $\Tilde{\beta}$ in this context, I do not focus on this approach and instead recommend choosing a penalty function that provides an upper bound on the asymptotic variance. I do allow for the possibility that the function $f(\theta,\eta)$ is unknown and must be estimated with the data.\footnote{Examples that fit this case are discussed in section \ref{s: extensions} and in Appendix B of the Supplementary Materials.} Since $S_{0,n}$ is also unknown, the nuisance parameters are chosen to minimize an estimated penalty function $\hat{f}$ among all parameters that come close to setting the sample versions of the moment conditions equal to zero. We can define the feasible set for the regularized parameters as 
$$\hat{S}_0 = \{ (\theta,\eta) \in \Theta_n \times H : ||\hat{g}(\theta)||_\infty \leq \lambda_\delta,\; ||\partial_\delta \hat{M}(\theta, \eta)||_\infty \leq \lambda_\eta \},$$
where $\lambda_\delta$ and $\lambda_\eta$ are tuning parameters whose choice is discussed below. Then we can estimate the parameters using

\begin{equation}\label{eq: regularized estimator}
    (\hat{\beta},\hat{\delta},\hat{\eta}) = \argmin_{(\theta,\eta) \in \hat{S}_0} \hat{f}(\theta,\eta).
\end{equation}

\subsection{Rate of Convergence}\phantomsection\label{sub: RateOfConvergence}

In the case for which $J$ is growing and $\hat{g}$ is linear in $\theta$, Assumption \hyperref[Ass2.1]{2.1} only requires that $||\hat{\delta} - \delta_{0,n}||_{\mathcal{E}} + ||\hat{\eta} - \eta_{0,n}||_{\mathcal{E}} = o_p(1/(r_J\log(n)))$ and $||\hat{\eta}\partial_\delta \hat{g}(\beta_{0,n},\delta)||_\infty = O_p(r_J \log(n) n^{-1/2}) $.  Since $||\hat{\eta}\partial_\delta \hat{g}(\hat{\theta})||_\infty \leq \lambda_\eta$,
in the linear case, the second condition can be directly achieved by choosing $\lambda_\eta$ to be $O_p(r_J \log(n) n^{-1/2}))$. 

\textbf{Assumption 3.1}\phantomsection\label{Ass3.1} Let $S_n^\zeta \coloneqq \{(\theta,\eta) \in \Theta_{n} \times H :f(\theta,\eta) \leq f(\theta_{0,n},\eta_{0,n}) + \zeta\}$ and $S_{0,n}^\zeta \coloneqq \{(\theta,\eta) \in S_{0,n} :f(\theta,\eta) \leq f(\theta_{0,n},\eta_{0,n}) + \zeta\}$ for each $\zeta > 0$. Suppose that
\begin{enumerate}
    \item $g$ and $ \hat{g} $ are continuously differentiable on $\Theta_n$. For all $\zeta > 0$, $S_n^\zeta$ and $S_{0,n}^\zeta$ are compact and convex and $f$ is continuous on $S_n^\zeta$ for all $n$. 
    \item For some sequences of positive constants $\{a_n\}_{n \in \mathbb{N}}$ and $\{b_n\}_{n \in \mathbb{N}}$, for each $q \in \{1,...,Q\}$,  $$\sup_{(\theta,\eta) \in S_n}|\hat{g}_q(\theta) - g_q(\theta) | = O_p(a_n),\; \sup_{(\theta,\eta) \in S_{0,n}}|\hat{g}_q(\theta) - g_q(\theta) | = O_p(b_n),$$ $$\sup_{(\theta,\eta) \in S_n}||\partial_\delta \hat{g}_q(\theta) - \partial_\delta g_q(\theta) ||_\infty = O_p(a_n),\; \sup_{(\theta,\eta) \in S_{0,n}}||\partial_\delta \hat{g}_q(\theta) - \partial_\delta g_q(\theta) ||_\infty = O_p(b_n).$$ 
    \item For some sequences of non-negative constants $\{c_n\}_{n \in \mathbb{N}}$ $$\sup_{(\theta,\eta) \in \hat{S}_0}|\hat{f}(\theta,\eta) - f(\theta,\eta)| = O_p(c_n).$$
    
    \item For all $\zeta > 0$, there exists constants $C_1, C_2 > 0$ such that for all $(\theta,\eta) \in S_n^\zeta$,\footnote{I use $||(\theta,\eta) - S_{0,n}||_2$ to denote   
    $\inf_{(\theta^*,\eta^*) \in S_{0,n}} ||\theta - \theta^*||_2 + ||\eta - \eta^*||_2$.}

    $$||g(\theta)||_\infty + ||\partial_\delta \eta g(\theta)||_\infty  \geq C_2 \min \{ ||(\theta,\eta) - S_{0,n}||_{\mathcal{E}} , C_1\}.$$

    \item There exists constants $C_3,C_4,C_5,C_6 >0$ and $\gamma_1,\gamma_2 > 0$, such that for all $(\theta,\eta) \in S_{0,n}$ $|f(\theta,\eta) - f(\theta_{0,n},\eta_{0,n})| \geq C_4 \min \{((||\theta - \theta_{0,n}||_{\mathcal{E}} + ||\eta - \eta_{0,n}||_{\mathcal{E}})^{\gamma_1},C_3\}$ and for all $(\theta_1,\eta_1),(\theta_2,\eta_2) \in S_n^{C_5}$, $||\theta_1 - \theta_2||_{\mathcal{E}} + ||\eta_1 - \eta_2||_{\mathcal{E}} \geq C_6 |f(\theta_1,\eta_1) - f(\theta_2,\eta_2)|^{\gamma_2}$.
\end{enumerate}

\par 
Assumption \hyperref[Ass3.1]{3.1} is phrased to hold with either $\mathcal{E} = 1$ or $\mathcal{E} = 2$, depending on whether we want to establish a rate of convergence for the nuisance parameters in the $L_1$ or $L_2$ norm. For the estimator defined by equation \eqref{eq: regularized estimator}, both the objective function and feasible set may be stochastic, so there can be uncertainty coming from both the estimated penalty function and the sample moment conditions. As a result, the rate of convergence for $\hat{\delta}$ and $\hat{\eta}$ depends both on the rate of convergence of the sample moments to the population moments $a_n$ and the rate of convergence of the estimated penalty function $c_n$. However, $f$ can be chosen to be a known function, in which case Assumption \hyperlink{Ass3.1}{3.1.3} holds trivially with $\hat{f}(\theta,\eta) = f(\theta,\eta)$, so $c_n = 0$. I focus on using $f(\theta,\eta) = ||\delta||_2^2 + ||\eta||_2^2$ as discussed in more detail in subsection \ref{sub: IV-SCE continued}.
\par 
Assumption \hyperref[Ass3.1]{3.1.1} allows the parameter spaces not to be compact, as long as the feasible values of the parameters that make the penalty sufficiently small are compact. Assumption \hyperref[Ass3.1]{3.1.4} provides a strong identification condition for the identified set $S_{0,n}$. Strong partial identification conditions of this form are common in the literature on estimating identified sets and are imposed by others such as \cite{ChernozhukovHongTamer2007}. It holds under a linear factor model data-generating process, provided that there are no weak factors. Assumption \hyperref[Ass3.1]{3.1} weakens the assumptions imposed by \cite{ChernozhukovHongTamer2007} by not requiring that the parameter space or the identified sets be compact, and it allows for the dimension of $\theta$ to grow.\footnote{Part of the reason these conditions can be weakened is that I do not need to show that $\hat{S}_0$ converges to $S_{0,n}$. Instead, I only need to show convergence of the feasible set on the subset of the parameter space where $f$ is small.}
\par 
Assumption \hyperref[Ass3.1]{3.1.2} imposes a uniform rate of convergence of the sample moment conditions. I impose a rate of convergence for the estimated penalty function in Assumption \hyperref[Ass3.1]{3.1.3} and a condition relating $|f(\beta_{0,n},\delta,\eta) - f(\theta_{0,n},\eta_{0,n})|$ to $||\delta - \delta_{0,n}||_1$ and $||\eta - \eta_{0,n}||_1$ in Assumption \hyperref[Ass3.1]{3.1.5}. Note that for Assumption \hyperref[Ass3.1]{3.1.5}, it is only necessary that among elements of identified set that are close to $\theta_{0,n}$ and $\eta_{0,n}$, $(||\delta - \delta_{0,n}||_1 + ||\eta - \eta_{0,n}||_1)^{\gamma_1}$ can be bounded by $f(\beta_{0,n},\delta,\eta) - f(\beta_{0,n}, \delta_{0,n},\eta_{0,n})$. This allows us to guarantee that if elements of the identified sets achieve close to the minimum value of $f$, then they must be close to $\theta_{0,n}$ and $\eta_{0,n}$. It may often be the case that $(\delta_{0,n},\eta_{0,n})$ is on the boundary of the identified set $S_{0,n}$ and is not a local minimum of $f(\beta_{0,n},\delta,\eta)$ on the whole parameter space. This is why the second part of Assumption \hyperref[Ass3.1]{3.1.5} is needed, because it allows us to place a bound on the rate of change of $f(\theta,\eta)$ for values of the nuisance parameter that give a penalty value which is close to $f(\theta_{0,n},\eta_{0,n})$. As a result, values just outside $S_{0,n}$ should not make $f$ much lower than $f(\theta_{0,n},\eta_{0,n})$. I show that in the traditional SCE setting, when $f(\theta,\eta) = ||\delta||_2^2 + ||\eta||_2^2$, the properties of the identified set allow for Assumption \hyperref[Ass3.1]{3.1.5} to hold with $\gamma_1 =2$ and $\gamma_2 = 1$.

\textbf{Lemma 3.1 (Rate of Convergence of Regularized Estimates)}\phantomsection\label{L3.1} Suppose that Assumption \hyperref[Ass3.1]{3.1} holds, and $\lambda_\delta, \lambda_\eta \overset{p}{\rightarrow} 0$ such that $b_n/\min\{\lambda_\delta,\lambda_\eta\} \overset{p}{\rightarrow} 0$. Then 
$$||\hat{\delta} - \delta_{0,n}||_{\mathcal{E}} + ||\hat{\eta}-\eta_{0,n}||_{\mathcal{E}} = O_p((\max \{\lambda_\delta^{\gamma_2},\lambda_\eta^{\gamma_2},a_n^{\gamma_2},c_n\})^{1/\gamma_1}).$$
\par 
This result has a similar form as Theorem 3.1 of \cite{ChernozhukovHongTamer2007} for the rate of convergence of the estimated identified set, except here we have the terms $c_n$, $\gamma_1$, and $\gamma_2$, which capture how the behavior of the penalty function influences the convergence of the penalized estimator. Lemma \hyperref[L3.1]{3.1} requires that $\lambda_\delta$ and $\lambda_\eta$ are shrinking more slowly than $b_n$. This guarantees that $S_{0,n} \subseteq \hat{S}_0$ with probability approaching one. We can therefore satisfy this condition by having $\lambda_\eta$ and $\lambda_\delta$ shrinking at a slightly slower rate than $b_n$ (e.g., if $b_n = \log(J)/\sqrt{n}$ then $ \lambda_\delta,\lambda_\eta \asymp_p O_p(\log(J)\log (n)/\sqrt{n})$). So when $a_n = \log(J)/\sqrt{n}$ as well and when $c_n =0$, $\gamma_1 = 2$, and $\gamma_2 = 1$ due to the $L_2$ norm penalty, this gives $||\hat{\delta} - \delta_{0,n}||_{\mathcal{E}} + ||\hat{\eta}-\eta_{0,n}||_{\mathcal{E}} = O_p((\sqrt{\log(J)\log (n)}/n^{1/4})$. This is notably slower than the standard parametric rate, but still fast enough to satisfy the linear case of Assumption \hyperref[Ass2.1]{2.1}, depending on how fast $J$ is growing and whether $\mathcal{E}  =1$ or $\mathcal{E} = 2$. Having constraints on $\delta$ is beneficial for guaranteeing $a_n$ and $b_n$ grow slowly in $J$.

\textbf{Remark 3.1 (Simplex Constraints)}\phantomsection\label{R3.1} It is common in SC applications to constrain $\delta$ so that $D_n = \{ \delta \geq 0 : ||\delta||_1 =1 \}$. Suppose the moment conditions have the form $\hat{g}(\theta) = \sum_{t \in \mathcal{T}_0} (X_{0,t} - X_{\mathcal{J},t}\delta)/T_0$ or $\hat{g}(\theta) = \sum_{t \in \mathcal{T}_1} (X_{0,t} - \beta - X_{\mathcal{J},t}\delta)/T_1$ like in the linear-IV application. Then for the uniform convergence of the $q$-th sample moment condition, 
$$\sup_{\theta \in \Theta_n}|\hat{g}_q(\theta) - g_q(\theta)| \leq ||\sum_{t \in \mathcal{T}_{b_q}}( E[X_{0,t}] - X_{0,t})/T_{b_q}| +$$
$$  \max_{1 \leq j \leq J}|\sum_{t \in \mathcal{T}_{b_q}} (E[X_{jt}] - X_{jt})/T_{b_q}|.$$ Note that the last term is also a bound on $\sup_{\theta \in \Theta_n}||\partial_\delta \hat{g}_q(\theta) - \partial_\delta g_q(\theta)||_\infty$. In many cases, the maximum in the last term grows slowly with $J$.  Additionally, if these constraints make $\delta_{0,n}$ sparse, we can achieve a faster rate of convergence for the estimated nuisance parameters.
\par 
The conditions placed on $\lambda_\delta$ and $\lambda_\eta$ in Lemma \hyperref[L3.1]{3.1} allow for them to be stochastic, and therefore potentially chosen in a data-driven way, but do not tell us how to choose them in practice. It is important to have an algorithm for determining these tuning parameters to reduce researchers' room for specification searching. For the empirical application and simulations in section \ref{s: EmpiricalApplication}, I use a computationally simple method which guarantees that $\hat{S}_0$ is non-empty but also that $\lambda_\delta$ and $\lambda_\eta$ are shrinking at the appropriate rate. This method involves finding the minimum values of $\lambda_\delta$ and $\lambda_\eta$ that make $\hat{S}_0$ non-empty, and then multiplying them by a nonrandom function of the number of time periods and the number of controls. 

\subsection{IV Synthetic Control Example Continued}\label{sub: IV-SCE continued}

Returning to the linear-IV example, we can choose the penalty function to equal an upper bound on the asymptotic variance as a function of the nuisance parameters, which is tight in special cases. Because $\hat{g}$ is linear in $\delta$, each of the elements of the asymptotic variance of the original moment conditions $V_g$ involves a quadratic form $\delta' \Omega \delta$ for some positive definite matrix $\Omega$. Also, since $V(\theta,\eta) = \eta V_g(\theta) \eta'/\eta_Q^2$ where $V_g(\theta)$ is positive definite for any fixed $\theta \in \Theta_n$, choosing $\hat{f}(\theta,\eta) = f(\theta,\eta) = ||\delta||_2^2 + ||\eta||_2^2/\eta_Q^2$  minimizes an upper bound on $V(\theta,\eta)$ and does not involve $\hat{V}(\theta,\eta)$. However, $||\eta||_2^2/\eta_Q^2 = \sum_{q=1}^{Q-1} (\eta_q/\eta_Q)^2 + 1$ may not have a unique minimum on $$ \{ \eta \in \mathbb{R}^Q : \eta \begin{pmatrix}
   \sum_{t \in \mathcal{T}_0} E[Z_t f_t \mu_{\mathcal{J}}]/T_0 \\
   \sum_{t \in \mathcal{T}_1}E[f_t \mu_{\mathcal{J}}]/T_1
\end{pmatrix}  = 0 \},$$ 
which is the reason for normalizing $\eta_Q = 1$. I show in the proof of Corollary \hyperlink{C4.1}{4.1} that this allows $\eta_{0,n}$ to be identified. Then the penalty can simply be set to $\hat{f}(\theta,\eta) = f(\theta,\eta) = ||\delta||_2^2 + ||\eta||_2^2$ and have the parameter space for $\eta$ be equal to $H = \{ \eta \in \mathbb{R}^Q : \eta_Q = 1 \}$. Because the penalty is additively separable and the sets of constraints which involve $\delta$ and $\eta$ are disjoint, we estimate $\delta$ and $\eta$ separately using the quadratic programming problems:
$$\hat{\delta} = \argmin_{\delta \in \Delta^J} ||\delta||_2^2 \text{ s.t. } ||\sum_{t \in \mathcal{T}_0} Z_t (Y_{0t} - Y_{\mathcal{J},t}\delta)/T_0||_\infty \leq \lambda_\delta$$
$$\text{ and } \hat{\eta} = \argmin_{\eta \in H } ||\eta||_2^2 \text{ s.t. } ||\sum_{t \in \mathcal{T}_1} Y_{\mathcal{J},t}/T_1 - \sum_{q=1}^{Q-1} \eta_q \sum_{t \in \mathcal{T}_0}  Z_{qt}' Y_{\mathcal{J},t}/T_0||_\infty \leq \lambda_\eta.$$ 
In Appendix B of the Supplementary Materials, I show when the upper bound on the asymptotic variance being minimized is tight. This will happen when idiosyncratic shocks are homogeneous across units, conditional on the instruments, and the instruments are not redundant and have similar variability. In order for Assumption \hyperref[Ass3.1]{3.1} to be satisfied, I impose the following condition:

\textbf{Condition 2}\phantomsection\label{Ass2} As $T_0,T_1,J \rightarrow \infty$ while $Q$ is fixed, 
\begin{enumerate}
    \item $E[Z_t \epsilon_{it}] = 0$ for all $i,t$ and  $\max_{ 0 \leq i \leq J} \{ ||\frac{1}{T_0} \sum_{t \in \mathcal{T}_0} Z_t\epsilon_{it}||_1 \} = O_p(\log (J)/\sqrt{T_0})$.
    \item $Z^{pre'}f^{pre}/T_0 = E[Z^{pre'}f^{pre}/T_0] + O_p(1/\sqrt{T_0})$ and the positive singular values of \\ $E[Z^{pre'}f^{pre}/T_0]\mu_{\mathcal{J}}$ are bounded away from zero.
    \item $E[Z^{pre'}f^{pre}/T_0]'E[Z^{pre'}f^{pre}/T_0] \rightarrow \Omega_0$ and $E[\sum_{t \in \mathcal{T}_1} f_t/T_1] \rightarrow \Omega_1$ where $\Omega_0$ is full rank. The sequence of factor loadings $\mu_i$ is uniformly bounded.
    \item $\max_{0 \leq i \leq J} \{| \frac{1}{T_1} \sum_{t \in \mathcal{T}_1} \epsilon_{it} | \} = O_p(\log (J)/\sqrt{T_1})$ and \\ $\sum_{t \in \mathcal{T}_1}f_t/T_1  = E[\sum_{t \in \mathcal{T}_1}f_t/T_1] + O_p(1/\sqrt{T_1})$.
    \item $D_{0,n}$ is non-empty for all $J > C$ for some integer $C \geq 1$.
\end{enumerate}

Condition \hyperref[Ass2]{2.1} ensures that the instruments satisfy the exclusion restriction, and Condition \hyperref[Ass2]{2.3} guarantees that the instruments are relevant and that there are enough of them to identify $D_{0,n}$. Furthermore, following the reasoning discussed in Remark \hyperref[R3.1]{3.1}, by setting $D_n = \Delta^J$, the rate of convergence conditions in Condition \hyperref[Ass2]{2} guarantee that the sample moment conditions converge to the population moment conditions at a rate of $\log (J)/\sqrt{\min \{T_0,T_1\}}$ uniformly in $\delta \in D_n$. The second part of Condition \hyperref[Ass2]{2.2} guarantees that the set $D_{0,n}$ is strongly partially identified. Condition \hyperref[Ass2]{2.5}, imposes that once enough control units are added, the factor loadings of the treated unit $\mu_0$ fall in the convex hull of the factor loadings of the control units. Note that the factors may not have the same average values before and after treatment, which allows for cases where there is a dependence between the values of the latent factors and the timing of treatment.

\section{Variance Estimation and Inference}\phantomsection\label{s: VarianceEstimation}

As mentioned earlier, the orthogonalized moment conditions can usually be used to estimate $\beta$ by method of moments, or more generally via a GMM estimator:
\begin{equation}\phantomsection\label{eq: GMM Estimator}
    \Tilde{\beta} = \argmin_{\beta \in B} \hat{M}(\beta,\hat{\delta},\hat{\eta})' W_n \hat{M}(\beta,\hat{\delta},\hat{\eta}),
\end{equation}
where $W_n$ is a $m \times m$ weighting matrix. In many cases, it is possible to show that $\sqrt{n}\hat{M}(\theta_{0,n},\eta_{0,n})$ is asymptotically normal since it is a linear function of a vector of averages. This allows for modified versions of standard arguments for the asymptotic normality of GMM estimators.

\textbf{Assumption 4.1}\phantomsection\label{Ass4.1}  Suppose that
\begin{enumerate}
   \item $\sup_{\theta \in \Theta_n} ||\hat{g}(\theta) - g(\theta)||_2 = o_p(1)$ .
     \item For all $\epsilon >0$, there exists $\gamma_\epsilon$ such that \\ $P(\sup_{\theta \in \Theta_n : ||\theta - \theta_{0,n}||_1 < \gamma_\epsilon}||\partial_\beta \hat{g}(\theta) - \partial_\beta \hat{g}(\theta_{0,n})||_2 > \epsilon ) \rightarrow 0$.
    \item $\sqrt{n}\hat{M}(\theta_{0,n},\eta_{0,n}) \overset{d}{\rightarrow} N(0,V_M)$ for some sequence of positive definite matrix $V_M$.
    \item $\partial_\beta M(\beta_{0,n},\delta_{0,n},\eta_{0,n}) \rightarrow M_\beta$ for some matrix $M_\beta$ with  $rank(M_\beta) = p$ and \\ $||\beta_1 - \beta_2||_2 \leq C ||M(\beta_1,\delta_{0,n},\eta_{0,n}) - M(\beta_2,\delta_{0,n},\eta_{0,n})||_2$ for all $\beta_1,\beta_2 \in B$ and all $n$ for some $C > 0$.
    \item $W_n - W \overset{p}{\rightarrow} 0$ for some positive definite matrix $W$ and the sequence $\eta_{0,n}$ is bounded.
    \item There exists $\epsilon >0$ such that $\{\beta : ||\beta - \beta_{0,n}||_1 < \epsilon\} \subseteq B$ for all $n$.
\end{enumerate}
\par 
 $\partial_\beta \hat{g}(\theta)$ generally does not vary with $\delta$, but if it does Assumption \hyperref[Ass4.1]{4.1.2} is needed to ensure that $\partial_\beta \hat{g}(\beta_{0,n},\hat{\delta})$ converges to $\partial_\beta \hat{g}(\beta_{0,n},\delta_{0,n})$ as $\hat{\delta}$ converges to $\delta_{0,n}$. Assumption \hyperref[Ass4.1]{4.1.3} can be directly combined with the adaptivity condition to show the asymptotic normality of $\sqrt{n}\hat{M}(\beta_{0,n},\hat{\delta},\hat{\eta})$. Assumption \hyperref[Ass4.1]{4.1.4} guarantees the identification of $\beta$ and holds trivially with $M_\beta = I_p$ when $\eta$ is constrained so that $\partial_\beta \hat{M}(\theta,\eta) = I_p$. Because $\hat{M}(\theta_{0,n},\eta_{0,n})$ involves taking sample averages, it can be satisfied by fairly standard conditions that allow for a Central Limit Theorem to be applied when $\delta_{0,n}$ is sparse. Assumption \hyperref[Ass4.1]{4.1.6} is generally trivial unless we are interested in constraining are estimates of average treatment effects.

\textbf{Proposition 4.1 (Asymptotic Normality)}\phantomsection\label{P4.1}   Suppose $(\beta_{0,n},\delta_{0,n}) \in  \Theta_{0,n}$, $\eta_{0,n}$ satisfies equation \eqref{eq: Orthogonality Condition}, and Assumptions \hyperref[Ass2.1]{2.1} and \hyperref[Ass4.1]{4.1} hold. Then, 
$$\sqrt{n}(\Tilde{\beta} - \beta_{0,n}) \overset{d}{\rightarrow} N(0, V),$$
where $V = (M_\beta' W M_\beta)^{-1} M_\beta' W V_M W M_\beta (M_\beta' W M_\beta)^{-1}$.
\par 
Note that when $W = I_m$ and $\eta$ is constrained so that $M_\beta = I_p$, $V$ simplifies to $V_M$. In this case, we can obtain an upper bound for $V(\theta,\eta)$ by bounding the variance of the sample moment conditions. More generally, the choice of the nuisance parameters $\delta$ and $\eta$ would influence the precision of $\Tilde{\beta}$ both through changing how much variability there is in the sample moment conditions (i.e., $V_M$) and through how sensitive the population moment conditions are to $\beta$ (i.e., $M_\beta$). We can use 
$$\hat{V}(\hat{\theta},\hat{\eta}) = (\hat{M}_\beta' W_n \hat{M}_\beta)^{-1} \hat{M}_\beta' W_n \hat{V}_M(\hat{\theta},\hat{\eta}) W_n \hat{M}_\beta (\hat{M}_\beta' W_n \hat{M}_\beta)^{-1},$$
where $\hat{M}_\beta = \partial_\beta \hat{M}(\hat{\theta},\hat{\eta})$ and $\hat{V}_M(\hat{\theta},\hat{\eta}) = \hat{\eta} \hat{V}_g(\hat{\theta})\hat{\eta}'$ where $\hat{V}_g(\hat{\theta})$ is an estimate of the asymptotic variance of the moment conditions. 
\par 
When $V_g$ corresponds to the long-run variance of the sample moment conditions and the structure of the dependence across observations is unknown, it is common to use estimators in the class of quadratic Heteroscedastic Autocorrelation Consistent (HAC) variance estimators. In this context, where the moment conditions may involve using different blocks on time periods, the $q,l$ element of $\hat{V}_g(\theta)$ will be given by
\begin{equation}\label{eq: HAC formula}
    \hat{V}_{g,ql}(\theta) = n \sum_{i \in \mathcal{T}_{b_q}} \sum_{s \in \mathcal{T}_l }  Q_K(\frac{i}{n},\frac{s}{n}) (g_{q,i}(\theta) - \hat{g}_q (\theta))(g_{l,s}(\theta) - \hat{g}_l (\theta))'/(|\mathcal{T}_{b_q}|  |\mathcal{T}_{b_l} |),
\end{equation}
where $g_{q,i}(\theta)$ is the $i$th observation of the $q$th moment condition and $Q_K(i,s)$ is a weighting function that depends on a smoothing parameter $K$. This includes kernel variance estimators such as those of \cite{Andrews1991} and \cite{NeweyWest1987} as well as the orthonormal series variance estimators such as that of \cite{Phillips_2005}. For conventional Kernel HAC estimators, $Q_K(i, s) = \mathcal{K} ((i  -s)/b)$ where $\mathcal{K}$ is a kernel and $b= 1/K$ is the bandwidth. For Series HAC estimators, $Q_K(i,s) = \sum_{k=1}^K \phi_k (i) \phi_k(s)/K$ where $ \{ \phi_k(s) \}_{k=1}^K$ are orthonormal basis functions taking values in $[0,1]$.
\par 
Asymptotic results for non-parametric long-run variance estimators that rely on $K \rightarrow \infty$ as $n \rightarrow \infty$ can often provide poor approximations in practice, particularly when the degree of temporal dependence is high relative to the sample size. Intuitively, this is because the uncertainty in our estimation of $V_M$ significantly contributes to our uncertainty in our test statistic in these cases, but this is not captured by increasing-smoothing asymptotic results. SCEs are often used in settings with small to moderate sample sizes and with data that display a high degree of dependence over time, making the use of increasing-smoothing asymptotic results particularly questionable. This is illustrated by \cite{t-test2022}, who show that their inference procedure for SCE performs very poorly when they calculate standard errors using a HAC estimator and rely on asymptotic results for increasing-smoothing to obtain critical values.
\par 
The notion of fixed-smoothing asymptotics was first introduced by \cite{VogelsangKiefer2002a}. As the name suggests, it involves keeping the smoothing parameter $K$ fixed as the sample size grows. For both Kernel and Series HAC estimators, this results in $\hat{V}_M(\hat{\theta},\hat{\eta})$ converging to a stochastic matrix. The fact that $\hat{V}_M(\hat{\theta},\hat{\eta})$, and therefore $\hat{V}(\hat{\theta},\hat{\eta})$, are converging to something stochastic has several implications for this procedure. If $\hat{V}$ is not converging to $V$, then $\hat{f}$ should not be a function of $\hat{V}$ in order to satisfy Assumption \hyperref[Ass3.1]{3.1.3}. Therefore, it is infeasible to choose the nuisance parameters that minimize the asymptotic variance of the parameter of interest. For using $\hat{V}_M(\hat{\theta},\hat{\eta})$ to weight the moment conditions, \cite{HwangSun2018oneVsTwoStep} compare the performance of the One-Step and Two-Step GMM estimator under a fixed-smoothing asymptotic framework. They show that whether the Two-Step GMM procedure outperforms the One-Step GMM procedure depends on the values of long-run correlation coefficients. Because these long-run correlations can also not be consistently estimated under the fixed-smoothing asymptotic framework, it is generally unclear whether the One-Step or Two-Step GMM estimators perform better. As shown by \cite{Sun2014TwoStep}, under fixed-smoothing asymptotics, while the One-Step GMM estimator is still asymptotically normal, the Two-Step GMM estimator is asymptotically mixed normal. For these reasons, I focus on using $W_n = I_m$ and have the penalty function be based on the upper bound of the asymptotic variance when fixed-smoothing asymptotic results are relevant. 
\par 
However, common test statistics can still have nonstandard limiting distributions even when $\Tilde{\beta}$ is asymptotically normal. For example, the Wald test statistic, rather than converging to a chi-squared distribution, converges to a distribution that depends on the kernel or basis function and the smoothing parameter. For Kernel HAC estimators, \cite{Sun2014Fixedb} provides conditions under which an adjusted Wald statistic and an adjusted $t$-statistic have asymptotic distributions that can be approximated by $F$ and $t$ distributions, but do not converge exactly to $F$ and $t$ distributions. On the other hand, \cite{Sun2013} gives versions of the Wald statistic and $t$-statistic that converge exactly to $F$ and $t$ distributions when $W_n = I_m$ and a Series HAC estimator is used. Furthermore, \cite{LazarusLewisStock2021} characterize the size-power frontier for Kernel and Series HAC estimators under a fixed-smoothing framework and find that there is little cost to restricting attention to tests that converge exactly to $t$ and $F$ distributions. I therefore focus on verifying that the conditions of \cite{Sun2013} for the GMM estimator with $W_n = I_m$ and $\hat{V}_M$ being a Series HAC estimator, with the test statistic being the standard Wald statistic defined as
\begin{equation*}
    \mathbb{W}_n =  n(\Tilde{\beta} - \beta_{0,n})'\hat{V}(\hat{\theta},\hat{\eta})^{-1}(\Tilde{\beta} - \beta_{0,n}),
\end{equation*}
and the $t$-statistic is defined as 
\begin{equation*}
    t_n =  \sqrt{n}(\Tilde{\beta} - \beta_{0,n})/\sqrt{\hat{V}(\hat{\theta},\hat{\eta})}
\end{equation*}
when $p =1$. In order to conduct valid and standard inference in a fixed-smoothing asymptotic framework, I impose conditions given in Assumption \hyperref[Ass4.2]{4.2} on the Series estimator used and on the partial sums of the moment conditions. For notational simplicity when referencing these partial sums, I used $\mathcal{T}_b^t$ to denote the first $t$ indices in the $b$-th block of time periods.  
\par 
\textbf{Assumption 4.2}\phantomsection\label{Ass4.2} Suppose that with $K$ fixed, we have that
\begin{enumerate}
    \item For all $\epsilon > 0$ and each $q$, there exists $\gamma_\epsilon$, such that uniformly over $1 \leq t \leq T_{b_q}$,
    
    $P(\sup_{ \theta \in \Theta_n : ||\theta - \theta_{0,n}||_1 < \gamma_\epsilon} ||\sum_{i \in  \mathcal{T}_{b_q}^t} (\partial_\beta g_{q,i}(\beta,\delta) - \partial_\beta g_{q,i}(\beta,\delta_{0,n}))/T_{b_q} ||_2 > \epsilon) \rightarrow 0$.
    \item $K \geq p$ and $\{ \phi_k (x) \}_{k=0}^K$ with $\phi_0(x) = 1$ is a sequence of continuously differentiable and orthonormal basis functions in $\mathcal{L}^2[0,1]$ satisfying $\int_0^1 \phi_k(x) dx = 0$ for each $k$.
    \item For each $q$, uniformly in $1 \leq t \leq T_{b_q}$ and $\lambda_n \in [0,1]$, $$\sum_{i \in \mathcal{T}_{b_q}^t} \partial_\beta g_{q,i}(\beta_{0,n} + \lambda_n (\Tilde{\beta} - \beta_{0,n}),\delta_{0,n})/n -  r \partial_\beta g_q(\theta_{0,n})\overset{p}{\rightarrow} 0.$$
    \item $$ V_M^{-1/2} \sqrt{\min \{T_0,...,T_Q\}} \eta_{0,n}  \begin{pmatrix}
\sum_{t \in \mathcal{T}_{b_1}} \phi_k(\frac{t}{T_{b_1}}) g_{1,t}(\theta_{0,n})/T_{b_1} \\ 
... \\
\sum_{t \in \mathcal{T}_{b_{Q-1}}} \phi_k(\frac{t}{T_{b_{Q-1}}}) g_{Q-1,t}(\theta_{0,n})/T_{b_{Q-1}}
 \\
    \sum_{t \in \mathcal{T}_{b_{Q}}} \phi_k(\frac{t}{T_{b_{Q}}}) g_{Q,t}(\theta_{0,n})/T_{b_{Q}}
\end{pmatrix} \overset{d}{\rightarrow} \xi_k $$ jointly for $k = 0,...,K$ with $\xi_k \sim iid N(0,I_m)$. 
\end{enumerate}

Assumptions \hyperref[Ass4.2]{4.2.2}-\hyperref[Ass4.2]{4.2.4} contain the conditions imposed by \cite{Sun2013} adjusted to this setting. Assumption \hyperref[Ass3.1]{3.1.2} is satisfied for commonly used basis functions.\footnote{For example, $\phi_k(x) = \sqrt{2}\sin (2\pi kx)$ and $\phi_k(x) = \sqrt{2}\cos (2\pi kx)$ satisfy the condition. However, there are basis functions such as $\phi_k(x) = \sqrt{2} \sin (\pi (0.5 - k)x)$ that do not satisfy it because it does not satisfy the mean-zero condition.} Assumption \hyperref[Ass3.1]{3.1.4} is standard in the literature on fixed-smoothing asymptotics (see, for example, \cite{VogelsangKiefer2005}).\footnote{In Lemma A2, I show how this holds when the function of partial sums is converging weakly to a Gaussian process.}  One additional complication that is not present in \cite{Sun2013}, or other previous fixed-smoothing results, is the plugged-in values of the nuisance parameters $\hat{\delta}$ and $\hat{\eta}$. This is why Assumption \hyperref[Ass4.2]{4.2.1} is imposed, as it serves the same role of bounding how sensitive $\partial_\beta g_i (\theta)$ is to $\delta$. Both Assumptions \hyperref[Ass4.2]{4.2.1} and \hyperref[Ass4.2]{4.2.3} hold trivially when $\partial_\beta \hat{g}(\theta) = I_p$. 
\par 
Additionally, similarly to with the adaptivity condition in equation \eqref{eq: Adaptivity Condition}, we want $ \hat{V}(\beta_{0,n},\hat{\delta},\hat{\eta})$ to be asymptotically equivalent to $\hat{V}(\beta_{0,n},\delta_{0,n},\eta_{0,n})$. In order for this to be the case, I impose Assumption \hyperref[Ass2.1*]{2.1*}, which slightly strengthens some of the conditions of Assumption \hyperref[Ass2.1]{2.1} to hold with partial sums.

\textbf{Assumption 2.1*}\phantomsection\label{Ass2.1*} Suppose that
\begin{enumerate}
   \item  Uniformly over $(1,1,..,1) \leq (t_1,t_2,...,t_Q) \leq (T_{b_1},T_{b_2},...,T_{b_Q})$,
   $$|| \partial_\delta \eta_{0,n} \begin{pmatrix}
       E[ \sum_{i \in \mathcal{T}_{b_1}^{t_1}} g_{1,i}(\theta_{0,n})/T_{b_1}] \\
        ... \\
        E[\sum_{i \in \mathcal{T}_{b_Q}^{t_Q}} g_{Q,i}(\theta_{0,n})/T_{b_Q}]
    \end{pmatrix} ||_{\mathcal{D}} = O_p(r_J/\sqrt{n}).$$
    \item For each $q \in \{1,...,Q\}$, uniformly over $1 \leq t \leq T_{b_q}$, 
    $$||\partial_\delta E[ \sum_{i \in \mathcal{T}_{b_q}^{t}} g_{q,i}(\theta_{0,n})/T_{b_q}] -  \partial_\delta \sum_{i \in \mathcal{T}_{b_q}^{t}} g_{q,i}(\theta_{0,n})/T_{b_q}||_{\mathcal{D}} = O_p(r_J/\sqrt{n}). $$
    \item There exists $\epsilon > 0$ such that for each $q \in \{ 1,...,Q\}$,
    
    $\sup_{1 \leq t \leq T_{b_q}, \delta: ||\delta - \delta_{0,n}||_{\mathcal{E}} \leq \epsilon} \max eig(\partial_\delta^2 \sum_{i \in \mathcal{T}_{b_q}^t} g_{q,i}(\beta_{0,n},\delta)/T_{b_q}) = O_p(r_J) $.

\end{enumerate}

Under Assumptions \hyperref[Ass2.1]{2.1} and \hyperref[Ass2.1*]{2.1*}, when $\hat{V}_M$ is a Series HAC estimator, 

$$\hat{V}_M(\beta_{0,n},\hat{\delta},\hat{\eta}) - \hat{V}_M(\beta_{0,n},\delta_{0,n},\eta_{0,n}) \overset{p}{\rightarrow} 0.$$
Since the partial sums include fewer terms than the full sum, they are generally smaller asymptotically, so the bounds on partial sums in Assumptions \hyperref[Ass2.1*]{2.1*} have similar sufficient conditions to before. Together with Assumption \hyperref[Ass4.2]{4.2}, we can show that the Wald and $t$ statistics converge to $F$ and $t$ distributions.

\textbf{Proposition 4.2 (Fixed-Smoothing Results) }\phantomsection\label{P4.2}  Suppose $\Tilde{\beta}$ is estimated using equation \eqref{eq: GMM Estimator} with $W_n = I_m$ and Assumptions \hyperref[Ass2.1]{2.1*}, \hyperref[Ass2.1]{2.1}, \hyperref[Ass4.1]{4.1}, and \hyperref[Ass4.2]{4.2} hold with $K \geq p$. When $K$ is fixed, 
\begin{equation*}
    \frac{K - p + 1}{pK}\mathbb{W}_{n} \overset{d}{\rightarrow} F_{p,K-p+1} \text{ and } t_n \overset{d}{\rightarrow} t_{K} \text{ when }p=1,
\end{equation*}
where $F_{p,K-p+1}$ is an $F$ distribution with $p,K-p+1$ degrees of freedom and $t_K$ is a $t$-distribution with $K$ degrees of freedom.
\par
To the best of my knowledge, this is the first extension of fixed-smoothing results to high-dimensional settings. If we were instead to focus on asymptotics with $K \rightarrow \infty$ where $V(\theta_{0,n},\eta_{0,n})$ is consistently estimated, then we have that $\mathbb{W}_n \overset{d}{\rightarrow} \chi^2_p/p$, where $\chi^2_p$ is chi-squared distribution with $p$ degrees of freedom, and $t_n \overset{d}{\rightarrow} N(0,1)$. Note that $F_{p,K-p+1} \overset{d}{\rightarrow} \chi^2_p/p$ and $t_K \overset{d}{\rightarrow} N(0,1)$ as $K \rightarrow \infty$, so critical values obtained from the fixed-smoothing asymptotic results are approximately the same as the critical values from increasing smoothing asymptotic results when $K$ is large. For choosing $K$, I recommend using the Coverage Probability Error (CPE)-optimal choice of \cite{Sun2013}, which generally will choose a smaller value of $K$ than the Mean Squared Error (MSE)-optimal choice of \cite{Phillips_2005}, in order to have better small-sample coverage. The values of the smoothing parameter $K$ are often small in applications. For example, in the empirical application below, $K$ is equal to $4$ when the method of \cite{Sun2013} is used. This translates to a 95\% confidence interval that's about 41\% wider than if critical values from a standard normal distribution were used.

\subsection{IV Synthetic Control Example Continued}

Under Condition \hyperref[Ass2]{2} and Condition \hyperref[Ass3]{3} below, the $l$-th row and $q$-th column of the asymptotic variance of the original moment conditions $V_g$ is given by
\begin{equation}\label{Asymptotic Variance of SCE}
    \lim_{T_0,T_1 \rightarrow \infty} \min \{T_0,T_1\} \sum_{t \in \mathcal{T}_{I_{l=Q}}}\sum_{s \in \mathcal{T}_{I_{q=Q}}} E[g_{l,t}(\theta_{0,n}) g_{q,s}(\theta_{0,n})']/(|\mathcal{T}_{I_{l=Q}} \mathcal{T}_{I_{q=Q}}|),
\end{equation}
where $I_{q=Q}$ is an indicator function that is equal to $1$ if and only if $q = Q$. I define $\hat{V}_g$ to be a Series HAC estimator based on equation \eqref{eq: HAC formula} and $\hat{V}(\hat{\theta},\hat{\eta}) = \hat{\eta}\hat{V}_g(\hat{\theta})\hat{\eta}'/\hat{\eta}_Q^2 = \hat{\eta}\hat{V}_g(\hat{\theta})\hat{\eta}'$. Also, for testing a null hypothesis $H_0 : \beta_{0,n} = \Bar{\beta}$, I use the test statistic $\sqrt{\min \{T_0,T_1\}} (\Tilde{\beta} - \Bar{\beta})/\sqrt{\hat{V}(\hat{\theta},\hat{\eta})}$ and critical values from a $t$ distribution with $K$ degrees of freedom. 
\par 
In order for the conditions of Propositions \hyperref[P4.1]{4.1} and \hyperref[P4.2]{4.2} to hold, I impose the following additional condition:

\textbf{Condition 3}\phantomsection\label{Ass3} As $T_0,T_1,J \rightarrow \infty$ while $Q$ and $K$ are fixed,
\begin{enumerate}
    \item For $q,l \in  \{1,...,Q-1\}$, 
    $$E[\sum_{t \in \mathcal{T}_0 } Z_{qt} Z_{lt}(\epsilon_{\mathcal{J},t}\delta_{0,n})^2/T_0] \rightarrow \Sigma_\delta^{q,l}\text{ and }E[\sum_{t \in \mathcal{T}_0} Z_{qt}Z_{lt}\epsilon_{0t}^2/T_0]] \rightarrow \Sigma_0^{q,l};$$
    $$ E[\sum_{t \in \mathcal{T}_1}( \epsilon_{\mathcal{J},t} \delta_{0,n})^2/T_1] \rightarrow \Sigma_\delta \text{ and } E[(\sum_{t \in \mathcal{T}_1}\epsilon_{it})^2/T_1] \rightarrow \Sigma_0,$$
    for some positive constants $\Sigma_0^{q,l}$, $\Sigma_\delta^{q,l}$, $\Sigma_0$, and $\Sigma_\delta.$
    \item $\{\phi_k(x)\}_{k =1}^K$ are chosen to satisfy Assumption \hyperref[Ass4.2]{4.2.2}.
    \item $||\delta_{0,n}||_0 = s$ where $s$ may be growing with $J$. 
    \item The sequence $(Z_t,\epsilon_t)_{t \in \mathbb{N}}$ is $\alpha$-mixing with mixing coefficients $\alpha(t)$.
    \item There exists $\gamma > 2$ such that $\sup_{t \in \mathcal{T}_0} E[||Z_t(\epsilon_0 - \epsilon_{\mathcal{J},t})\delta_{0,n}]||^\gamma]$ and $\sup_{t \in \mathcal{T}_1}E[|\epsilon_{0t} - \epsilon_{\mathcal{J},t}\delta_{0,n}|^\gamma]$ are bounded and $\sum_{t \in \mathbb{N}} \alpha(t)^{1 - 2/\gamma} < \infty$.
\end{enumerate}

Condition \hyperref[Ass3]{3.1} helps us specify what the asymptotic variance of the sample moment conditions is. Along with the mixing condition and requiring a bound on the moments of $\hat{g}(\beta_{0,n},\delta_{0,n})$, this is sufficient to apply a Functional Central Limit Theorem to the partial sums of the sample moment conditions, which allows for the estimator to be asymptotically normal and our $t$-statistic to have a $t$-distribution asymptotically. As shown by equation \eqref{eq: IV-SCE Adaptivity}, the asymptotic distribution of the estimator depends only on a finite number of averages of the idiosyncratic shocks and averages of the product of the idiosyncratic shocks with the instruments, so we need sufficient conditions for a linear combination of these sample averages to satisfy a Central Limit Theorem.
\par 
Condition \hyperref[Ass3]{3.3} is a sparsity condition where $s$ denotes the number of non-zero elements of $\delta_{0,n}$, which will be assumed to grow slowly with $J$. In this context, sparsity of the control weights is plausible for two reasons. First, because there are only $R$ latent factors, it is plausible that there exists $\delta \in \Delta^J$ such that $\mu_0 = \mu_{\mathcal{J}}\delta$ and $||\delta||_0$ is equal to or only slightly larger than $R$. Second, the simplex constraints tend to select sparse solutions, which is why most empirical applications with many controls find sparse weights. On the other hand, the $L_2$ penalty tends to select one of the more dense elements of $D_{0,n}$, so even if there exists a sparse set of weights in $D_{0,n}$, this does not guarantee that $\delta_{0,n}$ is sparse. This is similar to the trade-off present in other settings involving both non-uniqueness and high dimensionality. For example, when performing high-dimensional least squares regression with regressors that are (close to) collinear, introducing an $L_1$ penalty induces sparsity but may be unstable due to there not being a unique linear combination of the regressors with the minimum $L_1$ norm. Whereas introducing an $L_2$ penalty would induce uniqueness but not sparsity. The estimator proposed for the control weights here aims to use the combination of the simplex constraints and the $L_2$ penalty to balance the need for uniqueness with the desirability of selecting a sparse set of control weights.

\textbf{Corollary 4.1}\phantomsection\label{C4.1} Suppose Conditions \hyperref[Ass1]{1}, \hyperref[Ass2]{2}, and \hyperref[Ass3]{3} hold, $\lambda_\delta$ and $\lambda_\eta$ satisfy $$\lambda_\delta,\lambda_\eta \asymp_p \log(J)\log(\min\{T_0,T_1\})/\sqrt{\min\{T_0,T_1\}} $$
$$\text{ and } \sqrt{ \max \{\lambda_\delta, \lambda_\eta\}s} \log (J) \log(\min \{ T_0,T_1\})  \overset{p}{\rightarrow} 0,$$
and $T_1/T_0 \rightarrow a$ for some $a > 0$ as $T_0,T_1 \rightarrow \infty$, then we have that
$$\sqrt{\min \{T_0,T_1 \}}(\Tilde{\beta} - \beta_{0,n})/\sqrt{\hat{V}(\hat{\theta},\hat{\eta})} \overset{d}{\rightarrow} t_{K}.$$
For empirical applications, I use an algorithm for selecting $\lambda_\delta$ and $\lambda_\eta$ that finds the minimum values of these tuning parameters that makes the feasible sets non-empty $\lambda_\delta^{min}$ and $\lambda_\eta^{min}$ and lets $\lambda_\delta = \lambda_\delta^{min}\log(\min\{T_0,T_1\})$ and $\lambda_\eta = \lambda_\eta^{min}\log(\min\{T_0,T_1\})$. Since the minimum feasible values will satisfy:
 $$\lambda_\delta^{min},\lambda_\eta^{min} \asymp_p \log(J)/\sqrt{\min\{T_0,T_1\}},$$ 
 the first condition in Corollary \hyperref[C4.1]{4.1} holds. The second condition on $\lambda_\delta$ and $\lambda_\eta$ is satisfied as long as 
$$ s\log(J)^3\log(\min\{T_0,T_1\})^3/\sqrt{\min\{T_0,T_1\}} \rightarrow 0.$$ 
This allows $J$ to grow faster than $T_0$ and $T_1$. If we instead removed the sparsity condition, we could still allow $J$ to grow, but at a rate slower than that of $T_0$ and $T_1$.

\section{Empirical Application}\phantomsection\label{s: EmpiricalApplication}

I now explore how my method works in practice by replicating the work of \cite{Andersson2019} and then examining its performance in simulations fitted to their data. \cite{Andersson2019} evaluate the impact of Sweden's carbon tax on CO2 emissions from transport per capita in the country. The carbon tax was introduced at US \$30 per ton of CO2 in 1990 and increased slightly during the 1990s to US \$44 in 2000. Then, from 2001 to 2004, the rate was increased to US \$109, and as of 2023 it is around \$125. When the carbon tax was implemented, it complemented an existing energy tax and there was also an addition of a Value-Added-Tax (VAT) of 25 percent in 1990. The primary treatment effect they are interested in is the combined effect of the carbon tax and VAT starting in 1990 on CO2 per capita on transport. While we could view this as a case with two continuous treatment variables (the carbon tax rate and the VAT tax rate), if we are only interested in estimating the average difference between actual CO2 per capita and CO2 per capita without the carbon tax and VAT, then we can view the carbon tax and VAT together as a single binary treatment.
\par 
\cite{Andersson2019} uses pre-treatment time periods of 1960 through 1989 and the post-treatment periods of 1990 through 2005. The units they use as controls are the 14 OECD countries: Australia, Belgium, Canada, Denmark, France, Greece, Iceland, Japan, New Zealand, Poland, Portugal, Spain, Switzerland, and the United States. They arrived at this set of 14 control countries by starting with 24 other OECD countries for which data were available. They excluded 10 of these countries: Ireland, Finland, Norway, the Netherlands, Germany, Italy, the United Kingdom, Austria, Turkey, and Luxembourg. In the case of Finland, Norway, the Netherlands, Germany, Italy, and the United Kingdom, their justification was based on these countries implementing a carbon tax or making significant changes to their fuel taxes. Because these units were experiencing similar interventions, using them as controls could lead us to underestimate the effect of the policies in Sweden. However, due in part to Sweden being one of the first countries to adopt a carbon tax, these other interventions occurred in the post-treatment period from 1990 to 2005. Since the post-treatment data for the instruments are not used, these policy changes do not necessarily present a problem for using these countries' CO2 per capita from transport data as instruments. In the case of Austria and Luxembourg, their justification for excluding them was based on concerns of “fuel tourism". They exclude Turkey because its CO2 emissions data were significantly different from the other OECD countries throughout the entire sample, and they exclude Ireland based on economic shocks that happened in the post-treatment time periods, which did not also occur in Sweden.\footnote{More specifically, they cite the Celtic Tiger expansion period in Ireland.}  I estimate the Orthogonalized SCE using the same set of controls as \cite{Andersson2019} and the set of instruments being Ireland, Finland, Norway, the Netherlands, Germany, Italy, the United Kingdom, and a constant, although I find similar results when also excluding Ireland from the set of instruments.

    \begin{figure}[ht]  \caption{Country Catagorization \label{F: CC}} 
        \centering
        \includegraphics[scale=0.25]{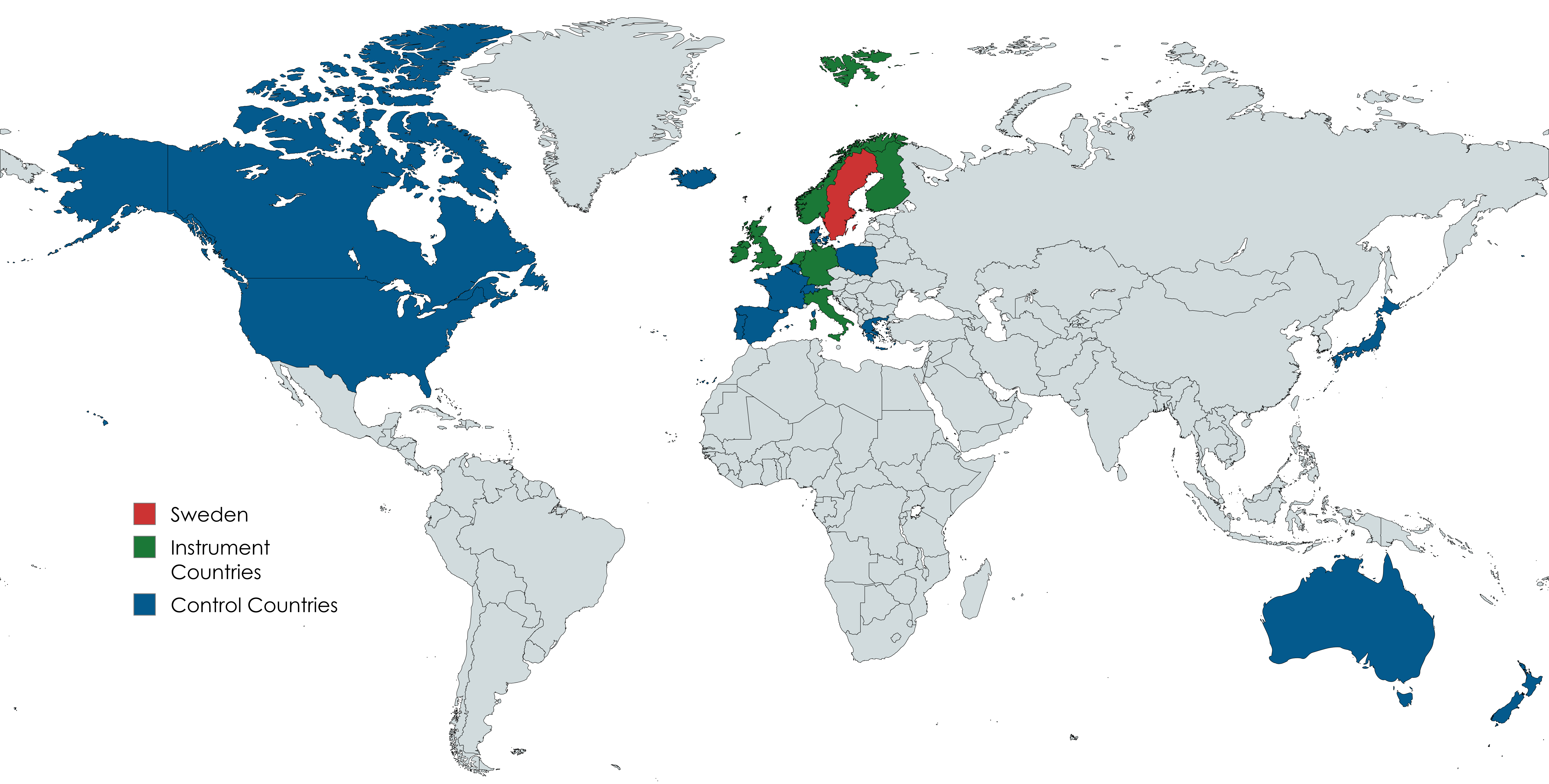}
\end{figure}

\par
In the main specification of \cite{Andersson2019}, their predictor variables are CO2 from transport per capita in 1970, 1980, and 1989, as well as GDP per capita, motor vehicles (per 1,000 people), gasoline consumption per capita, and urban population averaged for the period 1980–1989. They weight these predictors using the approach of \cite{Abadie2010}. As is common, the control weights they find end up being rather sparse, with only 6 of the 14 control countries included receiving weight greater than 1\%: Belgium (0.195), New Zealand (0.177), Denmark (0.384), Greece (0.090), Switzerland (0.061), and the United States (0.088). Using this SC as their counterfactual, they estimate an effect of -0.29 metric tons of CO2 emissions per capita per year, a 10.9 percent reduction, for the 1990–2005 period. Aggregating over the total population and the 1990-2005 period, the total cumulative reduction in emissions for the post-treatment period is 40.5 million tons of CO2. They perform several placebo tests and robustness checks, including the popular placebo test of \cite{AandG} and \cite{Abadie2010}. This method involves additionally estimating a synthetic unit for every control unit using the other control units. A test statistic is then constructed for the treated unit and every control unit by calculating the MSPE (mean squared prediction error) of each synthetic unit in the post-treatment time periods, and then either dividing by the synthetic unit's pre-treatment MSPE or excluding certain synthetic units with especially large pre-treatment MSPE. A p-value can then be calculated by examining what quantile the treated unit's test statistic falls in. In \cite{Andersson2019}, when they exclude the synthetic units with a pre-treatment MSPE at least 20 times larger than Synthetic Sweden’s pre-treatment MSPE, it leaves 9 control countries. The gap in emissions for Sweden in the post-treatment period is the largest of all remaining countries, giving a p-value of 1/10 = .1. When using the ratio of post-treatment MSPE to pre-treatment MSPE, the p-value is 1/15 = .067. In both cases, the test statistic is most extreme for the actually treated unit, but the p-value does not fall below the most common thresholds for statistical significance due to the small number of control units.
\par
When re-estimating the average treatment effect, I use the same set of pre- and post-treatment time periods. The weights of Synthetic Sweden are somewhat less sparse than before, with the countries receiving positive weight being: Australia (0.087), Belgium (0.113), Denmark (0.322), Greece (0.089), Japan (0.0190), New Zealand (0.105), Switzerland (0.201), and the United States (0.064). It is unsurprising that the weights are still sparse but with slightly more countries receiving non-zero weight due to the combination of the simplex constraints and $L_2$ penalty. That said, the weights are quite similar: all 6 countries that previously received positive weight still do, and Denmark still receives the most weight. The weights on the moment conditions $\hat{\eta}$ are fairly spread out across the pre-treatment moment conditions with the largest weight being placed on the one using the United Kingdom as an instrument and the smallest weight being placed on the one using the Netherlands as an instrument.\footnote{More specifically, the weights on the pre-treatment moment conditions are:   Finland (-6.166), Germany (-10.590), Ireland (7.883), Italy (7.912), Netherlands (-0.662), Norway (11.617), United Kingdom (-22.402), and the constant (17.882).} The average effect of the carbon tax and VAT from 1990 to 2005 is also approximately a decrease of 0.29 metric tons of CO2 per capita each year. Where the method introduced here allows for a notable difference is in terms of inference. Using the t-test described above, the p-value for the null hypothesis that these taxes had no average effect on per capita CO2 emissions from transport in the post-treatment periods is 0.00018.\footnote{When estimating the long-run variance of the moment conditions, the smoothing parameter $K$ is estimated to be 4, illustrating the empirical relevance of the fixed-smoothing asymptotics. For the series $\phi_k(x)$, I choose $\phi_k(x) = \sqrt{2}\sin(2 \pi x k)$ for even $k$ and $\phi_k(x) = \sqrt{2}\cos (2\pi x k)$ for odd $k$. Therefore, Assumption \hyperref[Ass4.2]{4.2.2} is satisfied.} This supports the results of the original paper by showing that, under plausible assumptions, the results would be very surprising if these policies had no effect on average. However, in addition to the method of \cite{Abadie2010} not necessarily providing statistically significant results, if the inference methods of \cite{ConformalInference} or \cite{CaoDowd2019} discussed below are used, we calculate p-values greater than .1 causing us to fail to reject the null of no effect at common levels of statistical significance.\footnote{Using the subsampling method with 300 iterations and a subsample size of 10 gives a p-value of .14. Using the conformal inference method with moving block permutations gives a p-value of 0.39. Using the End-of-Sample Instability test gives a p-value of 0.43. Using the t-test cross-fitting method of \cite{t-test2022} with $K = 3$ gives a p-value of .009.} This may be due to these methods having lower power in this case.

\subsection{Simulations Based on the Empirical Application}\phantomsection\label{sub: Simulations}

The placebo method used by \cite{Andersson2019} is commonly used in practice and, as previously noted by \cite{Abadie2010} and \cite{Abadie2015}, it corresponds to a traditional Fisher Randomization Test when treatment is randomly assigned. While this would mean that this test has an exact size from a design-based perspective, this condition is unrealistic in most current SCE applications. Several other methods of inference for SCEs have recently been proposed in addition to the method of \cite{Abadie2010}. I discuss the differences between these methods and the $t$-test using the Orthogonalized SCE and then compare their performance in simulations.
\par 
The method that is most comparable to mine is the cross-fitting t-test procedure of \cite{t-test2022}, where the pre-treatment time periods are split into $K$ blocks and control weights $\hat{\delta}_{k}$ are estimated using OLS withholding the $k$-th block, $H_k$. They perform a debiasing step by subtracting $\sum_{t \in H_k} (Y_{0t} - Y_{\mathcal{J},t}' \hat{\delta}_{k})/|H_k|$ from $\sum_{t \in \mathcal{T}_1} (Y_{0t} - Y_{\mathcal{J},t}' \hat{\delta}_{k})/T_1$, giving $K$ different estimates $\hat{\beta}_k$ which can be averaged: $\hat{\beta}= \frac{1}{K} \sum_{k=1}^K \hat{\beta}_k$. Their test statistic then relies on standardizing $\hat{\beta}$ using the variation across the estimates $\hat{\beta}_k$.\footnote{Following the suggestion of \cite{t-test2022}, I use $K = 3$ in the simulations.}  While \cite{ImperfectFit} show that using OLS to estimate $\delta$ can lead the SCE to be asymptotically biased, \cite{t-test2022}'s debiasing approach can fix this while avoiding the need for instruments. However, \cite{t-test2022}'s result relies on each of the sets of control weights $\hat{\delta}_k$ being approximately independent of the shocks that occur in $\mathcal{T}_1$ and $H_k$ and the bias of the OLS estimator being constant over time.\footnote{In cases where the timing of treatment is influenced by the values of the latent factors, we would usually expect the bias in the post-treatment time periods to be different from the bias in the pre-treatment time periods.} While \cite{t-test2022} do not explicitly orthogonalize their estimated ATT with respect to the control weights, the debiasing step serves a similar role and I further explore this connection in section \ref{s: extensions}. Another similarity between our approaches is that their test statistics asymptotically follow a t-distribution and do not require consistent estimation of the asymptotic variance of the estimated ATT.  
\par 
The Synthetic Difference-in-Difference (SDID) method of \cite{Arkhangelskyetal2021} involves estimating both weights on pre-treatment time periods as well as control units. These weights, along with even weights on treated units and post-treatment time periods, are then used in a Difference-in-Differences estimate of the ATT. They provide an asymptotic normality result, which I compare further to mine in section \ref{s: extensions}, and several methods for calculating standard errors. Here, I use the Placebo method as it is the only one that can be used in settings with a single treated unit. Unlike the Placebo method of \cite{Abadie2010}, it doesn't assume random assignment, but does rely on the asymptotic normality of the ATT as well as homogeneity of the error terms. \cite{Li2020} proposes a subsampling method that they show has asymptotically correct size when both $T_0$ and $T_1$ are large. However, they use an I.I.D. subsampling method rather than a block subsampling method which requires stronger independence conditions and \cite{ProblemWithSubampling} show that subsampling and m out of n bootstrap methods can have incorrect asymptotic size when the parameter is close to the boundary of the parameter space and the asymptotic results rely on keeping the number of control units fixed as the number of time periods grows, which can provide a poor approximation in applications where the number of controls is not small relative to the number of time periods. \cite{CARVALHO2018}'s Artificial Counterfactual (ArCo) method uses a LASSO-based estimator of the ATT that is asymptotically normal, but their inference method relies on consistently estimating the long-run variance, unlike my method and the $t$-test of \cite{t-test2022}. 
\par 
  \cite{ConformalInference} provides a method for conformal inference that can be used with different SCEs provided that the estimator satisfies certain consistency conditions. The End-of-Sample Instability test was originally introduced by \cite{Andrews2003}, suggested for SCE by \cite{SCandInference}, and formally analyzed and extended by \cite{CaoDowd2019}. It involves testing for a structural break in the sequence of differences between the treated unit and SC. \cite{ConformalInference}'s and \cite{CaoDowd2019}'s methods require stronger conditions on the idiosyncratic shocks, but they both have the potential advantage that the sizes of their tests are asymptotically correct when $T_1$ is fixed and only $T_0 \rightarrow \infty$. This suggests that they may be preferable when $T_1$ is very small. On the other hand, these methods and the placebo method are designed to test the sharp null hypothesis of no effect in every post-treatment time period, rather than testing a null hypothesis about the ATT. While it depends on context, usually testing the sharp null hypothesis is of less interest. Since the comparison is more straightforward, I focus on comparing my method with the other inference methods designed to test the null hypothesis about the ATT. In Appendix C of the Supplementary Materials, I also include results for \cite{ConformalInference}, \cite{CaoDowd2019}, and \cite{Abadie2010}'s methods.\footnote{When conducting the simulations for the sizes of the tests, $\beta_t = 0$ for all $t \in \mathcal{T}_1$, so both the sharp null and the null hypothesis of $\beta_{0,n} = 0$ are true.} 
 \par 
 I conduct the simulations by fitting a linear factor model to the pre-treatment CO2 emissions from transport per capita data from \cite{Andersson2019}.\footnote{I first estimate the number of factors using the Singular Value Thresholding method of \cite{OptimalSVThresholding}, so the number of factors is chosen to be the number of singular values greater than the median singular value times 2.858. Using this method, I estimate that there are five factors. I then estimate the factor loadings and factor realizations using Principal Components Analysis.} To allow the number of time periods to vary, I fit the estimated values of the factors $\hat{f}$ and the residuals $\hat{\epsilon}_{it} = Y_{it} - \sum_{r =1}^5 \hat{f}_{tr}\hat{\mu}_{ri}$ to models and use these models for a parametric bootstrap. I fit each factor to an ARIMA model,\footnote{This is done using the auto.arima() function in the forecast package in R, where AIC is used for model selection and QMLE is used to estimate the parameters.} and I sample the idiosyncratic shocks from mean-zero normal distributions that are independent across both unit and time. I set each of the variances of the idiosyncratic shocks to be the same over time but allow for heteroskedasticity across units by using the sample variance of $\{\hat{\epsilon}_{it}\}_{t=1960}^{1989}$ for each $i$. The factor loadings are fixed across the simulation draws.
\begin{figure}
\centering
\begin{subfigure}{.4\textwidth}
  \centering
  \includegraphics[width=\linewidth]{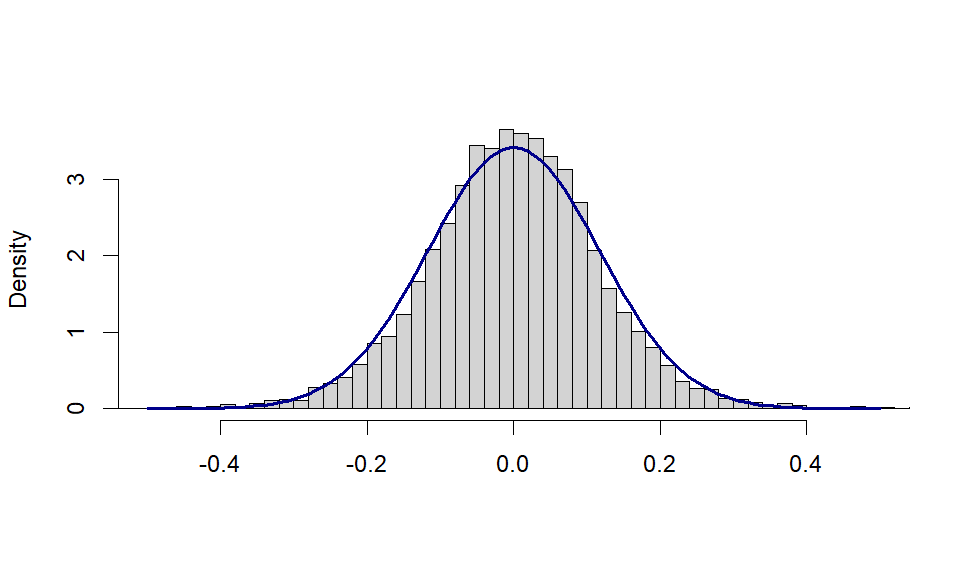}
  \caption{Histogram}
  \label{fig:sub1}
\end{subfigure}%
\begin{subfigure}{.4\textwidth}
  \centering
  \includegraphics[width=.8\linewidth]{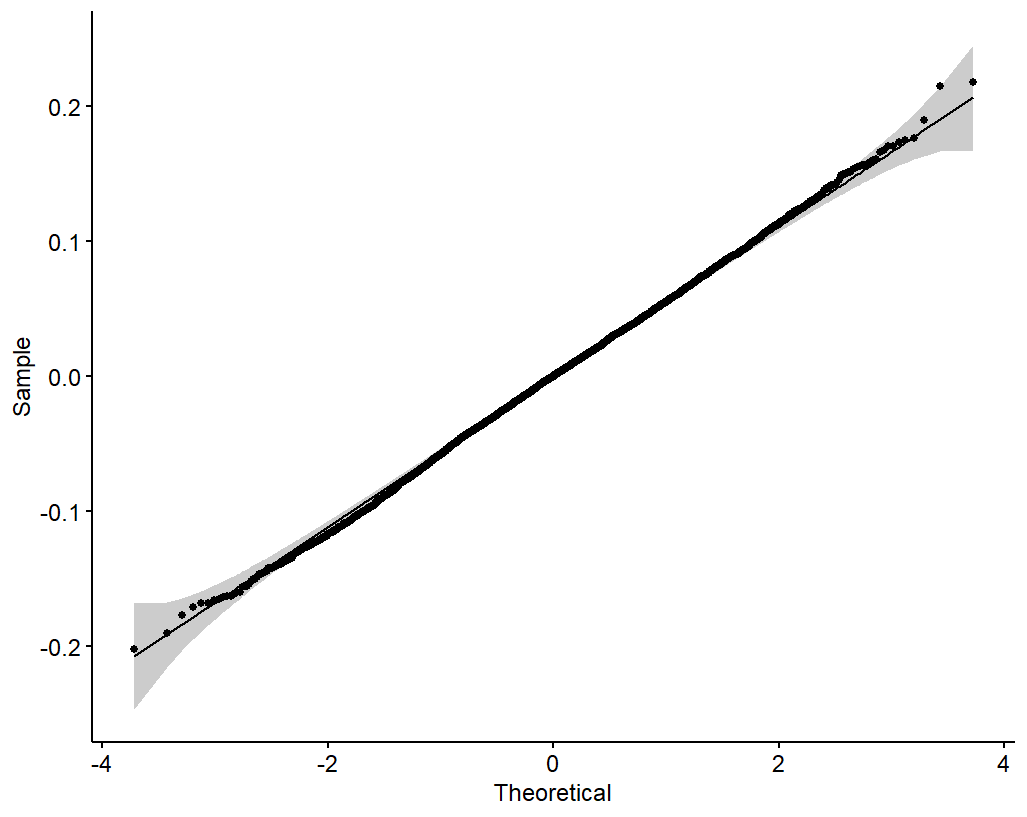}
  \caption{QQ-Plot}
  \label{fig:sub2}
\end{subfigure}
\caption{Normality of the Orthogonalized SCE when $T_0 = 30$ and $T_1 = 16$}
\label{F2}
\end{figure}
\par 
Figure \ref{F2} shows the histogram and Quantile-Quantile plot for the estimates of $\beta_{0,n}$ for the Orthogonalized SCE from 5000 draws when the sample size is the same as in \cite{Andersson2019}. We can see that, even with this relatively small sample size, the normal approximation holds relatively well, with slightly greater concentration near zero and slightly more outliers than expected. Using the Anderson-Darling test for normality gives a p-value 0.05092. 

\begin{table}[htb]\centering\caption{Main Size Results \label{Table22}}\scalebox{.75}{
\begin{threeparttable}
\begin{tabular}{l *{6}{c}} \hline
                         & \multicolumn{1}{c}{Orthogonalized} & \multicolumn{1}{c}{IV-SCE} & \multicolumn{1}{c}{Subsampling} & \multicolumn{1}{c}{Cross-Fitting} & \multicolumn{1}{c}{SDID} & \multicolumn{1}{c}{ArCo} \\
                         & \multicolumn{1}{c}{SCE t-test} & \multicolumn{1}{c}{t-test} & \multicolumn{1}{c}{Method} & \multicolumn{1}{c}{t-test} & \multicolumn{1}{c}{Placebo test} & \multicolumn{1}{c}{Method} \\
\toprule 
\multicolumn{5}{l}{\textbf{Rejection Rates with $\alpha = .05$}} \\ 
\midrule
$T_0 = 30$,$T_1 = 4$  & 0.099 & 0.332 & 0.000 & 0.127 & 0.061 & 0.167 \\
$T_0 = 30$,$T_1 = 16$  & 0.032 & 0.090 & 0.002 & 0.118 & 0.066 & 0.135 \\
$T_0 = 60$,$T_1 = 16$  & 0.042 & 0.074 & 0.009 & 0.135 & 0.047 & 0.103 \\
$T_0 = 30$,$T_1 = 32$  & 0.035 & 0.138 & 0.034 & 0.151 & 0.076 & 0.140 \\
$T_0 = 60$,$T_1 = 32$  & 0.051 & 0.099 & 0.035 & 0.135 & 0.061 & 0.085 \\ 
\hline
\end{tabular}
\begin{tablenotes}
      \small
      \item Notes: All simulations are conducted with a thousand replications.
\end{tablenotes}
    \end{threeparttable}}
\end{table}

Table \ref{Table22} contains the size results for the inference methods discussed above when the nominal size is $.05$. I also include a version of my proposed estimator that skips the orthogonalization step, denoted IV-SCE. Looking at the results, we can see that the rejection rates for the Orthogonalized SCE are generally below the nominal levels, except when the number of post-treatment time periods is very small. Since the asymptotic results involve both $T_0$ and $T_1$ growing, it is unsurprising that the method can fail to control size when either $T_0$ or $T_1$ is very small. When comparing this to the version that skips the orthogonalization step in column 2, we see that it consistently overrejects, suggesting that orthogonalizing with respect to the control weights plays an important role in controlling for size. \footnote{Here, for the t-test of the naive version of the method which skips the orthogonalization step, I calculate its standard errors using an analogous method, where the same Series HAC estimator is used to calculate its asymptotic variance, ignoring the uncertainty in the estimated nuisance parameters. The degrees of freedom for the t-distribution are also chosen the same way.} The Placebo method of \cite{Arkhangelskyetal2021} tends to experience only a small amount of over-rejection, while the subsampling method controls size (albeit by being overly conservative). On the other hand, the cross-fitting t-test and the ArCo method exhibit greater degrees of over-rejection. 

\begin{table}[htb]\centering\caption{Main Power Results \label{Table32}}\scalebox{.75}{
\begin{threeparttable}
\begin{tabular}{l *{6}{c}} \hline
                         & \multicolumn{1}{c}{Orthogonalized} & \multicolumn{1}{c}{IV-SCE} & \multicolumn{1}{c}{Subsampling} & \multicolumn{1}{c}{Cross-Fitting} & \multicolumn{1}{c}{SDID} & \multicolumn{1}{c}{ArCo} \\
                         & \multicolumn{1}{c}{SCE t-test} & \multicolumn{1}{c}{t-test} & \multicolumn{1}{c}{Method} & \multicolumn{1}{c}{t-test} & \multicolumn{1}{c}{Placebo test} & \multicolumn{1}{c}{Method} \\
\toprule 
\multicolumn{5}{l}{\textbf{Rejection Rates with $\alpha = .05$}} \\ 
\midrule
$T_0 = 30$,$T_1 = 4$  & 0.532 & 0.784 & 0.000 & 0.504 & 0.478 & 0.601 \\
$T_0 = 30$,$T_1 = 16$  & 0.858 & 0.969 & 0.101 & 0.882 & 0.790 & 0.940 \\
$T_0 = 60$,$T_1 = 16$  & 0.943 & 0.971 & 0.109 & 0.917 & 0.858 & 0.991 \\
$T_0 = 30$,$T_1 = 32$  & 0.956 & 1.000 & 0.682 & 0.955 & 0.880 & 0.988 \\
$T_0 = 60$,$T_1 = 32$  & 0.992 & 1.000 & 0.721 & 0.990 & 0.942 & 1.000 \\ 
\hline
\end{tabular}
\begin{tablenotes}
      \small
      \item Notes: All simulations are conducted with a thousand replications.
\end{tablenotes}
    \end{threeparttable}}
\end{table}

\par 
Table \ref{Table32} contains the power for the same set of inference methods. The simulations are done under the same conditions as before, but now under the alternative hypothesis of $\beta_{t} = -.25$ in each post-treatment time period, so $\beta_{0,n}$ is similar to its estimated value in the empirical application. In Appendix C, I also include the size-adjusted power.\footnote{The power is adjusted for size by finding the threshold for the p-value that makes the method's actual size equal to the nominal size under the null, and then seeing how often the p-value falls below this threshold under the alternative. While this adjustment is infeasible to do in practice, it is useful for comparing the power of inference methods that are over-rejecting under the null.} The cross-fitting t-test and the ArCo method have similar, or in some cases larger, power than the Orthogonalized SCE t-test, while the SDID placebo has slightly lower power, and the subsampling method has the lowest power. Overall, however, the Orthogonalized SCE t-test generally has the highest power of the tests that control for size in these simulations. 
\par 
One potential point of caution is that in small samples, the p-values may be fairly sensitive to the choice of the smoothing parameter $K$ for the Series HAC estimator. In the simulations, I use the method of \cite{Sun2013} to choose $K$ and it appears to perform quite well. In applications, it may be worth checking the robustness of the statistical significance of results to small changes in $K$. Also, one obvious drawback of using the set of moment conditions I have focused on here is that it requires a set of units to be used as instruments, while the others do not. This motivates further investigation of alternative sets of moment conditions that can identify ATT, which could mean an alternative choice of instruments, such as using shift-share instruments as suggested by \cite{SyntheticIV} or using lagged values of outcomes as instruments, or one of the non-instrument-based approaches to identification discussed in section \ref{s: extensions}.

\section{Variations and Connection to Existing Methods}\label{s: extensions}

\subsection{Staggered Adoption}

In the working example, I focused on the setting with a single treated unit. However, the method can be extended to other versions of the SCE, provided that there is a set of moment conditions that identify whatever function of treatment effects is of interest. For example, if there is a set of treated units with indices in $\mathcal{N}_1$ who become treated at the same time, then the same moment equations could be used with $Y_{0t}$ replaced with $\sum_{i \in \mathcal{N}_1} Y_{it}/|\mathcal{N}_1|$. This case would have very similar sufficient conditions as the single treated unit case.\footnote{As noted by others such as \cite{Arkhangelskyetal2021}, one difference is that the estimator of the ATT can still be consistent when $T_1$ is fixed as long as $|\mathcal{N}_1| = N_1 \rightarrow \infty$ and $||\hat{\delta}||_2 \overset{p}{\rightarrow} 0$ sufficiently fast.} 
\par 
In a staggered adoption setting, this can be extended by estimating separate control weights for each treatment block while using units in other treatment blocks as instruments. Suppose there are $G$ treatment groups, and there are time periods $\{1,...,T\}$ where treatment starts for the $g$th treatment group after $T_g$. Furthermore, assume we have a set of never-treated units with indices $j \in \mathcal{J}$ and units in the $g$th treatment group of indices $i \in \mathcal{G}_g$. To identify the overall ATT, the moment condition
 $$\sum_{g=1}^G
    (\sum_{t > T_g} (\Bar{Y}_{gt} - \beta - Y_{\mathcal{J},t}\delta^{g})/(T-T_g)G)$$
may be used, where $\Bar{Y}_{gt} = \sum_{i \in \mathcal{G}_g} Y_{gt}/|\mathcal{G}_g|$ is the average outcome in group $g$ in time period $t$ and $\delta = (\delta^1,...,\delta^G)$ is now the $GJ$ vector of nuisance parameters with $\delta^g \in \Delta^J$. In order to partially identify the weights, for each pair of treated groups $k,g \in \{1,...,G\}$ with $k \ne g$, we may use the moment condition
$$ \sum_{t \leq \min \{T_g,T_k\} }\Bar{Y}_{kt}(\Bar{Y}_{gt} - Y_{\mathcal{J},t}\delta^g )/\min \{T_g,T_k\}.$$
\par 
Since units need to be left untreated during the periods they are used as instruments, it is important to use only time periods when both groups are untreated. As long as there is an initial block of time periods in which all units are untreated $\mathcal{T}_0 = \{t : t < \min \{T_1,...,T_G\}\}$, this approach can be used for estimating the weights of all treatment groups. The estimated ATT will now be given by
     \begin{equation*}
  \Tilde{\beta} =  \sum_{g=1}^G
    (\sum_{t > T_g} (\Bar{Y}_{gt} - Y_{\mathcal{J},t}\hat{\delta}^{g})/(T-T_g)G)
\end{equation*}
\begin{equation*}
     - \sum_{g=1}^G  \sum_{k\ne g} \hat{\eta}_{g(G-1)+k}
    \sum_{t \leq \min \{T_g,T_k\} }\Bar{Y}_{k t}(\Bar{Y}_{gt} - Y_{\mathcal{J},t}\hat{\delta}^g )/\min \{T_g,T_k\}.
\end{equation*}
Note that under a linear factor model, there will often exist values of $\eta$ that orthogonalize $ M(\beta,\delta^1,...,\delta^G)$ with respect to all of the control weights. For example, with stationary factors, this will hold when $\eta$ satisfies:
$$ E[f_t] \mu_{\mathcal{J}} - \sum_{k=1}^G  \sum_{g \ne k} \eta_{g(G-1)+k} \mu_{\mathcal{G}_g} E[f_t'f_t] \mu_{\mathcal{J}} =0,$$
and there will exist such an $\eta$ when $G(G-1) > R$. Furthermore, if we wanted to estimate a separate average treatment effect for each group, we could instead let $\beta \in \mathbb{R}^G$ and have a separate moment condition to identify the ATT for each treatment group, so we could swap out the moment condition for the overall ATT for the $G$ moment conditions:
\begin{equation*}
\begin{pmatrix}  
  \sum_{t > T_1} (\Bar{Y}_{1t} - \beta_1 - Y_{\mathcal{J},t}\delta^{1})/(T-T_1) \\
    ... \\
     \sum_{t > T_G} (\Bar{Y}_{Gt} - \beta_G - Y_{\mathcal{J},t}\delta^{G})/(T-T_G)
\end{pmatrix}.
\end{equation*}
The general formal results allow for $\beta$ to be multidimensional, as long as we use the F-test for inference as described in section \ref{s: VarianceEstimation}. However, because they rely on the dimension of $\beta$ being fixed, we may only expect good finite sample performance from this version when the number of groups is fairly small.\footnote{Relatedly, the formal results also assume the number of moment conditions is fixed, so they will not be applicable to cases where the number of moment conditions is large due to the size of $G$. In this case, it may be sensible to modify the estimation of $\eta$ to take into account that it is high-dimensional.}

\subsection{Connection to Traditional Synthetic Control}

In the traditional SCE first introduced by \cite{AandG}, the SC weights are estimated by fitting to a set of predictor variables during pre-treatment time periods, and these predictor variables often include time-invariant covariates. If we let $X_{i}$ be an $L \times 1$ vector of time-invariant predictor variables, then suppose the outcomes follow the model:
\begin{equation*}
    Y_{it} = \beta_{it}D_{it} + X_{i}'\gamma_t + \epsilon_{it}
\end{equation*}
where $\gamma_t$ is the an $L \times 1$ capturing the effect of these predictors in the $t$-th time periods. Then, if we use the moment conditions 
\begin{equation*}
    g(\beta,\delta) = \begin{pmatrix}
        E[X_{0} - X_{\mathcal{J}}\delta] \\
        E[\sum_{t \in \mathcal{T}_1} (Y_{0t} - \beta - Y_{\mathcal{J},t}\delta)/T_1]
    \end{pmatrix}
\end{equation*}
will identify the ATT if $\epsilon_{it}$ are mean-zero. Then in this case, the Orthogonalized SCE is given by
\begin{equation*}
    \Tilde{\beta} =  \sum_{t \in \mathcal{T}_1}(Y_{0t} - Y_{\mathcal{J},t}\hat{\delta})/T_1 - \hat{\eta}(X_0 - X_{\mathcal{J}}\hat{\delta}) 
\end{equation*}
where the additional term can be thought of as adjusting for the bias due to failing to perfectly match on the variables in $X_{i}$.
\par 
However, there are likely unobserved confounding variables, in which case an unbiased SC cannot be estimated simply by matching on observable covariates. This is often the justification for also fitting on pre-treatment outcomes, although, as mentioned before, fitting directly on pre-treatment outcomes generally does not properly adjust for bias from the remaining omitted variables. Relating this model back to the linear factor model, by letting $ X_i' \gamma_t = \mu_i f_t$, we could think of this as the special case where the factor loadings are observable. This can be formalized with the factor model used by \cite{Abadie2010}, where
\begin{equation*}
    Y_{it} = \beta_{it}D_{it} + \rho_t + X_i'\gamma_t + \mu_i f_t + \epsilon_{it}
\end{equation*}
where $\rho_t$ is a shared common factor whose behavior can be unrestricted due to the fact that it will cancel out when $Y_{\mathcal{J},t}\delta$ is subtracted from $Y_{0t}$ for $\delta \in \Delta^J$. We can now use the moment conditions:
\begin{equation*}
    g(\beta,\delta) = \begin{pmatrix}
        E[X_0 - X_{\mathcal{J}}\delta] \\
        E[\sum_{t \in \mathcal{T}_0} Z_t (Y_{0t} - Y_{\mathcal{J},t}\delta)/T_0] \\
        E[\sum_{t \in \mathcal{T}_1} (Y_{0t} - Y_{\mathcal{J},t}\delta)/T_1]
    \end{pmatrix}
\end{equation*}
These moment conditions will identify $\beta_{0,n}$ when $X_0 = \sum_{j=1}^J X_j \delta_j$ and $\mu_0 = \mu_{\mathcal{J}}\delta$. As before, $\delta$ will generally be partially identified, but $\beta_{0,n}$ would still be point-identified under conditions analogous to those in Condition \hyperref[Ass1]{1}.

\subsection{Connection to Other Estimators}

As mentioned before, the estimator presented here is related to the cross-fitting t-test method of \cite{t-test2022}. In fact, the estimator of \cite{t-test2022} can be viewed as using $2 K$ moment conditions:

\begin{equation*}
    g(\beta,\delta^1,...,\delta^K,\alpha) = \begin{pmatrix}
        E[\sum_{t \in \mathcal{H}_1} (Y_{0t} - Y_{\mathcal{J},t}\delta^1)/|\mathcal{H}_1| ] - \alpha \\
        ... \\
         E[\sum_{t \in \mathcal{H}_K} (Y_{0t} - Y_{\mathcal{J},t}\delta^K)/|\mathcal{H}_K| ] - \alpha \\
         E[\sum_{t \in \mathcal{T}_1} (Y_{0t} - \beta - Y_{\mathcal{J},t}\delta^1)/T_1] - \alpha \\
         ... \\
          E[\sum_{t \in \mathcal{T}_1} (Y_{0t} - \beta - Y_{\mathcal{J},t}\delta^K)/T_1] - \alpha
    \end{pmatrix},
\end{equation*}
 where $\delta^,....,\delta^K$ are the different vectors of control weights used for each moment condition with $\delta^k \in \Delta^J$ for each $k$. The overall vector of nuisance parameters will then be $\delta = vec(\delta^1,...,\delta^K)$ with dimension $KJ$. The $k$ pre-treatment moment condition paired with the $k$th post-treatment moment condition will identify the ATT for a set of weights $\delta_0^1,...,\delta_0^K$ satisfying
$$E[\sum_{t \in \mathcal{T}_1} (Y_{0t}(0) - Y_{\mathcal{J},t}\delta_0^k)/T_1] =  E[\sum_{t \in \mathcal{H}_1} (Y_{0t} - Y_{\mathcal{J},t}\delta_0^k)/|\mathcal{H}_k| ], $$
where $Y_{0t}(0)$ is the untreated potential outcome of the treated unit. In other words, we would want to identify weight vectors that keep the SC's bias the same before and after treatment, so that the debiasing step works as intended. As noted by \cite{t-test2022}, if the outcomes follow a factor model 
\begin{equation*}
    Y_{it}(0) = \rho_t + \mu_i f_t + \epsilon_{it},
\end{equation*}
where $\rho_t$ is again a latent factor all units have a common level of exposure to, and the remaining factors $f_t$ are mean-invariant, then each pre-treatment moment condition paired with its corresponding post-treatment moment condition will identify the ATT for any set of weights with $\delta^k \in \Delta^J$.\footnote{\cite{t-test2022} also generalize this model somewhat by not requiring that $f_t$ be exactly mean-zero.} Because the control weights are completely unidentified by these moment conditions, it is entirely up to the penalty function to consistently estimate the weights. \cite{t-test2022} choose the control weights $\delta^k$ to minimize the MSE in all pre-treatment time periods, excluding $\mathcal{H}_k$, so
\begin{equation*}
    f(\delta) = \sum_{k=1}^K E[\sum_{t \in \mathcal{T}_0 \backslash \mathcal{H}_k} (Y_{0t} - Y_{\mathcal{J},t}\delta^k)^2/|\mathcal{T}_0 \backslash \mathcal{H}_k| ]
\end{equation*}
Note that this is consistent with the general results above since $\hat{f}$ is allowed to be stochastic. Also if we ignore the presence of $\rho_t$, which will cancel out for any $\delta^k \in \Delta^J$, the partial derivatives of $g_k(\beta,\delta)$ with respect to $\delta_j^k$ are given by
$$-E[\sum_{t \in \mathcal{T}_1} f_t/T_1]\mu_j +  E[\sum_{t \in \mathcal{H}_k} f_t/|\mathcal{H}_k| ]\mu_j,$$
and so the moment conditions will already be orthogonal with respect to the nuisance parameters, and there is no need to estimate a new linear combination of them to satisfy that property. However, with the framework of this paper in mind, a natural extension to consider is what if $f_t$ are not close to being mean-invariant. 

Following the constraint that $ \eta \partial_\beta g(\theta_{0,n}) - I_m = 0$, we would constrain $\eta$ so that 
$ \sum_{k=K+1}^{2K} \eta_{k} = 1$. For the partial derivatives of the intercept coefficients $\alpha = (\alpha_1...,\alpha_K)$, we would want $\eta_{K+k} = -\eta_{k}$ for each $k \in \{1,...,K\}$. By imposing these constraints, we can express the estimator as
\begin{equation*}
    \Tilde{\beta} = \sum_{k=1}^K \hat{\eta}_k( ( \sum_{t \in \mathcal{T}_1} (Y_{0t} - Y_{\mathcal{J},t}\hat{\delta}^k - \hat{\alpha}_k)/T_1  - \sum_{t \in \mathcal{H}_k} (Y_{0t} - Y_{\mathcal{J},t}\hat{\delta}^k - \hat{\alpha}_k)/|\mathcal{H}_k|)).
\end{equation*}
$$ = \sum_{k=1}^K \hat{\eta}_k( \sum_{t \in \mathcal{T}_1} (Y_{0t} - Y_{\mathcal{J},t}\hat{\delta}^k)/T_1 - \sum_{t \in \mathcal{H}_k} (Y_{0t} - Y_{\mathcal{J},t}\hat{\delta}^k)/|\mathcal{H}_k|).$$
Suppose it is possible to take a weighted average of the pre-treatment blocks such that
$$\sum_{k=1}^K \eta_k E[\sum_{t \in \mathcal{T}_1} f_t/T_1 ] = E[\sum_{t \in \mathcal{T}_1} f_t/T_1 ] = \sum_{k=1}^K \eta_k E[\sum_{t \in \mathcal{H}_k} f_t/ |\mathcal{H}_k|]. $$
This $\eta$ could be estimated by setting the $KJ$ partial derivatives equal to zero, so
$$E[\sum_{t \in \mathcal{T}_1} Y_{\mathcal{J},t}/T_1] = \sum_{k=1}^K \eta_k E[\sum_{t \in \mathcal{H}_k} Y_{\mathcal{J},t}/ |\mathcal{H}_k|].$$ 
This allows us to debias the estimator while weakening the assumption that the bias remains constant over time.\footnote{The t-test inference procedure using the Series variance estimator could be used, and it is not immediately clear whether the method of \cite{t-test2022} can be extended to this case. Either way, further work comparing these inference procedures is an interesting direction for future work.} On the other hand, because $\hat{\eta}$ is being estimated with data from all time periods, we should not expect it to have the same properties as the cross-fit estimates of the control weights.
\par 
 This estimator is also related to the SDID estimator of \cite{Arkhangelskyetal2021}, as it includes both weights over pre-treatment time periods and weights over control units. \cite{Arkhangelskyetal2021} place a different weight on each pre-treatment time period and additionally constrain the time weights to be non-negative so $\eta$ is also a weighted average. Constraining $\eta$ to be a weighted average is more helpful in this case, as $\eta$ is high-dimensional, so this serves as a method of regularization. Ideally, the SDID time-period weights would make the estimated ATT less sensitive to the control weights, and vice versa. However, the SDID estimates weights for both time periods and units by minimizing MSE rather than directly imposing the orthogonality condition, so the weights may not satisfy this property in general. Also, the fact that they assign different weights to each pre-treatment time period, rather than assigning weights to a small number of blocks of time periods, is relevant to whether the Adaptivity condition can be achieved. This can be seen by considering a version of the Orthogonalized SCE that uses moment conditions based on the average residual across several pre-treatment blocks being zero. Specifically, assume we split the pre-treatment time periods into $B$ blocks and use the moment conditions:
 \begin{equation*}
    g(\beta,\delta) = \begin{pmatrix}
        E[\sum_{t \in \mathcal{H}_1} (Y_{0t} - Y_{\mathcal{J},t}\delta)/|\mathcal{H}_1| ] \\
        ... \\
         E[\sum_{t \in \mathcal{H}_B} (Y_{0t} - Y_{\mathcal{J},t}\delta)/|\mathcal{H}_B| ] \\
         E[\sum_{t \in \mathcal{T}_1} (Y_{0t} - \beta - Y_{\mathcal{J},t}\delta)/T_1] .
    \end{pmatrix}
    \end{equation*}
In a setting with staggered treatment adoption, a natural way to split the pre-treatment time periods into $B$ blocks is to use when treatment starts in other groups (so there is a single block for the earliest treatment group, two blocks for the next, and so on). Since the orthogonality property relies on averaging over time periods within a pre-treatment block, this approach would be particularly suited to cases with a few treatment groups, rather than when the number of treatment groups is similar to the number of time periods. However, more work is needed to understand the properties of both this method and SDID in staggered adoption contexts with many treatment blocks. Under the linear factor model, the moments can identify the ATT if $B$ is at least as big as the number of latent factors and the factors are nonstationary, so their mean is different across the blocks. Using the same approach as before, this would give the orthogonalized estimator:
\begin{equation*}
    \Tilde{\beta} =  \sum_{t \in \mathcal{T}_1}(Y_{0t} - Y_{\mathcal{J},t}\hat{\delta})/T_1 - \sum_{b=1}^{B} \hat{\eta}_{b} \sum_{t \in \mathcal{H}_b} (Y_{0t} - Y_{\mathcal{J},t}\hat{\delta})/| \mathcal{H}_b|.
\end{equation*}
\par
For the Orthogonalized SCE presented here, because each element of $\eta$ is multiplied by a moment condition that involves averaging over many time periods, it is being multiplied by something which will be close to zero. Hence, the orthogonality of $\eta$ is automatically achieved. Because the SDID assigns different weights to each pre-treatment time period, the same property does not automatically hold. This is related to why \cite{Arkhangelskyetal2021} require that the $L_2$ norm of the weights converges to zero sufficiently fast to show asymptotic normality of their estimated ATT. On the other hand, this difference also allows the SDID estimator greater flexibility in weighting pre-treatment time periods based on how relevant they are to the post-treatment. Therefore, SDID is perhaps better suited to cases where a good fit can be achieved with weights spread out over pre-treatment time periods and control units.

{
\linespread{1}\selectfont
\singlespacing
\bibliographystyle{econ-aea}
\bibliography{references}}

\onehalfspacing

\section*{Appendix A}\phantomsection\label{ApA}

\textbf{Proof of Lemma \hyperref[L2.1]{2.1}:} Let $\gamma = (\delta,\eta_q)$ and $\hat{\gamma} = (\hat{\delta},\hat{\eta}_q)$ where $\eta_q$ is the $q$-th row of $\eta$. Since the $q$-th element on the orthogonalized sample moment conditions $\hat{M}_q$ for $q \in \{1,...,m\}$ are twice continuously differentiable in $\gamma$, for each $q$ there exists $\Bar{\gamma}$ with $||\Bar{\gamma} - \gamma_{0,n}||_{\mathcal{E}} \leq ||\hat{\gamma} - \gamma_{0,n}||_{\mathcal{E}}$ such that:

$$\sqrt{n}(\hat{M}_q(\beta_{0,n},\hat{\gamma}) - \hat{M}_q(\beta_{0,n}, \gamma_{0,n}))$$
$$ = \sqrt{n} \partial_\gamma \hat{M}_q(\beta_{0,n},\gamma_{0,n})(\hat{\gamma} - \gamma_{0,n}) + \sqrt{n}(\hat{\gamma} - \gamma_{0,n})'\partial^2_\gamma \hat{M}_q(\beta_{0,n},\Bar{\gamma})(\hat{\gamma} - \gamma_{0,n}).$$
The magnitude of the first term on the right-hand side is less than or equal to 
$$\sqrt{n} ||\partial_\gamma \hat{M}_q (\beta_{0,n},\gamma_{0,n})||_{\mathcal{D}} ||\hat{\gamma} - \gamma_{0,n}||_{\mathcal{E}} \leq \sqrt{n} O_p(r_J/\sqrt{n})o_p(1/r_J) = o_p(1).$$
The second term on the right-hand side is equal to $$\sqrt{n}(\hat{\delta} - \delta_{0,n})'\partial_\delta^2 \hat{M}_q(\beta_{0,n},\Bar{\gamma})(\hat{\delta} - \delta_{0,n}) + 2\sqrt{n}(\hat{\eta}_q - \eta_{0,n,q})\partial_{\delta} \hat{g}(\beta_{0,n},\Bar{\delta})(\hat{\delta} - \delta_{0,n}).$$ 
In the linear case, $\sqrt{n}(\hat{\delta} - \delta_{0,n})'\partial_\delta^2 \hat{M}_q(\beta_{0,n},\Bar{\gamma})(\hat{\delta} - \delta_{0,n})  = 0$ and $$\partial_\delta \hat{g}(\beta_{0,n},\Bar{\delta}) = \partial_\delta \hat{g}(\theta_{0,n}) = \partial_\delta \hat{g}(\hat{\theta}).$$ Therefore, 
$$|(\hat{\eta}_q - \eta_{0,n,q})\partial_\delta \hat{g}(\beta_{0,n},\Bar{\delta})(\hat{\delta} - \delta_{0,n})| $$
$$ = |\hat{\eta}_q\partial_\delta \hat{g}(\hat{\theta})(\hat{\delta} - \delta_{0,n}) - \eta_{0,n,q} \partial_\delta \hat{g}(\theta_{0,n})(\hat{\delta} - \delta_{0,n})| $$
$$\leq (||\hat{\eta}_q\partial_\delta \hat{g}(\hat{\theta})||_{\mathcal{D}} + ||\eta_{0,n,q} \partial_\delta \hat{g}(\theta_{0,n})||_{\mathcal{D}}) ||\hat{\delta} - \delta_{0,n}||_{\mathcal{E}} $$
$$= (||\hat{\eta}_q \partial_\delta \hat{g}(\hat{\theta})||_{\mathcal{D}} + ||\partial_\delta \hat{M}_q(\theta_{0,n},\eta_{0,n})||_{\mathcal{D}}) ||\hat{\delta} - \delta_{0,n}||_{\mathcal{E}}$$
$$= (O_p(r_J\log(n)/\sqrt{n}) + O_p(r_J/\sqrt{n}))o_p(1/(r_J\log(n))) = o_p(1/\sqrt{n}).$$
Otherwise, for the non-linear case, there exists $\Bar{\delta}^*$ such that $||\Bar{\delta}^* - \delta_{0,n}||_{\mathcal{E}} \leq ||\hat{\delta} - \delta_{0,n}||_{\mathcal{E}}$ 
$$\text{and } \partial_\delta \hat{g}(\beta_{0,n},\Bar{\delta}) = \partial_\delta \hat{g}(\theta_{0,n}) + (\Bar{\delta} - \delta_{0,n})' \partial_\delta^2 \hat{g}(\beta_{0,n},\Bar{\delta}^*).$$
Therefore, using Assumptions 2.1.2 and 2.1.4,
$$\sqrt{n}||(\hat{\eta}_q - \eta_{0,n,q})\partial_\delta \hat{g}(\beta_{0,n}, \Bar{\delta})(\hat{\delta} - \delta_{0,n})||_{\mathcal{D}} \leq $$
$$\sqrt{n}||\hat{\eta}_q - \eta_{0,n,q}||_{\mathcal{E}}( ||\partial_\delta \hat{g}(\theta_{0,n})||_{\mathcal{D}} ||\hat{\delta} - \delta_{0,n}||_{\mathcal{E}}$$
$$+   \max_{s \in \{1,...,Q\}}|(\Bar{\delta} - \delta_{0,n})'\partial_\delta^2 \hat{g}_s (\beta_{0,n},\Bar{\delta}^*)(\hat{\delta} - \delta_{0,n})|)$$
$$\leq \sqrt{n} o_p(n^{-1/4}/\sqrt{r_J}) O_p(r_J)o_p(n^{-1/4}/\sqrt{r_J}) $$
$$+ \sqrt{n}o_p(n^{-1/4}/\sqrt{r_J}) ||\hat{\delta} - \delta_{0,n}||_{\mathcal{E}}^2 \max_{s\in \{1,...,Q\}} \max eig(\partial_\delta^2 \hat{g}_s (\beta_{0,n},\Bar{\delta}^*)).$$

Then, using $\epsilon > 0$ defined by Assumption 2.1.3, because $||\Bar{\delta}^* - \delta_{0,n}||_{\mathcal{E}} \leq ||\hat{\delta} - \delta_{0,n}||_{\mathcal{E}} < \epsilon$ wpa1, then

$$\max_{s\in \{1,...,Q\}} \max eig(\partial_\delta^2 \hat{g}_s (\beta_{0,n},\Bar{\delta}^*)) \leq \max_{s \in \{1,...,Q\}} \sup_{\delta: ||\delta - \delta_{0,n}||_{\mathcal{E}} < \epsilon} \max eig(\partial_\delta^2 \hat{g}_s(\beta_{0,n},\delta)) = O_p(r_J) $$
$$\text{ and }\max eig(\partial_\delta^2 \hat{M}_q (\beta_{0,n},\Bar{\gamma})) \leq \sup_{\gamma: ||\gamma-\gamma_{0,n}||_{\mathcal{E}} < \epsilon} \max eig(\partial_\delta^2 \hat{M}_q(\beta_{0,n},\gamma)) = O_p(r_J) \text{wpa1}.$$
Hence,
$$\sqrt{n}||(\hat{\eta}_q - \eta_{0,n,q})\partial_\delta \hat{g}(\beta_{0,n}, \Bar{\delta})(\hat{\delta} - \delta_{0,n})||_{\mathcal{D}} \leq $$
$$\leq \sqrt{n} o_p(n^{-1/4}/\sqrt{r_J}) O_p(r_J)o_p(n^{-1/4}/\sqrt{r_J}) $$
$$+ \sqrt{n}o_p(n^{-1/4}/\sqrt{r_J}) o_p(n^{-1/2}/\sqrt{r_J})O_p(r_J) = o_p(1).$$
Also,
$$|\sqrt{n}(\hat{\delta} - \delta_{0,n})'\partial_\delta^2 \hat{M}_q(\beta_{0,n},\Bar{\gamma})(\hat{\delta} - \delta_{0,n})| \leq \sqrt{n}||\hat{\delta} - \delta_{0,n}||_{\mathcal{E}}^2 \max eig(\partial_\delta^2 \hat{M}_q(\beta_{0,n},\Bar{\gamma}))$$
$$\leq \sqrt{n} o_p(n^{-1/2}/r_J)O_p(r_J) = o_p(1).$$
Therefore, 
$$\sqrt{n}(\hat{M}_q(\beta_{0,n},\hat{\gamma}) - \hat{M}_q(\beta_{0,n}, \gamma_{0,n})) = o_p(1),$$
for each $q \in \{1,...,m\}.$

\textbf{Proof of Lemma \hyperref[L3.1]{3.1}:} If $S_{0,n}$ is a singleton for all $n \geq N$ for some $N$, then the results is implied directly by Lemma A1. Suppose then that, for infinitely many $n$, $S_{0,n}$ is not a singleton. Let $\{ \epsilon_n \}_{n \in \mathbb{N}}$ be a decreasing sequence with $\epsilon_n > 0$ and $\epsilon_n \rightarrow 0$. Then let $N_{\epsilon_n}(\theta_{0,n},\eta_{0,n})$ be an open $\epsilon$-ball using the $||\cdot||_{\mathcal{E}}$ norm centered at $(\theta_{0,n},\eta_{0,n})$. For some fixed $\zeta > 0$ to be specified below, let $S_{0,n}^\zeta \coloneqq \{ (\theta,\eta) \in S_{0,n} : f(\theta,\eta) \leq f(\theta_{0,n},\eta_{0,n}) + \zeta\}$. Then let $\tau_n = d_H(S_{0,n}^\zeta, \{(\theta_{0,n},\eta_{0,n})\},||\cdot||_{\mathcal{E}})$, where $d_H(,,||\cdot||_{\mathcal{E}})$ denotes the Hausdorff distance using the $||\cdot||_{\mathcal{E}}$ norm. Note $N_{\epsilon_n}^c(\theta_{0,n},\eta_{0,n}) \cap S_{0,n}^\zeta = \emptyset$ exactly when $\tau_n < \epsilon_n$, so $N_{\min\{\epsilon_n,\tau_n/2\}}^c(\theta_{0,n},\eta_{0,n}) \cap S_{0,n}^\zeta \ne \emptyset$ for all $n$. Because $S_{0,n}^\zeta$ is compact and $f$ is continuous on $S_{0,n}^\zeta$ for each $n$ by Assumption \hyperref[Ass3.1]{3.1.1}, $S_{0,n}^\zeta \cap N^c_{\min\{\epsilon_n,\tau/2\}}(\theta_{0,n},\eta_{0,n})$ is also compact for each $n$. Then we can define 
$$\gamma_n := \min_{(\theta,\eta) \in S_{0,n}^\zeta \cap N^c_{\min\{\epsilon_n,\tau_n/2\} }(\theta_{0,n},\eta_{0,n})} f(\theta,\eta) - f(\theta_{0,n},\eta_{0,n}).$$ 
\par 
Note that by Assumption \hyperref[Ass3.1]{3.1.5}, $C_6 |f(\theta_1,\eta_1) - f(\theta_2,\eta_2)|^{\gamma_2} \leq ||\theta_1 - \theta_2||_{\mathcal{E}} + ||\eta_1 - \eta_2||_{\mathcal{E}}$ when $f(\theta_1,\eta_1),f(\theta_2,\eta_2) \leq f(\theta_{0,n},\eta_{0,n}) + C_5$. So if $\zeta$ is chosen such that $\zeta \leq C_5$, then for $\kappa_n = (\gamma_n/(4C_6))^{1/\gamma_2} > 0$, it holds that for all $(\theta_1,\eta_1),(\theta_2,\eta_2) \in S_n^\zeta \coloneqq \{(\theta,\eta) \in \Theta_n \times H \;:\; f(\theta,\eta) \leq f(\theta_{0,n},\eta_{0,n}) + \zeta \}$, if $||\theta_1 - \theta_2||_{\mathcal{E}} + ||\eta_1 - \eta_2||_{\mathcal{E}} < \kappa_n$, then $|f(\theta_1,\eta_1) - f(\theta_2,\eta_2)| < \gamma_n/4$. Letting $S_{0,n}^\zeta(\kappa)$ denote the closed $\kappa$ blow-up of $S_{0,n}$ using the $||\cdot||_{\mathcal{E}}$ norm, we obtain that:
$$\min_{(\theta,\eta) \in S_{0,n}^\zeta(\kappa_n) \cap N^c_{\min\{\epsilon,\tau_n/2\}}(\theta_{0,n},\eta_{0,n})} f(\theta,\eta) - f(\theta_{0,n},\eta_{0,n}) > 3/4\gamma_n.$$
\par 
Since $(\hat{\theta},\hat{\eta}) \in \hat{S}_0$, then $(\hat{\theta},\hat{\eta}) \in \hat{S}_0^\zeta \coloneqq \{(\theta,\eta) \in \hat{S}_0: f(\theta,\eta) \leq f(\theta_{0,n},\eta_{0,n}) + \zeta \}$ when $|f(\hat{\theta},\hat{\eta}) -f(\theta_{0,n},\eta_{0,n})| \leq 3/4\gamma_n$, since by the definition of $\gamma_n$ we have that $3/4 \gamma_n \leq \gamma_n \leq \zeta$. Therefore, since $d_H(\hat{S}_0^\zeta,S_{0,n}^\zeta;||\cdot||_{\mathcal{E}}) < \kappa_n$ implies $\hat{S}_0^\zeta \subset S_{0,n}^\zeta(\kappa_n)$, it follows that
$$P(||\hat{\theta} - \theta_{0,n}||_{\mathcal{E}} + ||\hat{\eta} - \eta_{0,n}||_{\mathcal{E}} < \epsilon) \geq P(||\hat{\theta} - \theta_{0,n}||_{\mathcal{E}} + ||\hat{\eta} - \eta_{0,n}||_{\mathcal{E}} < \min\{\epsilon,\tau_n/2\})$$
$$\geq P(|f(\hat{\theta},\hat{\eta}) - f(\theta_{0,n},\eta_{0,n})| \leq 3/4\gamma_n ; d_H(\hat{S}_0^\zeta,S_{0,n}^\zeta,||\cdot||_{\mathcal{E}}) < \kappa_n).$$
\par 
Let $(\theta_p,\eta_p) = \argmin_{(\theta,\eta) \in \hat{S}_0^\zeta} ||\theta - \theta_{0,n}||_{\mathcal{E}} + ||\eta - \eta_{0,n}||_{\mathcal{E}}$. If $d_H(\hat{S}_0^\zeta,S_{0,n}^\zeta,||\cdot||_{\mathcal{E}}) < \kappa_n$, then $||\theta_p - \theta_{0,n}||_{\mathcal{E}} + ||\eta_p - \eta_{0,n}||_{\mathcal{E}} < \kappa_n$ and therefore $f(\theta_p,\eta_p) < f(\theta_{0,n},\eta_{0,n}) + \gamma_n/4$. By definition of $\hat{\theta}$ and $\hat{\eta}$, $\hat{f}(\hat{\theta},\hat{\eta}) \leq \hat{f}(\theta_p,\eta_p)$. This implies $f(\hat{\theta},\hat{\eta}) - f(\theta_p,\eta_p) \leq |\hat{f}(\hat{\theta},\hat{\eta}) - f(\hat{\theta},\hat{\eta})| + |\hat{f}(\theta_p,\eta_p) - f(\theta_p,\eta_p)|.$
Thus $f(\theta_p,\eta_p) < f(\theta_{0,n},\eta_{0,n}) + \gamma_n/4$ and $|\hat{f}(\hat{\theta},\hat{\eta}) - f(\hat{\theta},\hat{\eta})| + |\hat{f}(\theta_p,\eta_p) - f(\theta_p,\eta_p)| < \gamma_n/4$ together imply $f(\hat{\theta},\hat{\eta}) - f(\theta_{0,n},\eta_{0,n}) < \gamma_n/2$. But since $d_H(\hat{S}_0^\zeta,S_{0,n}^\zeta,||\cdot||_{\mathcal{E}}) < \kappa_n$ implies $f(\theta_p,\eta_p) < f(\theta_{0,n},\eta_{0,n}) + \gamma_n/4$, we have that 
$$P(||\hat{\theta} - \theta_{0,n}||_{\mathcal{E}} + ||\hat{\eta} - \eta_{0,n}||_{\mathcal{E}} < \epsilon_n) $$
$$\geq P(|\hat{f}(\hat{\theta},\hat{\eta}) - f(\hat{\theta},\hat{\eta})| + |\hat{f}(\theta_p,\eta_p) - f(\theta_p,\eta_p)| < \gamma_n/4; d_H(\hat{S}_0^\zeta,S_{0,n}^\zeta,||\cdot||_{\mathcal{E}}) < \kappa_n).$$
$$\geq P(2 \sup_{(\theta,\eta) \in \hat{S}_0} |f(\theta,\eta) - \hat{f}(\theta,\eta)| < \gamma_n/4; d_H(\hat{S}_0^\zeta,S_{0,n}^\zeta,||\cdot||_{\mathcal{E}}) < \kappa_n).$$
By Assumption \hyperref[Ass3.1]{3.1.5}, for all $(\theta,\eta) \in S_{0,n}^\zeta$, $|f(\theta,\eta) - f(\theta_{0,n})| \geq C_4 \min \{ (||\theta - \theta_{0,n}||_{\mathcal{E}} + ||\eta - \eta_{0,n}||_{\mathcal{E}})^{\gamma_1}, C_3 \},$
and for sufficiently large $n$, $\epsilon_n^{\gamma_1} \leq C_3$. So for sufficiently large $n$, $\gamma_n \geq C_4 \epsilon_n^{\gamma_1}$. Since $\kappa_n = (\gamma_n/(4C_6))^{1/\gamma_2}$, this implies $\kappa_n^{\gamma_2}4C_6 \geq C_4 \epsilon_n^{\gamma_1}$ so $\kappa_n \geq \epsilon_n^{\gamma_1/\gamma_2} (C_4/(4C_6))^{1/\gamma_2} $. Therefore, for $n$ sufficiently large,
$$P(||\hat{\theta} - \theta_{0,n}||_{\mathcal{E}} + ||\hat{\eta} - \eta_{0,n}||_{\mathcal{E}} < \epsilon) $$
$$\geq P(2 \sup_{(\theta,\eta) \in \hat{S}_0} |f(\theta,\eta) - \hat{f}(\theta,\eta)| < C_4 \epsilon_n^{\gamma_1}/4; d_H(\hat{S}_0^\zeta,S_{0,n}^\zeta,||\cdot||_{\mathcal{E}}) < \epsilon_n^{\gamma_1/\gamma_2} (C_4/(4C_6))^{1/\gamma_2} ).$$
Then because
$$\sup_{(\theta,\eta) \in \hat{S}_0}|f(\theta,\eta) - \hat{f}(\theta,\eta)| = O_p(c_n) \text{ and } d_H(\hat{S}_0^\zeta,S_{0,n}^\zeta,||\cdot||_{\mathcal{E}}) = O_p(\max \{\lambda_\delta,\lambda_\eta,a_n \})$$
by Lemma A1, for any decreasing sequence $\{\epsilon_n\}_{n \in \mathbb{N}}$ with $\epsilon_n > 0$, $\epsilon_n \rightarrow 0$, and 
$$\epsilon_n /( \max \{\lambda_\delta^{\gamma_2},\lambda_\eta^{\gamma_2},a_n^{\gamma_2},c_n\})^{1/\gamma_1}) \rightarrow \infty,$$ 
we have that $P(||\hat{\theta} - \theta_{0,n}||_{\mathcal{E}} < \epsilon_n) \rightarrow 1$ and $P(||\hat{\eta} - \eta_{0,n}||_{\mathcal{E}} < \epsilon_n ) \rightarrow 1$. Therefore, 
$$||\hat{\delta} - \delta_{0,n}||_{\mathcal{E}} + ||\hat{\eta}-\eta_{0,n}||_{\mathcal{E}} \leq ||\hat{\theta} - \theta_{0,n}||_{\mathcal{E}} + ||\hat{\eta}-\eta_{0,n}||_{\mathcal{E}} = O_p((\max \{\lambda_\delta^{\gamma_2},\lambda_\eta^{\gamma_2},a_n^{\gamma_2},c_n\})^{1/\gamma_1}).$$

\textbf{Proof of Proposition \hyperref[P4.1]{4.1}:} I first show the consistency of $\Tilde{\beta}$. Note that 
$$||\hat{M}(\beta,\hat{\delta},\hat{\eta}) - \hat{M}(\beta,\delta_{0,n},\eta_{0,n})||_2 = ||\hat{\eta}\hat{g}(\beta,\hat{\delta}) - \eta_{0,n}\hat{g}(\beta,\delta_{0,n})||_2 \leq ||\hat{\eta}||_2 ||\hat{g}(\beta,\hat{\delta}) - g(\beta,\hat{\delta})||_2$$
$$+ ||\hat{\eta}||_2 ||g(\beta,\hat{\delta}) - g(\beta,\delta_{0,n})||_2 + ||\hat{\eta} - \eta_{0,n}||_2 ||g(\beta,\delta_{0,n}||_2 $$
$$+ ||\eta_{0,n}||_2 ||g(\beta,\delta_{0,n}) - \hat{g}(\beta,\delta_{0,n})||_2.$$
Note that since $||\eta_{0,n}||_2 = O(1)$ and $||\hat{\eta} - \eta_{0,n}||_2 = o_p(1)$, $||\hat{\eta}||_2 = O_p(1)$. Then $\sup_{\beta \in B}||g(\beta,\hat{\delta}) - g(\beta,\delta_{0,n})||_2 = o_p(1)$ by the continuous mapping theorem, and
$$\sup_{\beta \in B}||\hat{g}(\beta,\hat{\delta}) - g(\beta,\hat{\delta})||_2 = o_p(1) \text{ and } \sup_{\beta \in B}||g(\beta,\delta_{0,n}) - \hat{g}(\beta,\delta_{0,n})||_2 = o_p(1)$$ 
by Assumption \hyperref[Ass4.1]{4.1.1}. Hence,
$\sup_{\beta \in B} ||\hat{M}(\beta,\hat{\delta},\hat{\eta}) - \hat{M}(\beta,\delta_{0,n},\eta_{0,n})||_2 = o_p(1).$ Therefore, since $W_n \overset{p}{\rightarrow} W$ where $W$ is positive definite, we have that 
$$\sup_{\beta \in B}|\hat{M}(\beta,\hat{\delta},\hat{\eta})'W_n\hat{M}(\beta,\hat{\delta},\hat{\eta}) - \hat{M}(\beta,\delta_{0,n},\eta_{0,n})'W_n\hat{M}(\beta,\delta_{0,n},\eta_{0,n})| = o_p(1).$$
 Let $\hat{Q}(\beta)$ be equal to the objective function from equation \eqref{eq: GMM Estimator}. Combining this with Assumption \hyperref[Ass4.1]{4.1.1} and $W_n \overset{p}{\rightarrow} W$ gives $\sup_{\beta \in B} |\hat{Q}(\beta) - Q(\beta)| = o_p(1)$, where $Q(\beta) = M(\beta,\delta_{0,n},\eta_{0,n})'W M(\beta,\delta_{0,n},\eta_{0,n}).$ Let $\epsilon > 0$. By the definition of $\Tilde{\beta}$, $\hat{Q}(\Tilde{\beta}) < \hat{Q}(\beta_{0,n}) + \epsilon/3$. Then by the uniform convergence we have that $Q (\Tilde{\beta}) < \hat{Q}(\Tilde{\beta}) + \epsilon/3$ and $\hat{Q}(\beta_{0,n}) < Q(\beta_{0,n}) + \epsilon/3$ wpa1. Combining these inequalities gives $Q(\Tilde{\beta}) < Q(\beta_{0,n}) + \epsilon$ wpa1. By the strong identification condition of Assumption \hyperref[Ass4.1]{4.1.4} and $W$ being positive definite, for some $\Bar{C} > 0$, 
$$||\Tilde{\beta} - \beta_{0,n}||_2^2/\Tilde{C} \leq C^2 ||M (\Tilde{\beta},\delta_{0,n},\eta_{0,n})||_2^2/\Tilde{C} \leq  Q(\Tilde{\beta}) < \epsilon$$ 
wpa1. So we have that $\Tilde{\beta} - \beta_{0,n} \overset{p}{\rightarrow} 0$.
\par 
By Assumption \hyperref[Ass4.1]{4.1.4}, $\partial_\beta^2 Q(\beta_{0,n}) = \partial_\beta M(\theta_{0,n},\eta_{0,n})'W \partial_\beta M(\theta_{0,n},\eta_{0,n}) \rightarrow M_\beta' W M_\beta$ and $M_\beta' W M_\beta$ is positive definite. Using Assumption \hyperref[Ass4.1]{4.1.2} and the consistency of $\hat{\delta}$ and $\hat{\eta}$, $\partial_\beta \hat{g}(\beta_{0,n},\hat{\delta})'\hat{\eta}'W_n \hat{\eta} \partial_\beta \hat{g}(\beta_{0,n},\hat{\delta}) - \partial_\beta M(\theta_{0,n},\eta_{0,n})'W \partial_\beta M(\theta_{0,n},\eta_{0,n}) \overset{p}{\rightarrow} 0.$ So $\partial_\beta^2 \hat{Q}(\beta_{0,n}) \overset{p}{\rightarrow} M_\beta' W M_\beta$. Since $\hat{Q}(\beta)$ is twice continuously differentiable, 
$$\hat{Q}(\Tilde{\beta}) = \hat{Q}(\beta_{0,n}) + \partial_\beta \hat{Q}(\beta_{0,n})(\Tilde{\beta} - \beta_{0,n}) + (\Tilde{\beta} - \beta_{0,n})' \partial_\beta^2 \hat{Q}(\Bar{\beta}) (\Tilde{\beta} - \beta_{0,n})/2,$$
for some $\Bar{\beta}$ with $||\Bar{\beta} - \beta_{0,n}||_2 \leq ||\Tilde{\beta} - \beta_{0,n}||_2$. Then since $\hat{Q}(\Tilde{\beta}) \leq \hat{Q}(\beta_{0,n})$, we have that
$0 \geq \partial_\beta \hat{Q}(\beta_{0,n})(\Tilde{\beta} - \beta_{0,n}) + (\Tilde{\beta} - \beta_{0,n})' \partial_\beta^2 \hat{Q}(\Bar{\beta}) (\Tilde{\beta} - \beta_{0,n})/2.$ Using Assumption \hyperref[Ass4.1]{4.1.2} and the consistency of $\Tilde{\beta}$, 
$$||\partial_\beta^2 \hat{Q}(\Bar{\beta}) - \partial_\beta^2 \hat{Q}(\beta_{0,n})||_2 \leq \sup_{\beta : ||\beta - \beta_{0,n}||_2 < ||\Tilde{\beta} - \beta_{0,n}||_2} ||\partial_\beta^2 \hat{Q}(\beta) - \partial_\beta^2 \hat{Q}(\beta_{0,n})||_2 = o_p(1).$$
Then 
$0 \geq \partial_\beta \hat{Q}(\beta_{0,n})(\Tilde{\beta} - \beta_{0,n}) + (\Tilde{\beta} - \beta_{0,n})' M_\beta' W M_\beta (\Tilde{\beta} - \beta_{0,n})/2 + o_p(||\Tilde{\beta} - \beta_{0,n}||_2^2).$
Multiplying both sides by $\frac{n}{(1 + \sqrt{n}||\Tilde{\beta} - \beta_{0,n}||_2)^2}$ gives 
$$\frac{\sqrt{n}||\Tilde{\beta} - \beta_{0,n}||_2}{(1 + \sqrt{n}||\Tilde{\beta} - \beta_{0,n}||_2)^2} \sqrt{n} \partial_\beta \hat{Q}(\beta_{0,n}) $$
$$+ \sqrt{n}(\Tilde{\beta} - \beta_{0,n})'M_\beta'W M_\beta \sqrt{n}(\Tilde{\beta} - \beta_{0,n})/2 + o_p(1) \leq 0.$$
Then if $\sqrt{n}||\Tilde{\beta} - \beta_{0,n}||_2 \rightarrow \infty$, $ \sqrt{n}(\Tilde{\beta} - \beta_{0,n})'M_\beta'W M_\beta \sqrt{n}(\Tilde{\beta} - \beta_{0,n})/2 \leq o_p(1)$. But since $'M_\beta'W M_\beta$ is positive definite, this implies that $\sqrt{n}(\Tilde{\beta} - \beta_{0,n}) = o_p(1)$ which is a contradiction. Therefore, $\sqrt{n}(\Tilde{\beta} - \beta_{0,n}) = O_p(1)$. 
\par
I now show that asymptotic normality of $\sqrt{n}(\Tilde{\beta} - \beta_{0,n})$ when Assumption \hyperref[Ass4.1]{4.1.6} additionally holds. Since $\beta_{0,n}$ is bounded away from the boundary of $B$, wpa1 the first order condition is satisfied,
$\hat{M}(\Tilde{\beta},\hat{\delta},\hat{\eta})'W_n\partial_\beta \hat{M}(\Tilde{\beta},\hat{\delta},\hat{\eta}) = 0.$
Using the Mean Value Theorem, for some $\Bar{\beta}$ such that $||\Bar{\beta} - \beta_{0,n}||_2 \leq ||\Tilde{\beta} - \beta_{0,n}||_2$, $(\hat{M}(\beta_{0,n},\hat{\delta},\hat{\eta}) + \partial_\beta \hat{M}(\Bar{\beta},\hat{\delta},\hat{\eta})(\Tilde{\beta} - \beta_{0,n}))'W_n\partial_\beta \hat{M}(\Tilde{\beta},\hat{\delta},\hat{\eta}) = 0.$
Therefore,
$$\sqrt{n}(\Tilde{\beta} - \beta_{0,n}) = $$
$$(\partial_\beta \hat{M}(\Bar{\beta},\hat{\delta},\hat{\eta})'W_n\partial_\beta \hat{M}(\Tilde{\beta},\hat{\delta},\hat{\eta}))^{-1}\partial_\beta \hat{M}(\Tilde{\beta},\hat{\delta},\hat{\eta})'W_n
\sqrt{n}\hat{M}(\beta_{0,n},\hat{\delta},\hat{\eta}) + o_p(1).$$
Using the adaptivity condition from Lemma \hyperref[L2.1]{2.1},
$$\sqrt{n}\hat{M}(\beta_{0,n},\hat{\delta},\hat{\eta}) = \sqrt{n}\hat{M}(\theta_{0,n},\eta_{0,n}) + o_p(1) \overset{d}{\rightarrow} N(0, V_M).$$
Again using Assumption \hyperref[Ass4.1]{4.1.2}, $\partial_\beta \hat{M}(\Bar{\beta},\hat{\delta},\hat{\eta}) - \partial_\beta \hat{M}(\theta_{0,n},\eta_{0,n}) = o_p(1)$ and $$\partial_\beta \hat{M}(\Tilde{\beta},\hat{\delta},\hat{\eta}) - \partial_\beta \hat{M}(\theta_{0,n},\eta_{0,n}) = o_p(1).$$ Then since $\partial_\beta \hat{M}(\theta_{0,n},\eta_{0,n}) = \partial_\beta M(\theta_{0,n},\eta_{0,n}) + o_p(1) = M_\beta + o_p(1)$ and $W_n \overset{p}{\rightarrow} W$, we have that
$\sqrt{n}(\Tilde{\beta} - \beta_{0,n}) \overset{d}{\rightarrow} N(0,V),$ where $$(M_\beta' W M_\beta)^{-1} M_\beta' W V_M W M_\beta (M_\beta' W M_\beta)^{-1}.$$

\textbf{Proof of Proposition \hyperref[P4.2]{4.2}:} The proof proceeds by verifying Assumption 3.1 of \cite{Sun2013} and then replicating the argument in the proof of Theorem 3.1 of \cite{Sun2013} for the case with the sample moment conditions being $\hat{M}(\beta, \hat{\delta},\hat{\eta})$ and with drifting sequences of the true parameter $\beta_{0,n}$. Proposition \hyperref[P4.1]{4.1} holds so $\sqrt{n}(\Tilde{\beta} - \beta_{0,n}) = O_p(1)$. Assumption \hyperref[Ass2.1]{2.2} imposes that $\beta_{0,n}$ is an interior point of $B$ bounded away from the boundary and Assumption \hyperref[Ass2.1]{2.1} imposes that $\hat{g}$ is twice continuously differentiable in $\theta$ which implies that $\hat{g}(\beta,\hat{\delta})$ is twice continuously differentiable in $\beta$. Let $q \in \{1,..., Q\}$.

Without loss of generality, assume the time periods are indexed so $\mathcal{T}_{b_q} = \{1,...,T_{b_q}\}$.

Then note that for $\lambda_n,r \in [0,1]$,
$$\sum_{i=1}^{\lfloor r T_{b_q} \rfloor} \partial_\beta \hat{g}_{q,i} (\beta_{0,n} + \lambda_n (\Tilde{\beta} - \beta_{0,n}),\hat{\delta})/T_{b_q} = (\sum_{i=1}^{\lfloor r T_{b_q} \rfloor}\partial_\beta g_{q,i}(\beta_{0,n} + \lambda_n(\Tilde{\beta} - \beta_{0,n}),\hat{\delta})/T_{b_q} - $$
$$\sum_{i=1}^{\lfloor r T_{b_q} \rfloor}\partial_\beta g_{q,i}(\beta_{0,n} + \lambda_n(\Tilde{\beta} - \beta_{0,n}),\delta_{0,n}))/T_{b_q} + \sum_{i=1}^{\lfloor r T_{b_q} \rfloor} \partial_\beta g_{q,i}(\beta_{0,n} + \lambda_n(\Tilde{\beta} - \beta_{0,n}),\delta_{0,n})/T_{b_q}.$$
The difference between the first and second terms is converging in probability to zero uniformly in $\lambda_n,r \in [0,1]$ by Assumption \hyperref[Ass4.2]{4.2.1}. Therefore, by Assumption \hyperref[Ass4.2]{4.2.3} and $||\hat{\eta} - \eta_{0,n}||_1 = o_p(1)$, we have that
$$ \hat{\eta}
\begin{pmatrix}
    \sum_{t \in \mathcal{T}_{b_1}} \partial_\beta g_{1,t}(\beta_{0,n} + \lambda_n(\Tilde{\beta} - \beta_{0,n}),\hat{\delta})/T_{b_1} \\
    ... \\
    \sum_{t \in \mathcal{T}_{b_Q}} \partial_\beta g_{Q,t}(\beta_{0,n} + \lambda_n(\Tilde{\beta} - \beta_{0,n}),\hat{\delta})/T_{b_Q}
\end{pmatrix} \overset{p}{\rightarrow} r M_\beta,$$

$$\sum_{i=1}^{\lfloor rn \rfloor} \partial_\beta \hat{M}_i(\beta_{0,n} + \lambda_n (\Tilde{\beta} - \beta_{0,n}),\hat{\delta},\hat{\eta})/n \overset{p}{\rightarrow} rM_\beta,$$
uniformly in $\lambda_n,r \in [0,1]$. Also, $M_\beta$ is full rank by Assumption \hyperref[Ass4.1]{4.1}. By Lemma A3 and Assumption \hyperref[Ass4.2]{4.2.4},
$$V_M^{-1/2} \sqrt{\min \{T_{b_1},...,T_{b_Q}\}} \hat{\eta} \begin{pmatrix}
    \sum_{t \in \mathcal{T}_{b_1}} \phi_k (\frac{t}{T_{b_1}}) g_{1,t}(\beta_{0,n},\hat{\delta})/T_{b_1} \\
    ... \\
    \sum_{t \in \mathcal{T}_{b_Q}} \phi_k (\frac{t}{T_{b_Q}}) g_{Q,t}(\beta_{0,n}, \hat{\delta})/T_{b_Q}
\end{pmatrix} $$
$$ = V_M^{-1/2}\hat{\eta}
\begin{pmatrix}
    \sqrt{\min \{T_{b_1},...,T_{b_Q}\}/T_{b_1}} \sum_{t \in \mathcal{T}_{b_1}} ( \phi_k (\frac{t}{T_{b_1}}) - \phi_k (\frac{t+1}{T_{b_1}}) ) g_{1,t}(\beta_{0,n}, \hat{\delta})/\sqrt{T_{b_1}} \\
    ... \\
    \sqrt{\min \{T_{b_1},...,T_{b_Q}\}/T_{b_Q}} \sum_{t \in \mathcal{T}_{b_Q}} ( \phi_k (\frac{t}{T_{b_Q}}) - \phi_k (\frac{t+1}{T_{b_Q}}) ) g_{Q,t}(\beta_{0,n}, \hat{\delta})/\sqrt{T_{b_Q}}
\end{pmatrix}
$$
$$= V_M^{-1/2} \eta_{0,n}
\begin{pmatrix}
    \sqrt{\min \{T_{b_1},...,T_{b_Q}\}/T_{b_1}} \sum_{t \in \mathcal{T}_{b_1}} ( \phi_k (\frac{t}{T_{b_1}}) - \phi_k (\frac{t+1}{T_{b_1}}) ) g_{1,t}(\beta_{0,n}, \delta_{0,n})/\sqrt{T_{b_1}} \\
    ... \\
    \sqrt{\min \{T_{b_1},...,T_{b_Q}\}/T_{b_Q}} \sum_{t \in \mathcal{T}_{b_Q}} ( \phi_k (\frac{t}{T_{b_Q}}) - \phi_k (\frac{t+1}{T_{b_Q}}) ) g_{Q,t}(\beta_{0,n}, \delta_{0,n})/\sqrt{T_{b_Q}} 
\end{pmatrix}  $$
$$ + o_p(1) = V_M^{-1/2} \sqrt{\min \{T_{b_1},...,T_{b_Q}\}} \hat{\eta} \begin{pmatrix}
    \sum_{t \in \mathcal{T}_{b_1}} \phi_k (\frac{t}{T_{b_1}}) g_{1,t}(\beta_{0,n},\hat{\delta})/T_{b_1} \\
    ... \\
    \sum_{t \in \mathcal{T}_{b_Q}} \phi_k (\frac{t}{T_{b_Q}}) g_{Q,t}(\beta_{0,n}, \hat{\delta})/T_{b_Q}
\end{pmatrix} + o_p(1) \overset{d}{\rightarrow} \xi_k, $$
for each $k \in \{0,...,K\}$, where I have normalized $\phi_k(\frac{T_{b_q}+1}{T_{b_q}}) = 0$ for each $q$. I now extend the proof of Theorem 3.1 of \cite{Sun2013}. 
For the $b$-th block of time periods with indices in $\mathcal{T}_b$, let $\mathcal{T}_b(t)$ denote the $t$th time period in that block (so $\mathcal{T}_b = \{ \mathcal{T}_b(1),...,\mathcal{T}_b(T_b)\}$).
Then, for each $q$, define the partial sums  that use the first $t$ time periods as 
$$S_{q,t} (\beta) := \sum_{i = \mathcal{T}_{b_q}(1)}^{\mathcal{T}_{b_q(t)}} g_{q,i}(\beta,\hat{\delta}).$$

Then, since the moment conditions are twice continuously differentiable in $\beta$, 
$$ S_{q,t}(\beta)/\sqrt{T_{b_q}}  = S_{q,t}(\beta_{0,n})/\sqrt{T_{b_q}} + (\sum_{i = \mathcal{T}_{b_q}(1)}^{\mathcal{T}_{b_q(t)}} g_{q,i}(\Tilde{\beta}_t,\hat{\delta})/T_{b_q})\sqrt{n}(\Tilde{\beta} - \beta_{0,n})$$

for all $q$, where $\Tilde{\beta}_t = \beta_{0,n} + \lambda_n \pdot (\Tilde{\beta} - \beta_{0,n})$ for some $\lambda_n \in [0,1]^p$ where $\pdot$ denotes the element-wise product. From Proposition \hyperref[P4.1]{4.1} and $W = I_m$, 
$$\sqrt{n}(\Tilde{\beta} - \beta_{0,n}) = \sqrt{n}(M_\beta' M_\beta)^{-1} M_\beta' \hat{M}(\beta_{0,n},\hat{\delta},\hat{\eta}) + o_p(1)$$ 
$$ = (M_\beta' M_\beta)^{-1} M_\beta' \sqrt{n}
\begin{pmatrix}
    S_{1,T_{b_1}}(\beta_{0,n})/T_{b_1} \\
    ... \\
    S_{Q,T_{b_Q}}(\beta_{0,n})/T_{b_Q}
\end{pmatrix} + o_p(1).$$
Therefore, $S_{q,t}(\Tilde{\beta})/\sqrt{T_{b_q}} =$
$$ S_{q,t}(\beta_{0,n})/\sqrt{T_{b_q}} + (\sum_{i = \mathcal{T}_{b_q}(1)}^{\mathcal{T}_{b_q(t)}} g_{q,i}(\Tilde{\beta}_t,\hat{\delta})/T_{b_q}) (M_\beta' M_\beta)^{-1} M_\beta' \sqrt{n}
\begin{pmatrix}
    S_{1,T_{b_1}}(\beta_{0,n})/T_{b_1} \\
    ... \\
    S_{Q,T_{b_Q}}(\beta_{0,n})/T_{b_Q}
\end{pmatrix} + o_p(1)$$
$$ = S_{q,t}(\beta_{0,n})/\sqrt{T_{b_q}} - \frac{t}{n}S_{q,T_{b_q}}(\beta_{0,n})/\sqrt{T_{b_q}} + o_p(1) ,  \text{ uniformly over }t\text{. Then, }$$
$$\hat{\eta} \begin{pmatrix}
    \sum_{t \in \mathcal{T}_{b_1}} \phi_k(\frac{t}{T_{b_1}}) g_{1,t} (\Tilde{\beta},\hat{\delta})/\sqrt{T_{b_1}} \\
    ... \\
     \sum_{t \in \mathcal{T}_{b_Q}} \phi_k(\frac{t}{T_{b_Q}}) g_{Q,t} (\Tilde{\beta},\hat{\delta})/\sqrt{T_{b_Q}}
\end{pmatrix} = $$
$$\hat{\eta} \begin{pmatrix}
    \sum_{t \in \mathcal{T}_{b_1}} (\phi_k(\frac{t}{T_{b_1}}) - \phi_k(\frac{t+1}{T_{b_1}})) \frac{\sqrt{n}}{T_{b_1}}(S_{1,t}(\beta_{0,n}) - \frac{t}{T_{b_1}} S_{1,T_{b_1}}(\beta_{0,n})) \\ 
    ... \\
    \sum_{t \in \mathcal{T}_{b_Q}} (\phi_k(\frac{t}{T_{b_Q}}) - \phi_k(\frac{t+1}{T_{b_Q}})) \frac{\sqrt{n}}{T_{b_Q}}(S_{Q,t}(\beta_{0,n}) - \frac{t}{T_{b_Q}} S_{Q,T_{b_Q}}(\beta_{0,n}))
\end{pmatrix} + o_p(1)$$
$$\hat{\eta}  \begin{pmatrix}
   \frac{\sqrt{n}}{T_{b_1}} \sum_{t \in \mathcal{T}_{b_1}} \phi_k(\frac{t}{T_{b_1}}) (g_{1,t}(\beta_{0,n},\hat{\delta}) - \frac{1}{T_{b_1}} S_{1,T_{b_1}}(\beta_{0,n})) \\ 
    ... \\
     \frac{\sqrt{n}}{T_{b_Q}} \sum_{t \in \mathcal{T}_{b_Q}} \phi_k(\frac{t}{T_{b_Q}}) (g_{Q,t}(\beta_{0,n},\hat{\delta}) - \frac{1}{T_{b_Q}} S_{1,T_{b_Q}}(\beta_{0,n}))
\end{pmatrix} + o_p(1)$$
$$=  \sqrt{n}\hat{\eta} \begin{pmatrix}
    \sum_{t \in \mathcal{T}_{b_1}} \phi_k(\frac{t}{T_{b_1}}) (g_{1,t}(\beta_{0,n},\hat{\delta})/T_{b_1} \\
    ... \\
     \sum_{t \in \mathcal{T}_{b_Q}} \phi_k(\frac{t}{T_{b_Q}}) (g_{Q,t}(\beta_{0,n},\hat{\delta})/T_{b_Q} 
\end{pmatrix} + o_p(1)$$
Again, defining $\phi_k (\frac{T_{b_q}+1}{T_{b_q}}) = 0$ for each $q$, then
$$V_M^{-1/2} \sqrt{n}\hat{\eta} \begin{pmatrix}
    \sum_{t \in \mathcal{T}_{b_1}} \phi_k(\frac{t}{T_{b_1}}) (g_{1,t}(\beta_{0,n},\hat{\delta})/T_{b_1} \\
    ... \\
     \sum_{t \in \mathcal{T}_{b_Q}} \phi_k(\frac{t}{T_{b_Q}}) (g_{Q,t}(\beta_{0,n},\hat{\delta})/T_{b_Q} 
\end{pmatrix} \overset{d}{\rightarrow} \xi_k,$$
jointly for $k = 0,1,...,K$. Therefore,
$$\frac{K-p+1}{K}\mathbb{W}_n \overset{d}{\rightarrow} \frac{K-p+1}{K}\xi_0' (\sum_{k=1}^K \xi_k \xi_k'/K )^{-1}\xi_0 = F_{p,K-p+1}$$
$$\text{and }t_n \overset{d}{\rightarrow} \xi_0^2/\sqrt{\sum_{k=1}^K \xi_k^2/K} = t_K \text{ when }p =1.$$

\textbf{Proof of Corollary \hyperref[C4.1]{4.1}:} I first verify that Assumption \hyperref[Ass3.1]{3.1} holds when Conditions \hyperref[Ass1]{1} and \hyperref[Ass2]{2} hold. Assumption \hyperref[Ass3.1]{3.1.1} holds because $D_n = \Delta^J$ is compact for all $J$, $f(\theta,\eta) = ||\delta||_2^2 + ||\eta||_2^2$, and $\hat{g}$ and $g$ are linear in $\delta$. For Assumption \hyperref[Ass3.1]{3.1.2}, note that for each $q \in 1,...,Q-1$,
$$\sup_{\delta \in \Delta^J}|Z_q^{pre}(Y_0^{pre'} - Y_{\mathcal{J}}^{pre'}\delta)/T_0 - E[Z_q^{pre}(Y_{0}^{pre'} - Y_{\mathcal{J}}^{pre'}\delta)/T_0]|
$$
$$ = \sup_{\delta \in \Delta^J}|Z_q^{pre}(Y_0^{pre'} - Y_{\mathcal{J}}^{pre'}\delta)/T_0 - E[Z_q^{pre} f^{pre}/T_0](\mu_0 - \mu_{\mathcal{J}}\delta)|$$
$$\leq 
\sup_{\delta \in \Delta^J} \{|(Z_q^{pre}f^{pre}/T_0 - E[Z_q^{pre}f^{pre}/T_0])(\mu_0 - \mu_{\mathcal{J}}\delta)' | $$
$$+ |Z_q^{pre}(\epsilon_0^{pre'} - \epsilon_{\mathcal{J}}^{pre'}\delta)/T_0|\} \leq $$
$$\max_{ 1 \leq j \leq J} ||\mu_0 - \mu_j||_2 ||Z_q^{pre}f^{pre}/T_0 - E[Z_q^{pre}f^{pre}/T_0]||_2 + 2\max_{0 \leq j \leq J} |Z_q^{pre}\epsilon_j^{pre'}/T_0|$$
$$ = O(1) O_p(1/\sqrt{T_0}) + O_p(\log(J)/\sqrt{T_0}).$$
Similarly, 
$$\sup_{\beta,\delta \in \Delta^J}| \sum_{t \in \mathcal{T}_1}(Y_{0t} - Y_{\mathcal{J},t}\delta)/T_1 - \beta - E[ \sum_{t \in \mathcal{T}_1}(Y_{0t} - Y_{\mathcal{J},t}\delta)/T_1 - \beta] | = $$ 
$$\sup_{\beta,\delta \in \Delta^J}| \sum_{t \in \mathcal{T}_1}(Y_{0t} - Y_{\mathcal{J},t}\delta)/T_1 - \beta - (E[\sum_{t\in\mathcal{T}_1 } f_t/T_1](\mu_0 - \mu_{\mathcal{J}}\delta) + \beta_{0,n} - \beta)| $$
$$\leq \max_{0 \leq j \leq J}||\sum_{t \in \mathcal{T}_1} f_t/T_1 - E[\sum_{t \in \mathcal{T}} f_t/T_1]||_2||\mu_0 - \mu_j||_2 + 2\max_{0 \leq j \leq J}|\sum_{t \in \mathcal{T}_1}\epsilon_{jt}/T_1|$$
$$= O_p(1/\sqrt{T_1})O(1) + O_p(\log(J)/\sqrt{T_1}).$$
Therefore, $\sup_{\theta \in\Theta_n} ||\hat{g}(\theta) - g(\theta)||_\infty = O_p(\log (J)/\sqrt{\min\{T_0,T_1\}})$. This also shows that $$\sup_{\theta \in\Theta_n} ||\partial_\delta \hat{g}(\theta) - \partial_\delta g(\theta)||_\infty = O_p(\log (J)/\sqrt{\min\{T_0,T_1\}}),$$ so Assumption \hyperref[Ass3.1]{3.1.2} holds with $a_n, b_n = \log (J)/\sqrt{\min \{T_0,T_1\}}$. Because $\hat{f}(\theta,\eta) = f(\theta,\eta)$, Assumption \hyperref[Ass3.1]{3.1.3} holds trivially with $c_n = 0$.
\par 
For Assumption \hyperref[Ass3.1]{3.1.4}, for any $\delta' \in D_n$, let $\delta^* = \argmin_{\delta \in D_{0,n}} ||\delta' - \delta||_2$. Provided that $||\delta' - \delta^*||_2 > 0$, $(\delta' - \delta^*)$ will be orthogonal to the hyperplane $\{ \delta : E[Z^{pre}f^{pre}/T_0]\mu_{\mathcal{J}}\delta = 0\} $ because $D_{0,n} = \{ \delta \in D_n : E[Z^{pre}f^{pre}/T_0]\mu_{\mathcal{J}}(\delta - \delta^*) = 0 \}$. Hence, because $ E[Z^{pre}f^{pre}/T_0]\mu_{\mathcal{J}}$ has non-zero rank,
$$||g(\theta)||_\infty \geq ||g(\theta)||_2/\sqrt{Q} \geq ||E[Z^{pre} f^{pre}/T_0] \mu_J(\delta' - \delta^*)||_2/\sqrt{Q} \geq \sqrt{C/Q} ||\delta' - \delta^*||_2, $$
where $C$ is the minimum eigenvalue of $\mu_{\mathcal{J}}' E[Z^{pre} f^{pre}/T_0]'E[Z^{pre} f^{pre}/T_0] \mu_J$, which is bounded away from zero by Condition \hyperref[Ass2]{2}. 
\par 
For any $\eta \in H$, there exists a subset of $\mathcal{I}$ indices with $|\mathcal{I}| = R$ and the minimum singular value of $E[Z_{\mathcal{I}}^{pre}f^{pre}/T_0]$ is bounded below by some constant $C > 0$. Without loss of generality, assume that the row of $Z^{pre}$ are order so that the indices in $\mathcal{I}$ correspond to the first $R$ rows of $Z^{pre}$. Note that for any $\eta$, there exists $\Tilde{\eta}$ such that $\Tilde{\eta} \in H_{0,n} := \{\eta \in H: \partial_\delta M(\theta,\eta) = 0\}$ for all $n$ and  $\eta_q = \Tilde{\eta}_q$ for all $q > R$. Then $$||\eta \partial_\delta g(\delta)||_\infty
 = || (\eta_{-Q} - \Tilde{\eta}_{-Q})E[Z^{pre}F^{pre}/T_0]\mu_{\mathcal{J}}||_\infty $$
 $$\geq C ||(\eta_{-Q} - \Tilde{\eta}_{-Q} )E[Z_{\mathcal{I}}^{pre}f^{pre}/T_0]||_2 \geq C ||(\eta_{\mathcal{I}} - \Tilde{\eta}_{\mathcal{I}} )E[Z_{\mathcal{I}}^{pre}f^{pre}/T_0]||_2$$
 $$  \geq C'||\eta_{\mathcal{I}} - \Tilde{\eta}_{\mathcal{I}}||_2,$$
 for some $C,C' > 0$. Therefore, Assumption \hyperref[Ass3.1]{3.1.4} holds. 
\par 
Let $H_{0,n} := \{ \eta \in H : \eta g(\theta_{0,n}) \}$. For the first part of Assumption \hyperref[Ass3.1]{3.1.5}, note that, for each $n$, the identified set $D_{0,n} \times H_{0,n}$ and the set $$S = \{ (\delta,\eta) : ||\delta||_2^2 + ||\eta||_2^2 \leq ||\delta_{0,n}||_2^2 + ||\eta_{0,n}||_2^2 \}$$ are convex. By the separating hyperplane theorem, there exists a hyperplane described by the linear equations $Ax = b$ such that for all $\gamma = (\delta,\eta) \in S$ we have that $A\gamma \geq b$ and for all $\gamma \in D_{0,n} \times H_{0,n}$, $A\gamma \leq b$. Then since $(\delta_{0,n},\eta_{0,n}) \in S$ and $(\delta_{0,n},\eta_{0,n}) \in D_{0,n} \times H_{0,n}$,
 $A (\delta_{0,n},\eta_{0,n}) = b$ so $(\delta_{0,n},\eta_{0,n})' \gamma \geq 0$ for all $\gamma$ in the half space with $A \gamma \leq b$. Then it follows that $$||\gamma||_2^2 - ||(\delta_{0,n},\eta_{0,n})||_2^2 \geq ||\gamma - (\delta_{0,n},\eta_{0,n})||_2^2$$ for all $\gamma \in D_{0,n} \times H_{0,n}$. 
 
Therefore, we have that for all $n$, for all $\theta \in \Theta_{0,n}$ and $\eta \in H_{0,n}$, $$|f(\theta,\eta) - f(\theta_{0,n},\eta_{0,n})| \geq  C_4 (||\delta - \delta_{0,n}||_2 + ||\eta - \eta_{0,n}||_2)^2$$ for some constant $ C_4 > 0$ which does not depend on $\theta$, $\eta$, or $n$.
 \par 
 For the second part of Assumption \hyperref[Ass3.1]{3.1.5}, note that for any $\delta_1,\delta_2 \in D_n$, $$ \geq 2||\delta_1 - \delta_2||_2  \geq (||\delta_1||_2 + ||\delta_2||_2) |\;||\delta_1||_2 - ||\delta_2||_2| = 2 | \;||\delta_1||_2^2 - ||\delta_2||_2^2|.$$ Similarly, for any $\delta_1,\delta_2 \in D_n$ and $\eta_1,\eta_2 \in H$, if $C_5 \geq ||\delta_1||_2^2 + ||\eta_1||_2^2 \geq ||\eta_1||_2^2$  and $C_5 \geq ||\delta_2||_2^2 + ||\eta_2||_2^2 \geq ||\eta_2||_2^2$ for some $C_5 > 0$, then, $$ 2 C_5 ||\eta_1 - \eta_2||_2 \geq (||\eta_1||_2 + ||\eta_2||_2)|\;||\eta_1||_2 - ||\eta_2||_2\;| = |\; ||\eta_1||_2^2 - ||\eta_2||_2^2\;|.$$ Then for any $n$ and any $\theta_1,\theta_2 \in \Theta_{n}$ and $\eta_1,\eta_2 \in H$ with $f(\theta_1,\eta_1),f(\theta_2,\eta_2) \leq C_5$, $$||\theta_1 - \theta_2||_2 + ||\eta_1 - \eta_2||_2 \geq ||\delta_1 - \delta_2||_2 + ||\eta_1 - \eta_2||_2 \geq |f(\theta_1,\eta_1) - f(\theta_2,\eta_2)|/\max \{2 C_5, 2\}.$$ Therefore, Assumption \hyperref[Ass3.1]{3.1.5} holds with $\gamma_1 = 2$ and $\gamma_2 = 1$.
\par 
I now verify the conditions of Assumption \hyperref[Ass2.1]{2.1}. Since Assumption \hyperref[Ass3.1]{3.1} holds, by Lemma \hyperref[L3.1]{3.1} we have that $$||\hat{\delta} - \delta_{0,n}||_2 = O_p(\max \{\log (J)/\sqrt{\min \{T_0,T_1\}},\lambda_\delta,\lambda_\eta\}^{\frac{1}{2}})$$ 
Let $\mathcal{P}$ denote the set of indices in which $\delta_{0,n}$ is taking non-zero values, where $|\mathcal{P}| = s$ by Condition \hyperref[Ass3]{3.3}. Then since $||\delta_{0,n}||_1 = ||\hat{\delta}||_1 = 1$, we have that $$||\delta_{0,n} - \hat{\delta}||_1 = ||\delta_{0,n,\mathcal{P}}' - \hat{\delta}_{\mathcal{P}}||_1 + ||\hat{\delta}_{\mathcal{P}^c}||_1 = ||\delta_{0,n,\mathcal{P}} - \hat{\delta}_{\mathcal{P}}||_1 + (1 - ||\hat{\delta}_{\mathcal{P}}||_1) $$
 $$= ||\delta_{0,n,\mathcal{P}} - \hat{\delta}_{\mathcal{P}}||_1 + (||\delta_{\mathcal{P}}^*||_1 - ||\delta_{\mathcal{P}}'||_1) \leq 2 ||\delta_{0,n,\mathcal{P}} - \hat{\delta}_{\mathcal{P}}||_1 \leq $$
$$2\sqrt{s} ||\delta_{0,n,\mathcal{P}} - \hat{\delta}_{\mathcal{P}}||_2 \leq 2 \sqrt{s} ||\delta_{0,n} - \hat{\delta}||_2.$$
Therefore,
$$||\hat{\delta} - \delta_{0,n}||_1 = O_p(\sqrt{s \max \{\log (J)/\sqrt{\min \{T_0,T_1\}},\lambda_\delta,\lambda_\eta\}})$$
$$O_p(\sqrt{s \max \{\lambda_\delta,\lambda_\eta\}})$$
$$ = o_p(1/(\log(J)\log(\min \{T_0,T_1\}))),$$
so Assumption \hyperref[Ass2.1]{2.1.1} holds. For Assumption \hyperref[Ass2.1]{2.1.4}, $\hat{g}$ is linear in $\theta$ and $$||\hat{\eta}\partial_\delta \hat{g}(\theta)||_\infty \leq \lambda_\eta = O_p(\log(J) \log(\min\{T_0,T_1\})/\sqrt{\min\{T_0,T_1\}}).$$ Assumption \hyperref[Ass2.1]{2.1.3}  holds trivially since $\hat{g}$ is linear in $\theta$ and Assumption \hyperref[Ass2.1]{2.1.2} is shown above. To verify Assumption \hyperref[Ass2.1*]{2.1*}, first note that $\eta_{0,n}$ satisfies,
$$\sum_{t \in \mathcal{T}_0} \sum_{q=1}^{Q-1} E[\eta_{0,n,q} Z_{qt} Y_{\mathcal{J},t}]/T_0 + \sum_{t \in \mathcal{T}_1} E[Y_{\mathcal{J},t}]/T_1 =  $$
$$(\sum_{t \in \mathcal{T}_0} \sum_{q=1}^{Q-1} E[\eta_{0,n,q} Z_{qt}f_t] /T_0 + \sum_{t \in \mathcal{T}_1} E[f_{t}]/T_1)\mu_{\mathcal{J}} = 0.$$
Therefore,
$$\sum_{t \in \mathcal{T}_0} \sum_{q=1}^{Q-1} E[\eta_{0,n,q} Z_{qt}f_t] /T_0 + \sum_{t \in \mathcal{T}_1} E[f_{t}]/T_1 = 0,$$
for each $\eta_{0,n}$. As a result, Condition \hyperref[Ass2]{2.3} implies that $\eta_{0,n} = \eta^*$ for some $\eta^*$ when $n$ is sufficient large. 
Then, for partial sums of the population moment conditions using the first $t_0$ pre-treatment time periods and $t_1$ post-treatment time periods,
$$||\sum_{i \in \mathcal{T}_0^{t_0}} \sum_{q=1}^{Q-1} E[\eta_{0,n,q} Z_{qi} Y_{\mathcal{J},i}]/T_0 + \sum_{i \in \mathcal{T}_1^{t_1}} E[Y_{\mathcal{J},i}]/T_1 ||_\infty = O(\log(J)/\sqrt{n})$$
as $T_0,T_1 \rightarrow \infty$, uniformly over $1 \leq t_0 \leq T_0$ and $1 \leq t_1 \leq T_1$. Hence Assumption \hyperref[Ass2.1*]{2.1.1} holds. For Assumption \hyperref[Ass2.1*]{2.1.2*},
 $$||\sum_{i \in \mathcal{T}_0^{t_0}} \sum_{q=1}^{Q-1} \eta_{0,n,q} Z_{qi} Y_{\mathcal{J},i}/T_0 + \sum_{i \in \mathcal{T}_1^{t_1}} Y_{\mathcal{J},i}/T_1  $$
 $$- \sum_{i \in \mathcal{T}_0^{t_0}} \sum_{q=1}^{Q-1} E[\eta_{0,n,q} Z_{qi} Y_{\mathcal{J},i}]/T_0 - \sum_{i \in \mathcal{T}_1^{t_1}} E[Y_{\mathcal{J},i}]/T_1 ||_\infty $$
$$\leq ||\eta_{0,n} ||_1 (|| \sum_{i \in \mathcal{T}_0^{t_0}} \sum_{q=1}^{Q-1} Z_{qi} \epsilon_{\mathcal{J},i}/T_0||_\infty + || \sum_{i \in \mathcal{T}_0^{t_0}} \sum_{q=1}^{Q-1} (E[Z_{qi}f_i] - Z_{qi}f_i)\mu_{\mathcal{J}}/T_0||_\infty $$
$$+ ||\sum_{i \in \mathcal{T}_1^{t_1}} (E[f_i] - f_i)\mu_{\mathcal{J}}/T_1||_\infty + ||\sum_{i \in \mathcal{T}_1^{t_1}} \epsilon_{\mathcal{J},i}/T_1||_\infty)  $$
$$ = O(1) O_p(\log(J)/\sqrt{T_0}) + O(1) O_p(\log(J)/\sqrt{T_1}) = O_p(\log(J)/\sqrt{\min \{T_0,T_1\}}) $$
uniformly over $1\leq t_0\leq T_0$ and $1\leq t_1 \leq T_1$ as $T_0,T_1 \rightarrow \infty$. Assumption \hyperref[Ass2.1*]{2.1.3*} holds trivially because $\hat{g}$ is linear in $\theta$. 
\par 
I now verify the conditions of Assumption \hyperref[Ass4.1]{4.1}. Since $$\sup_{\theta \in \Theta_n} ||\hat{g}(\theta) - g(\theta)||_2 = O_p(\log (J) /\sqrt{\min \{T_0,T_1\}})$$ and  $\log(J) /\sqrt{\min \{T_0,T_1\}} \rightarrow 0$, Assumption \hyperref[Ass4.1]{4.1.1} holds. Because $g(\theta)$ and $\hat{g}(\theta)$ are linear in $\theta$, Assumption \hyperref[Ass4.1]{4.1.2} holds. Also $\partial_\beta M(\beta_{0,n},\delta_{0,n},\eta_{0,n}) = -1 \ne 0$ and $||\beta_1 - \beta_2||_2 = ||M(\beta_1,\delta_{0,n},\eta_{0,n}) -M(\beta_2,\delta_{0,n},\eta_{0,n})||_2$ for all $\beta_1,\beta_2 \in B$ so Assumption \hyperref[Ass4.1]{4.1.4} holds. Then for Assumption \hyperref[Ass4.1]{4.1.5}, since there is single moment condition we can set $W_n = W = 1$ and $$\hat{M}(\beta , \delta_{0,n},\eta_{0,n}) - M(\beta,\delta_{0,n},\eta_{0,n}) = \eta_{0,n} (\hat{g}(0,\delta_{0,n}) - g(0,\delta_{0,n})) \overset{p}{\rightarrow} 0$$ as shown above.
\par 
I prove Assumption \hyperref[Ass4.1]{4.1.3} along with Assumption \hyperref[Ass4.2]{4.2}. Condition \hyperref[Ass3]{3} directly guarantees that Assumption \hyperref[Ass4.2]{4.2.2} is satisfied. For Assumption \hyperref[Ass4.2]{4.2.1}, it holds trivially because $g_i(\theta)$ is linear in $\theta$. Assumption \hyperref[Ass4.2]{4.2.3} holds because $\partial_\beta \hat{M}(\beta,\delta,\eta) = \partial_\beta M(\beta,\delta,\eta) = 1$. 
 \par 
Then for Assumption \hyperref[Ass4.2]{4.2.4} and \hyperref[Ass4.1]{4.1.3}, first note that for $q \in \{1,..,Q-1\}$ and $k \in \{0,1,...,K\}$,
$$\sum_{t \in \mathcal{T}_0} \phi_k(\frac{t}{T_0}) g_{q,t}(\theta_{0,n})/\sqrt{T_0} = \sum_{t \in \mathcal{T}_0} \phi_k(\frac{t}{T_0}) Z_{qt}(\epsilon_{0t} - \sum_{j \in \mathcal{J}} \delta_{0,n,j} \epsilon_{jt})/\sqrt{T_0} + o_p(1),$$
$$= \sum_{t \in \mathcal{T}_0} \phi_k(\frac{t}{T_0}) Z_{qt} \epsilon_{0t}/\sqrt{T_0} + \sum_{j \in \mathcal{J}} \delta_{0,n,j} \sum_{t \in \mathcal{T}_0} \phi_k(\frac{t}{T_0})Z_{qt} \epsilon_{jt}/\sqrt{T_0} + o_p(1)$$
and similarly for $q = Q$, $\sum_{t \in \mathcal{T}_1} \phi_k(\frac{t}{T_1}) g_{q,t}(\theta_{0,n})/\sqrt{T_1} = $
$$ \sum_{t \in \mathcal{T}_1} \phi_k(\frac{t}{T_1})(\beta_t - \beta_{0,n}) + \sum_{t \in \mathcal{T}_1} \phi_k(\frac{t}{T_1}) \epsilon_{0t} + \sum_{j \in \mathcal{J}} \delta_{0,n,j} \sum_{t \in \mathcal{T}_1} \phi_k(\frac{t}{T_1}) \epsilon_{jt} /\sqrt{T_1} + o_p(1)$$
because $\mu_0 = \mu_{\mathcal{J}}\delta_{0,n}$. Because of Condition \hyperref[Ass3]{3.1} and $T_1/T_0 \rightarrow a > 0$, the limit of $V(\theta_{0,n})$ does exist and is equal to $V_g$, for some fixed positive definite matrix $V_g$. Therefore, the conditions of Lemma A2 are satisfied for the sequence of random vectors
\begin{equation*}
    \begin{pmatrix}
        \sqrt{\frac{\min \{T_0,T_1\}}{T_0}}g_{1t}(\theta_{0,n}) \\
        ... \\
        \sqrt{\frac{\min \{T_0,T_1\}}{T_0}}g_{q-1.t}(\theta_{0,n})        
    \end{pmatrix}
\end{equation*}
with $t \in \mathcal{T}_0$ and the sequence of random variables $\sqrt{\frac{\min \{T_0,T_1\}}{T_1}}g_{Qt}(\theta_{0,n})$ with $t \in \mathcal{T}_1$. Then, by Lemma A2 and the mixing conditions in Condition \hyperref[Ass3]{3}, we have that
$$V_g^{-1/2}\sqrt{\min \{T_0,T_1\}}\begin{pmatrix}
    \sum_{t \in \mathcal{T}_0} \phi_k(\frac{t}{T_0}) g_{1,t}(\theta_{0,n})/T_0 \\
    ... \\
    \sum_{t \in \mathcal{T}_0} \phi_k(\frac{t}{T_0}) g_{Q-1,t}(\theta_{0,n})/T_0 \\
    \sum_{t \in \mathcal{T}_1} \phi_k(\frac{t}{T_1})  g_{Q,t}(\theta_{0,n})/T_1    
\end{pmatrix}$$
$$= V_g^{-1/2}\begin{pmatrix}
    \sum_{t \in \mathcal{T}_0} \phi_k(\frac{t}{T_0}) \sqrt{\frac{\min\{T_0,T_1\}}{T_0}} g_{1,t}(\theta_{0,n})/\sqrt{T_0} \\
    ... \\
    \sum_{t \in \mathcal{T}_0} \phi_k(\frac{t}{T_0}) \sqrt{\frac{\min\{T_0,T_1\}}{T_0}} g_{Q-1,t}(\theta_{0,n})/\sqrt{T_0} \\
    \sum_{t \in \mathcal{T}_1} \phi_k(\frac{t}{T_1}) \sqrt{\frac{\min\{T_0,T_1\}}{T_1}} g_{Q,t}(\theta_{0,n})/\sqrt{T_1}    
\end{pmatrix} \overset{d}{\rightarrow} \zeta_k, $$
jointly for $k \in \{0,1,...,K\}$ with $\zeta_k \sim iid N(0,I_Q)$. Then since $V_M(\theta_{0,n},\eta_{0,n}) = \eta_{0,n}V_g(\theta_{0,n})\eta_{0,n}$ and $\hat{M}(\theta_{0,n},\eta_{0,n}) = \eta_{0,n}\hat{g}(\theta_{0,n})$ we also have that

$$V_M^{-1/2}\sqrt{\min \{T_0,T_1\}} \eta_{0,n} \begin{pmatrix}
    \sum_{t \in \mathcal{T}_0} \phi_k(\frac{t}{T_0}) g_{1,t}(\theta_{0,n})/T_0 \\  
    ...
    \\
     \sum_{t \in \mathcal{T}_0} \phi_k(\frac{t}{T_0}) g_{Q-1,t}(\theta_{0,n})/T_0 \\
     \sum_{t \in \mathcal{T}_1} \phi_k(\frac{t}{T_1}) g_{Q,t}(\theta_{0,n})/T_1
\end{pmatrix} \overset{d}{\rightarrow} \xi_k,$$
jointly for $k \in \{0,1,...,K\}$ with $\xi_k \sim iid N(0,1)$ so Assumptions \hyperref[Ass4.1]{4.1.3} and \hyperref[Ass4.2]{4.2.4} hold.

\textbf{Lemma A1 (Rate of Convergence of the Estimated Identified Set)}\phantomsection\label{LA1} Suppose that the conditions of Lemma 3.1 hold. Then for any $\zeta > 0$, $d_H(\hat{S}_0^\zeta,S_{0,n}^\zeta, ||\cdot||_{\mathcal{E}}) = O_p(\max \{ a_n \lambda_\delta, \lambda_\eta \})$ where $\hat{S}_0^\zeta = \{ (\theta,\eta) \in \hat{S}_0 : f(\theta,\eta) \leq f(\theta_{0,n},\eta_{0,n}) + \zeta\}$.

\textbf{Proof:} Note that by the identification condition on $g$ in Assumption 3.1.4, for any $(\theta,\eta) \in \hat{S}_0^\zeta$,

$$C_2 \min \{ ||(\theta,\eta) - S_{0,n}^\zeta||_{\mathcal{E}} , C_1 \} \leq ||g(\theta)||_\infty + ||\eta \partial_\delta g(\theta)||_\infty \leq $$
$$||\hat{g}(\theta)||_\infty + ||g(\theta) - \hat{g}(\theta)||_\infty + ||\eta \partial_\delta \hat{g}(\theta)||_\infty + ||\eta (\partial_\delta g(\theta) - \partial_\delta \hat{g}(\theta))||_\infty$$
$$\leq \lambda_\delta + ||g(\theta) - \hat{g}(\theta)||_\infty + \lambda_\eta +  ||\eta||_1 \sum_{q=1}^Q ||\partial_\delta g_q(\theta) - \partial_\delta \hat{g}_q(\theta)||_\infty.$$
By Assumption 3.1.1, $\sup_{(\theta,\eta) \in \hat{S}_0^\zeta} ||\eta||_1 \leq \sup_{(\theta,\eta) \in S_n^\zeta} ||\eta||_1 = O(1)$. 

Then because $\lambda_\delta,\lambda_\eta, \sup_{\theta \in\Theta_n} ||g(\theta) - \hat{g}(\theta)||_\infty,\sup_{\theta \in\Theta_n} ||\partial_\delta g(\theta) - \partial_\delta \hat{g}(\theta)||_\infty = o_p(1) $, this implies that $\sup_{(\theta, \eta) \in \hat{S}_0^\zeta} ||(\theta,\eta) - S_{0,n}||_{\mathcal{E}}  < C_1$ wpa1. Then wpa1, $\sup_{(\theta,\eta) \in \hat{S}_0^\zeta} ||(\theta,\eta) -S_{0,n}^\zeta ||_{\mathcal{E}} \leq $

$$\lambda_\delta + \lambda_\eta + \sup_{\theta \in \Theta_n} ||g(\theta) - \hat{g}(\theta)||_\infty + \sup_{(\theta, \eta) \in \hat{S}_0^\zeta }||\eta\|_1 ||\partial_\delta g(\theta) - \partial_\delta \hat{g}(\theta)||_\infty$$
$$ = \lambda_\delta + \lambda_\eta + O_p(a_n) + O(1)O_p(a_n) = O_p(\max \{\lambda_\delta,\lambda_\eta,a_n\}).$$

Also note that since $\sup_{(\theta,\eta) \in S_{0,n}^\zeta} ||\hat{g}(\theta)||_\infty \leq \sup_{\theta \in \Theta_{0,n}} ||\hat{g}(\theta)||_\infty = O_p(b_n)$, 
$$\sup_{(\theta,\eta) \in S_{0,n}^\zeta} ||\partial_\delta \eta g(\theta)||_\infty \leq \sup_{(\theta,\eta) \in S_{0,n}^\zeta} ||\eta||_1 ||\partial_\delta \hat{g}(\theta) - \partial_\delta g(\theta)||_\infty \ = O(1) O_p(b_n) = O_p(b_n),$$ 
and $ b_n/\min \{ \lambda_\delta, \lambda_\eta \} \overset{p}{\rightarrow} 0$, we have that 

$$\sup_{(\theta,\eta) \in S_{0,n}^\zeta} ||\hat{g}(\theta)||_\infty < \lambda_\delta \text{ and } \sup_{(\theta,\eta) \in S_{0,n}^\zeta} ||\partial_\delta \eta \hat{g}(\theta) ||_\infty < \lambda_\eta$$ 

wpa1. Hence, $S_{0,n}^\zeta \subset \hat{S}_0^\zeta$ wpa1. Therefore, 
$$d_H(S_{0,n}^\zeta , \hat{S}_0^\zeta,||\cdot||_{\mathcal{E}}) = O_p( \max \{a_n, \lambda_\delta,\lambda_\eta\}).$$

\textbf{Lemma A2 (Applying a Functional Central Limit Theorem)}\phantomsection\label{LA2} Suppose $\{ \phi_k(x)\}_{k=0}^K$ satisfies Assumption 3.2.2 and there is a stochastic process $\{X_i \}_{i \in \mathcal{N}}$ with $E[X_i] = 0$, $E[X_i^2] < \infty$ for all $i \in \mathbb{N}$, and $E[(\sum_{i=1}^n X_i)^2/n] \rightarrow \sigma^2$ for some $\sigma^2 > 0$. Further suppose that the sequence is $\alpha$-mixing with mixing coefficients $\alpha(k)$ and there exists $\gamma > 2$ such that $\sup_{i \in \mathbb{N}} E[|X_i|^\gamma] < \infty$ and $\sum_{k=1}^\infty \alpha(k)^{1- 2/\gamma} < \infty$. Then

$$\sigma^{-1/2} \sum_{i=1}^n \phi_k(\frac{i}{n}) X_i/\sqrt{n} \overset{d}{\rightarrow} \xi_k, \text{ jointly for }k \in \{0,1,...,K\}\text{ with }\xi_k \sim iid N(0,1).$$

\textbf{Proof:} The conditions of Theorem 0 of \cite{Herrndorf1985} are satisfied, which provides a Functional Central Limit Theorem for partial sums for $\alpha$-mixing processes. Therefore, the partial sums function $\hat{B}(r) = \sum_{i=1}^{\lfloor rn \rfloor} X_i/(\sigma \sqrt{n})$ converges weakly to the standard Wiener measure. As pointed out by \cite{Phillips_2005} (see page 119), when Assumption 3.2.2 holds and $\hat{B}(r)$ is converging weakly to the standard Wiener measure, then standard functional limit arguments and Wiener integration show that

$$\sigma^{-1/2} \sum_{i=1}^n \phi_k(\frac{i}{n}) X_i/\sqrt{n} \overset{d}{\rightarrow} \int_0^1 \phi_k(r)dB(r) = \xi_k,$$

where $\xi_k \sim N(0, 1)$, jointly for $k \in \{0,1,...,K\}$. Then, due to the orthogonality property of the basis functions, these $\xi_k$ are uncorrelated and since they are jointly normal, they are independent.

\textbf{Lemma A3 (Partial Sums Adaptivity Condition)} Let $t = (t_1,...,t_Q)$ be $Q$-dimensional vector of time periods indices with $1_Q \leq t \leq T$, where $1_Q $ is a $Q$-dimensional vector of ones and $T = (T_{b_1},...,T_{b_Q})$. Then we can denote the $m$-dimensional vector of orthogonalized moment conditions using partial sums as 
$$ p(\theta,\eta,t) = \eta \begin{pmatrix}
    \sum_{i \in \mathcal{T}_{b_1}^{t_1}} g_{1,i}(\theta)/T_{b_1} \\
    ... \\
     \sum_{i \in \mathcal{T}_{b_Q}^{t_Q}} g_{Q,i}(\theta)/T_{Q_1}
\end{pmatrix}.$$
 Suppose $(\beta_{0,n},\delta_{0,n}) \in  \Theta_{0,n}$, $\eta_{0,n}$ satisfies the orthogonality condition, and Assumptions 2.1 and 2.1* hold. Then, uniformly over $t$ with $1_Q \leq t \leq T$,  
$$\sqrt{n}(p(\beta_{0,n},\hat{\delta},\hat{\eta},t) - p(\beta_{0,n}, \delta_{0,n},\eta_{0,n},t)) = o_p(1).$$

\onehalfspacing\noindent \textbf{Proof:}

Let $\gamma = (\delta,\eta_l)$ and $\hat{\gamma} = (\hat{\delta},\hat{\eta}_l)$ where $\eta_l$ is the $l$-th row of $\eta$. Since the $l$-th element $p_l(\theta,\eta,t)$ of $p(\theta,\eta,t)$ for $l \in \{1,...,m\}$ is twice continuously differentiable in $\gamma$, for each $l$ there exists $\Bar{\gamma}_t$ for each vector $t$ with $1_Q \leq t \leq T$ and $||\Bar{\gamma}_t - \gamma_0||_{\mathcal{E}} \leq ||\hat{\gamma} - \gamma_0||_{\mathcal{E}}$ such that:

$$\sqrt{n}(p_l(\beta_{0,n},\hat{\gamma},t) - p_l(\beta_{0,n},\gamma_0,t) = \sqrt{n}\partial_\gamma p_l(\beta_{0,n},\gamma_0,t) (\hat{\gamma} - \gamma_0)$$
$$ + \sqrt{n}(\hat{\gamma} - \gamma_0)'\partial^2_\gamma p_l(\beta_{0,n},\Bar{\gamma}_t,t)(\hat{\gamma} - \gamma_0). $$
The magnitude of the first term on the right-hand side is less than or equal to 

$$\sqrt{n} ||\partial_\gamma p_l(\beta_{0,n},\gamma_0,t)||_{\mathcal{D}} ||\hat{\gamma} - \gamma_0||_{\mathcal{E}} \leq \sqrt{n} O_p(r_J/\sqrt{n})o_p(1/r_J) = o_p(1).$$
The second term on the right-hand side is equal to $$\sqrt{n}(\hat{\delta} - \delta_{0,n})'\partial_\delta^2 p_l(\beta_{0,n},\Bar{\gamma}_t,t)(\hat{\delta} - \delta_{0,n}) + 2\sqrt{n}(\hat{\eta}_l - \eta_{0,n,l})\partial_{\delta} h (\beta_{0,n},\Bar{\delta},t)(\hat{\delta} - \delta_{0,n}),$$
where 
$$ h(\theta,t) = \begin{pmatrix}
    \sum_{i = \mathcal{T}_{b_1}^{t_1}} g_{1,i}(\theta)/T_{b_1} \\
    ... \\
     \sum_{i = \mathcal{T}_{b_Q}^{t_Q}} g_{Q,i}(\theta)/T_{Q_1}
\end{pmatrix}.$$
In the linear case, $\sqrt{n}(\hat{\delta} - \delta_{0,n})'\partial_\delta^2 p_l(\beta_{0,n},\Bar{\gamma}_t,t)(\hat{\delta} - \delta_{0,n})  = 0$ and $$\partial_\delta h(\beta_{0,n},\Bar{\delta},t) = \partial_\delta h (\theta_{0,n},t) = \partial_\delta h(\hat{\theta},t).$$ Therefore, 
$$|(\hat{\eta}_l - \eta_{0,n,l})\partial_\delta  h(\beta_{0,n},\Bar{\delta},t)(\hat{\delta} - \delta_{0,n})| = $$
$$|\hat{\eta}_l \partial_\delta h(\hat{\theta},t)(\hat{\delta} - \delta_{0,n}) - \eta_{0,n,l} \partial_\delta  h(\theta_{0,n},t)(\hat{\delta} - \delta_{0,n})| \leq$$
$$(||\hat{\eta}_l \partial_\delta  h(\hat{\theta},t)||_{\mathcal{D}} + ||\eta_{0,n,l} \partial_\delta h(\theta_{0,n},t)||_{\mathcal{D}}) ||\hat{\delta} - \delta_{0,n}||_{\mathcal{E}} $$
$$= (||\hat{\eta}_l \partial_\delta  h(\hat{\theta},t)||_{\mathcal{D}} + ||\partial_\delta p_l(\theta_{0,n},\eta_{0,n},t)||_{\mathcal{D}}) ||\hat{\delta} - \delta_{0,n}||_{\mathcal{E}}$$
$$= (O_p(r_J\log(n)/\sqrt{n}) + O_p(r_J/\sqrt{n}))o_p(1/(r_J\log(n))) = o_p(1/\sqrt{n}).$$

Otherwise, for the non-linear case, there exists $\Bar{\delta}_t^*$ for each $t$ with $1 \leq t \leq n$, such that 
$||\Bar{\delta}_t^* - \delta_{0,n}||_{\mathcal{E}} \leq ||\hat{\delta} - \delta_{0,n}||_{\mathcal{E}}$ and 

$$\partial_\delta h(\beta_{0,n},\Bar{\delta},t) = \partial_\delta  h(\theta_{0,n},t) + (\Bar{\delta} - \delta_{0,n})' \partial_\delta^2  h(\beta_{0,n},\Bar{\delta}_t^*,t).$$
Therefore, using Assumptions 2.1 and 2.1*,
$$\sqrt{n}|(\hat{\eta}_q - \eta_{0,n,q})\partial_\delta h(\beta_{0,n}, \Bar{\delta},t)(\hat{\delta} - \delta_{0,n})| \leq $$
$$\sqrt{n}||\hat{\eta}_q - \eta_{0,n,q}||_{\mathcal{E}}( ||\partial_\delta h(\theta_{0,n},t)||_{\mathcal{D}} ||\hat{\delta} - \delta_{0,n}||_{\mathcal{E}}$$
$$+   \max_{s \in \{1,...,Q\}}|(\Bar{\delta} - \delta_{0,n})'\partial_\delta^2 h_s (\beta_{0,n},\Bar{\delta}_t^*,t)(\hat{\delta} - \delta_{0,n})|)$$
$$\leq \sqrt{n} o_p(n^{-1/4}/\sqrt{r_J}) O_p(r_J)o_p(n^{-1/4}/\sqrt{r_J}) $$
$$+ \sqrt{n}o_p(n^{-1/4}/\sqrt{r_J}) ||\hat{\delta} - \delta_{0,n}||_{\mathcal{E}}^2 \max_{s\in \{1,...,Q\}} \max eig(\partial_\delta^2 h_s (\beta_{0,n},\Bar{\delta}_t^*,t)).$$

Then, using $\epsilon > 0$ defined by Assumption 2.1.4, because $||\Bar{\delta}_t^* - \delta_{0,n}||_{\mathcal{E}} \leq ||\hat{\delta} - \delta_{0,n}||_{\mathcal{E}} + ||\hat{\eta}_l - \eta_{0,n,l}||_{\mathcal{E}} < \epsilon$ wpa1, then wpa1 

$$\max_{s\in \{1,...,Q\}} \max eig(\partial_\delta^2 h_s(\beta_{0,n},\Bar{\delta}_t^*,t)) $$
$$\leq \max_{s \in \{1,...,Q\}} \sup_{\delta: ||\delta - \delta_{0,n}||_{\mathcal{E}} < \epsilon} \max eig(\partial_\delta^2  h_s(\beta_{0,n},\delta,t)) = O_p(r_J) \text{ and }$$
$$\max eig(\partial_\gamma^2 p_l(\beta_{0,n},\Bar{\gamma}_t,t)) $$
$$\leq \sup_{\gamma: ||\gamma-\gamma_0||_{\mathcal{E}} < \epsilon} \max eig(\partial_\gamma^2 p_l(\beta_{0,n},\gamma,t)) = O_p(r_J).$$
Hence,
$$\sqrt{n}|(\hat{\eta}_l - \eta_{0,n,l})\partial_\delta  h(\beta_{0,n}, \Bar{\delta},t)(\hat{\delta} - \delta_{0,n})| \leq $$
$$\leq \sqrt{n} o_p(n^{-1/4}/\sqrt{r_J}) O_p(r_J)o_p(n^{-1/4}/\sqrt{r_J}) $$
$$+ \sqrt{n}o_p(n^{-1/4}/\sqrt{r_J}) o_p(n^{-1/2}/r_J)O_p(r_J) = o_p(1).$$
Also,
$$|\sqrt{n}(\hat{\delta} - \delta_{0,n})'\partial_\delta^2 p_l(\beta_{0,n},\Bar{\gamma}_t,t)(\hat{\delta} - \delta_{0,n})/n| $$
$$\leq \sqrt{n}||\hat{\delta} - \delta_{0,n}||_{\mathcal{E}}^2 \max eig(\partial_\delta^2 p_l (\beta_{0,n},\Bar{\gamma}_t,t))$$
$$\leq \sqrt{n}O_p(r_J) o_p(n^{-1/2}/r_J) = o_p(1).$$
Therefore, uniformly over $t$ with $1_Q \leq t \leq T$,
$$\sqrt{n}(p_l(\beta_{0,n},\hat{\gamma},t) - p_l(\beta_{0,n},\gamma_0,t)) = o_p(1),$$
for each $l \in \{1,...,m\}$.

\end{document}